\documentclass[twocolumn,letterpaper,aps,prd,superscriptaddress,showpacs,nofootinbib,floatfix]{revtex4-1}

\usepackage{graphicx}
\usepackage{mathtools}
\usepackage{multirow}
\usepackage{makecell}
\usepackage[T1]{fontenc}
\usepackage{tikz}
\usepackage{adjustbox}
\usepackage[caption=false]{subfig}
\usepackage{hyperref}
\usepackage[utf8]{inputenc}
\usepackage{xcolor}

\newcommand{\mage}{\textsc{MaGe}}

\begin{document}

\title{Search for Double-Beta Decay of $^{76}$Ge to Excited States of $^{76}$Se with the \bfseries{\scshape{Majorana Demonstrator}}}

\newcommand{\ITEP}{National Research Center ``Kurchatov Institute'' Institute for Theoretical and Experimental Physics, Moscow, 117218 Russia}
\newcommand{\JINR}{Joint Institute for Nuclear Research, Dubna, 141980 Russia} 
\newcommand{\lbnl}{Nuclear Science Division, Lawrence Berkeley National Laboratory, Berkeley, CA 94720, USA}
\newcommand{\lbnle}{Engineering Division, Lawrence Berkeley National Laboratory, Berkeley, CA 94720, USA}
\newcommand{\lanl}{Los Alamos National Laboratory, Los Alamos, NM 87545, USA}
\newcommand{\queens}{Department of Physics, Engineering Physics and Astronomy, Queen's University, Kingston, ON K7L 3N6, Canada}
\newcommand{\uw}{Center for Experimental Nuclear Physics and Astrophysics, and Department of Physics, University of Washington, Seattle, WA 98195, USA}
\newcommand{\unc}{Department of Physics and Astronomy, University of North Carolina, Chapel Hill, NC 27514, USA}
\newcommand{\duke}{Department of Physics, Duke University, Durham, NC 27708, USA}
\newcommand{\ncsu}{Department of Physics, North Carolina State University, Raleigh, NC 27695, USA}	
\newcommand{\ornl}{Oak Ridge National Laboratory, Oak Ridge, TN 37830, USA}
\newcommand{\ou}{Research Center for Nuclear Physics, Osaka University, Ibaraki, Osaka 567-0047, Japan}
\newcommand{\pnnl}{Pacific Northwest National Laboratory, Richland, WA 99354, USA}
\newcommand{\ttu}{Tennessee Tech University, Cookeville, TN 38505, USA}
\newcommand{\sdsmt}{South Dakota School of Mines and Technology, Rapid City, SD 57701, USA}
\newcommand{\usc}{Department of Physics and Astronomy, University of South Carolina, Columbia, SC 29208, USA}
\newcommand{\usd}{Department of Physics, University of South Dakota, Vermillion, SD 57069, USA}  
\newcommand{\ut}{Department of Physics and Astronomy, University of Tennessee, Knoxville, TN 37916, USA}
\newcommand{\tunl}{Triangle Universities Nuclear Laboratory, Durham, NC 27708, USA}
\newcommand{\mpi}{Max-Planck-Institut f\"{u}r Physik, M\"{u}nchen, 80805, Germany}
\newcommand{\tum}{Physik Department and Excellence Cluster Universe, Technische Universit\"{a}t, M\"{u}nchen, 85748 Germany}
\newcommand{\williams}{Physics Department, Williams College, Williamstown, MA 01267, USA}
\newcommand{\ciemat}{Centro de Investigaciones Energ\'{e}ticas, Medioambientales y Tecnol\'{o}gicas, CIEMAT 28040, Madrid, Spain}

\affiliation{\pnnl}
\affiliation{\usc}
\affiliation{\ornl}
\affiliation{\ITEP}
\affiliation{\usd}
\affiliation{\ncsu}
\affiliation{\tunl}
\affiliation{\unc}
\affiliation{\duke}
\affiliation{\uw}
\affiliation{\lbnl}
\affiliation{\sdsmt}
\affiliation{\lanl}
\affiliation{\ciemat}
\affiliation{\ut}
\affiliation{\ou}
\affiliation{\williams}
\affiliation{\ttu}
\affiliation{\queens} 
\affiliation{\mpi}
\affiliation{\tum}
\affiliation{\JINR}

\author{I.J.~Arnquist}\affiliation{\pnnl} 
\author{F.T.~Avignone~III}\affiliation{\usc}\affiliation{\ornl}
\author{A.S.~Barabash}\affiliation{\ITEP}
\author{C.J.~Barton}\affiliation{\usd}	
\author{F.E.~Bertrand}\affiliation{\ornl}
\author{E. Blalock}\affiliation{\ncsu}\affiliation{\tunl} 
\author{B.~Bos}\affiliation{\unc}\affiliation{\tunl} 
\author{M.~Busch}\affiliation{\duke}\affiliation{\tunl}	
\author{M.~Buuck}\affiliation{\uw}\altaffiliation{SLAC National Accelerator Laboratory, Menlo Park, CA 94025, USA}  
\author{T.S.~Caldwell}\affiliation{\unc}\affiliation{\tunl}	
\author{Y-D.~Chan}\affiliation{\lbnl}
\author{C.D.~Christofferson}\affiliation{\sdsmt} 
\author{P.-H.~Chu}\affiliation{\lanl} 
\author{M.L.~Clark}\affiliation{\unc}\affiliation{\tunl} 
\author{C.~Cuesta}\affiliation{\ciemat}	
\author{J.A.~Detwiler}\affiliation{\uw}	
\author{A.~Drobizhev}\affiliation{\lbnl} 
\author{T.R.~Edwards}\affiliation{\lanl}\affiliation{\usd} 
\author{D.W.~Edwins}\affiliation{\usc} 
\author{Yu.~Efremenko}\affiliation{\ut}\affiliation{\ornl}
\author{H.~Ejiri}\affiliation{\ou}
\author{S.R.~Elliott}\affiliation{\lanl}
\author{T.~Gilliss}\affiliation{\unc}\affiliation{\tunl}\altaffiliation{Present address: Applied Physics Laboratory, Johns Hopkins University, Laurel, MD 20723, USA}  
\author{G.K.~Giovanetti}\affiliation{\williams}  
\author{M.P.~Green}\affiliation{\ncsu}\affiliation{\tunl}\affiliation{\ornl}   
\author{J.~Gruszko}\affiliation{\unc}\affiliation{\tunl} 
\author{I.S.~Guinn}\affiliation{\unc}\affiliation{\tunl} 
\author{V.E.~Guiseppe}\affiliation{\ornl}	
\author{C.R.~Haufe}\affiliation{\unc}\affiliation{\tunl}	
\author{R.~Henning}\affiliation{\unc}\affiliation{\tunl}
\author{D.~Hervas~Aguilar}\affiliation{\unc}\affiliation{\tunl} 
\author{E.W.~Hoppe}\affiliation{\pnnl}
\author{A.~Hostiuc}\affiliation{\uw} 
\author{M.F.~Kidd}\affiliation{\ttu}	
\author{I.~Kim}\affiliation{\lanl} 
\author{R.T.~Kouzes}\affiliation{\pnnl}
\author{A.M.~Lopez}\affiliation{\ut}	
\author{J.M. L\'opez-Casta\~no}\affiliation{\usd} 
\author{E.L.~Martin}\affiliation{\unc}\affiliation{\tunl}	
\author{R.D.~Martin}\affiliation{\queens}	
\author{R.~Massarczyk}\affiliation{\lanl}		
\author{S.J.~Meijer}\affiliation{\lanl}	
\author{S.~Mertens}\affiliation{\mpi}\affiliation{\tum}		
\author{J.~Myslik}\affiliation{\lbnl}		
\author{T.K.~Oli}\affiliation{\usd}  
\author{G.~Othman}\affiliation{\unc}\affiliation{\tunl} 
\author{L.S.~Paudel}\affiliation{\usd} 
\author{W.~Pettus}\affiliation{\uw}	
\author{A.W.P.~Poon}\affiliation{\lbnl}
\author{D.C.~Radford}\affiliation{\ornl}
\author{A.L.~Reine}\affiliation{\unc}\affiliation{\tunl}	
\author{K.~Rielage}\affiliation{\lanl}
\author{N.W.~Ruof}\affiliation{\uw}	
\author{B.~Sayk\i}\affiliation{\lanl} 
\author{M.J.~Stortini}\affiliation{\lanl} 
\author{D.~Tedeschi}\affiliation{\usc}		
\author{R.L.~Varner}\affiliation{\ornl}  
\author{S.~Vasilyev}\affiliation{\JINR}	
\author{J.F.~Wilkerson}\affiliation{\unc}\affiliation{\tunl}\affiliation{\ornl}    
\author{C.~Wiseman}\affiliation{\uw}		
\author{W.~Xu}\affiliation{\usd} 
\author{C.-H.~Yu}\affiliation{\ornl}
\author{B.X.~Zhu}\altaffiliation{Present address: Jet Propulsion Laboratory, California Institute of Technology, Pasadena, CA 91109, USA}\affiliation{\lanl}

\collaboration{{\sc{Majorana}} Collaboration}
\noaffiliation

%\ead{iguinn@email.unc.edu}

\begin{abstract}
  The \textsc{Majorana Demonstrator} is a neutrinoless double-beta decay search consisting of a low-background modular array of high-purity germanium detectors, $\sim2/3$ of which are enriched to 88\% in $^{76}$Ge. The experiment is also searching for double-beta decay of $^{76}$Ge to excited states (e.s.) in $^{76}$Se. $^{76}$Ge can decay into three daughter states of $^{76}$Se, with clear event signatures consisting of a $\beta\beta$-decay followed by the prompt emission of one or two $\gamma$-rays. This results with high probability in multi-detector coincidences. The granularity of the \textsc{Demonstrator} detector array enables powerful discrimination of this event signature from backgrounds. Using 41.9~kg~yr of isotopic exposure, the \textsc{Demonstrator} has set world leading limits for each e.s.\ decay of $^{76}$Ge, with 90\% CL lower half-life limits in the range of $(0.75-4.0)\times10^{24}$~yr. In particular, for the $2\nu$ transition to the first $0^+$ e.s.\ of $^{76}$Se, a lower half-life limit of $7.5\times10^{23}$~yr at 90\% CL was achieved.  
\end{abstract}

\maketitle

\section{Introduction}

\begin{figure*}[!htb]
  \centering
  \includegraphics[width=0.8\textwidth]{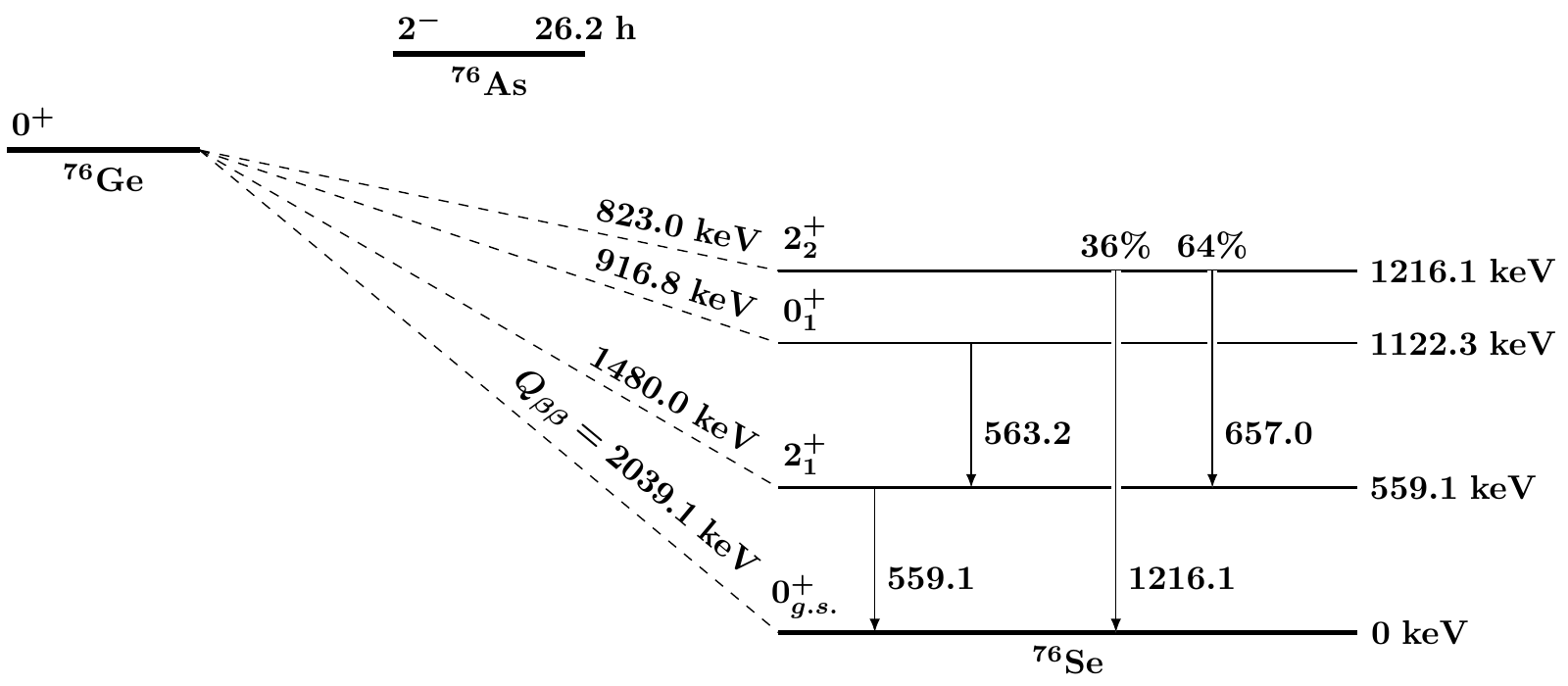}
  \caption{\label{leveldiagram}Level diagram of the $\beta\beta$-decay of $^{76}$Ge into $^{76}$Se.~{\begin{@fileswfalse} \cite{SINGH1995} \end{@fileswfalse}}}
\end{figure*}
Neutrinoless double-beta decay ($0\nu\beta\beta$) is a hypothetical lepton number violating process that, if discovered, would indicate the existence of physics beyond the Standard Model (BSM)~\cite{GoeppertMayer1935, Furry1939, Avignone2008, Vergados2012, Rodejohann2015, DelloOro2016, Dolinski2019}.
In particular, discovery would indicate that the neutrino is a Majorana fermion (i.e., its own anti-particle)~\cite{Majorana1937, Schechter1982}, and might provide a feasible mechanism for generation of the observed matter-antimatter asymmetry of the universe~\cite{Sakharov1967, FUKUGITA1986}.
For a set of BSM physics interactions generating $0\nu\beta\beta$ enumerated $i$, the half-life of $0\nu\beta\beta$ would follow
\begin{equation}
[T^{0\nu}_{1/2}]^{-1}=\big|\sum_i G^{0\nu}_i \cdot |M^{0\nu}_i|^2 \cdot \eta_i^2 \big|
\end{equation}
where $G^{0\nu}_i$ is the phase-space integral, $M^{0\nu}_i$ is the nuclear matrix element, and $\eta_i$ represents the amplitude of a general lepton number violating process.
The minimal extension to the Standard Model for providing Majorana neutrino mass adds a heavy right-handed neutrino, and generates a light mass for the Standard Model neutrino via the type~I see-saw mechanism.
Under this mechanism, the half-life would follow
\begin{equation}
[T^{0\nu}_{1/2}]^{-1}=G^{0\nu} \cdot (g^{eff, 0\nu}_{A})^4|M^{0\nu}|^2\langle m_{\beta\beta}\rangle^2
\end{equation}
where $\langle m_{\beta\beta}\rangle$ is the effective Majorana mass of the electron neutrino, and $g^{eff,0\nu}_{A}=q\cdot g_A~(g_A=1.27)$ is the axial vector coupling constant, with an empirical quenching term $q$ applied~\cite{Suhonen2017}.
In this case, a half-life measurement would provide information about the neutrino mass; while the phase-space factor can be precisely calculated~\cite{Kotila2012, Stoica2019}, an $m_{\beta\beta}$ measurement is subject to currently large uncertainties in calculations of $(g^{eff,0\nu}_A)^4|M^{0\nu}|^2$~\cite{Engel2017}.

Two-neutrino double-beta decay ($2\nu\beta\beta$) is a second-order weak process that has been directly observed in 11~isotopes, with half-lives ranging from $10^{18}-10^{24}~\mathrm{yr}$~\cite{barabash2015}.
The $2\nu\beta\beta$ half-life can be expressed as
\begin{equation}
  [T^{2\nu}_{1/2}]^{-1}=G^{2\nu} \cdot (g^{eff, 2\nu}_{A})^4|M^{2\nu}|^2
\end{equation}
Because this formula does not depend on unknown physics factors and the phase-space factor can be accurately calculated, a $2\nu\beta\beta$ half-life measurement allows direct measurement of $(g^{eff, 2\nu}_{A})^4|M^{2\nu}|^2$.
Furthermore, since nuclear matrix elements are calculated using similar techniques for $2\nu\beta\beta$ and $0\nu\beta\beta$ half-lives, such a measurement may help in evaluating $(g^{eff, 0\nu}_{A})^4|M^{0\nu}|$.

$\beta\beta$-decay, in both $0\nu$ and $2\nu$ modes, can produce daughter nuclei in either the ground state (g.s.) or an excited state (e.s.).
Transitions to an e.s.\ can be distinguished from g.s.\ transitions by a lower Q-value and the prompt emission of one or more $\gamma$ rays.
$\beta\beta$-decay transitions are allowed for transitions from parent $0^+$ g.s.\ to $0^+$ and $2^+$ states.
The half-lives of decays to excited states are heavily suppressed compared to the ground state decay.
The primary reason for this is energetic suppression in the phase space due to the reduced Q-values for decays to e.s.; in addition, decays to $2^+$ states experience further suppression due to conservation of angular momentum.

So far, the only $\beta\beta$ to e.s.\ observations have been $2\nu\beta\beta$ transitions to the first excited $0^+$ ($0^+_1$) daughter states, in two isotopes.
In $^{100}$Mo, this transition was first measured in 1995~\cite{BARABASH1995}, and the global average including additional measurements is $T^{2\nu}_{1/2}=(6.7^{+0.5}_{-0.4})\times10^{20}~\mathrm{yr}$~\cite{barabash2015}.
In $^{150}$Nd, this transition was first measured in 2004~\cite{Barabash2004, Barabash2009}, and the global average including additional measurements is $T^{2\nu}_{1/2}=(1.2^{+0.3}_{-0.2})\times10^{20}~\mathrm{yr}$~\cite{barabash2015}.
Searches have been performed in a variety of other isotopes as well~\cite{Barabash2017}.

$^{76}$Ge is a promising isotope for studying $\beta\beta$ decay, with the \textsc{Majorana Demonstrator}~\cite{mjd2014, mjd2019} and GERDA~\cite{gerda:2019, gerda:2020} experiments currently conducting sensitive searches for $0\nu\beta\beta$, and LEGEND-200 under construction~\cite{LEGEND200}.
In addition to the ground state, $^{76}$Ge can decay to three excited states of $^{76}$Se, as shown in Fig.~\ref{leveldiagram}.
Experiments have searched for these e.s.\ decay modes since 1977~~\cite{Fiorini1977}; current best limits have been set by GERDA phase~I~~\cite{gerda:ES}.
Among e.s.\ decay modes, the decay to the $0^+_1$~e.s.\ of $^{76}$Se is expected to dominate, with recent half-life predictions falling in the range $1.0\times10^{23}-7.1\times10^{24}$~yr ~\cite{Barabash2017}.

A measurement of the half-life of decay modes to various daughter states provides useful information beyond that provided from just g.s.\ measurements.
As in g.s.\ measurements, $2\nu\beta\beta$ to e.s.\ half-lives can be used to obtain direct measurements of $(g^{eff, 2\nu}_{A})^4|M^{2\nu}|^2$, providing a useful cross-check for the calculation techniques.
In addition, a measurement of the nuclear matrix element for $2\nu\beta\beta$ transitions to $2^+$ states would enable a sensitive test for a bosonic component to the neutrino wave function~\cite{Dolgov2005, Barabash2007}.
An experiment that measured $0\nu\beta\beta$ to both the ground state and $0^+_1$ state could use the ratio between these values to test the beyond the Standard Model physics mechanism generating these processes~\cite{Simkovic2002}.

\section{The MAJORANA DEMONSTRATOR}
The \textsc{Majorana Demonstrator} is studying $\beta\beta$-decay of $^{76}$Ge using high-purity germanium (HPGe) detectors.
The experiment consists of two modules, each consisting of an array of detectors operated in vacuum in separate cryostats.
Fifty-eight detectors totalling 44.1~kg are used, 29.7~kg of which are enriched to 88\% in $^{76}$Ge, allowing them to act as both the source and detector of $\beta\beta$-decay.
The remaining detectors have the natural isotopic abundance of 7.8\% $^{76}$Ge.
The HPGe detectors use the P-type Point Contact (PPC) detector geometry, which has advantages in energy resolution and sensitivity to event topology~\cite{Barbeau2007}.
The PPC geometry and the granularity of the detector array enable discrimination of single- and multi-site events~\cite{mjd:avse}.

\begin{figure}[tb]
  \centering
  \includegraphics[width=\linewidth]{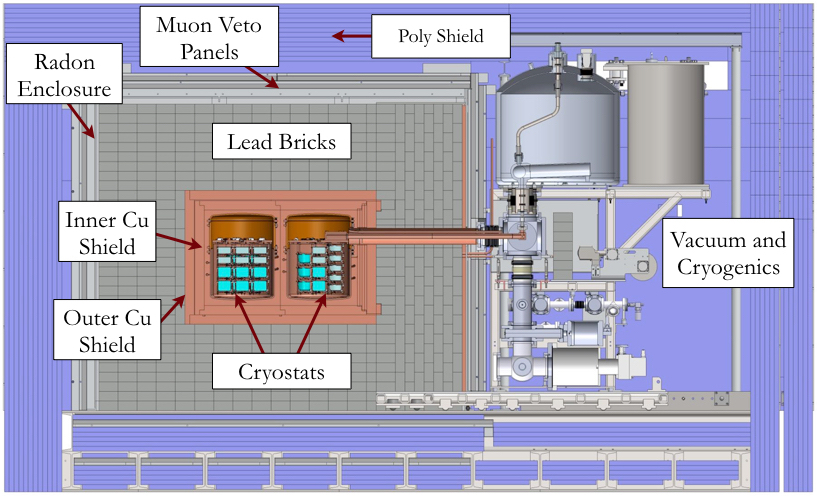}
  \caption{\label{fig:MJD}
    Diagram of the \textsc{Majorana Demonstrator} showing each shield layer, the inserted cryostats with their detector arrays, and the module hardware outside of the shielding.
  }
\end{figure}
In order to minimize backgrounds, the experiment is constructed using carefully selected low background materials~\cite{mjd:assay}.
The support material and cryostats housing the array primarily consists of underground electroformed copper (UGEFCu).
Carefully selected, commercially available materials were selected for the remaining materials, including cabling, cryostat seals, and insulating materials.
Low-background front-end electronics were developed for the experiment and are placed next to the detector contacts, enabling low noise measurements~\cite{mjd:LMFE}.
The modules are placed inside of a graded shield, with lower background shield materials used nearer to the detectors.
The inner most shield layers consist of 5~cm of UGEFCu followed by 5~cm of commercially available oxygen-free electronic (C10100) copper.
The next layer includes 45~cm of lead shielding.
These layers are contained in a stainless steel radon exclusion box that is constantly purged with liquid nitrogen boil-off gas.
The vacuum  hardware, cryogenic hardware, electronics, and calibration hardware sit outside of these layers, with a small gap carved out of the lead structure for a crossarm connecting these to the cryostats.
Surrounding this is 5~cm of borated polyethylene and 25~cm of un-borated polyethylene neutron shielding, and finally scintillating plastic veto panels surround the experiment, used to actively veto backgrounds caused by muons~\cite{mjd:muonveto}.
Each layer of shielding, with inserted cryostats, is shown in Fig.~\ref{fig:MJD}.
The experiment is housed at the 4850' level (4300 m.w.e) of the Sanford Underground Research Facility (SURF) in order to minimize exposure to cosmic ray muons.

HPGe detector waveforms are recorded by digitizers developed for the GRETINA experiment~\cite{Vetter2000}, with a sampling frequency of 100~MHz and 14~bits of resolution~\cite{Anderson2012}.
Each detector records on two channels, characterized by gains that differ by a factor of 3; the high gain channel has better signal to noise ratio and is preferred for detector hits with $<4$~MeV of energy.
Each detection channel triggers independently using an internal trapezoidal filter, with an energy threshold typically $<1$~keV.
Upon triggering, either a 20~$\mu$s waveform at the full sampling rate, or a 38~$\mu$s multi-sampled waveform with post-rising-edge using four pre-summed samples is read out and stored on disk, to be reanalyzed.
Waveforms are corrected for digitizer non-linearity~\cite{mjd:digitizer_nonlinearity} and energies are calculated using a charge-trapping corrected trapezoidal filter~\cite{mjd2019}.
This procedure produces an energy resolution for all combined detectors of $2.95$~keV at full-width half-max (FWHM) at the $^{208}$Th 2614~keV peak, leading the current generation of $\beta\beta$-decay experiments~\cite{mjd2019}.

Module~1 began operation in December~2015, and both \textsc{Demonstrator} modules have been in nearly continuous operation since August~2016.
The set of runs used for this analysis is a subset of those used in Ref.~\cite{mjd2019}, excluding runs recorded before the installation of the inner copper shield.
A statistical blinding scheme was applied to much of this data, with cycles of 93~hrs of blind data and 31~hrs of open data.
Approximately 50\% of the isotopic exposure used in this result was in blind runs.
The data is divided into eight datasets, labelled DS1, DS2, DS3, DS4, DS5a, DS5b, DS5c, and DS6a.
The datasets represent changes in the hardware and data taking configuration, summarized in Table~\ref{tab:datasets}.
A combined analysis is performed on all of these datasets.
\begin{table}
  \centering
  \caption{\label{tab:datasets} A summary of the start dates, key changes, and isotopic exposure of each data set. DS3 and DS4 were run simultaneously on separate DAQ systems, corresponding to Module~1 and Module~2, respectively.}
\begin{center}
  \begin{tabular}{lclrr}
    \hline
    \makecell{Data\\Set} & \makecell{Start\\Date} & \makecell{Data Set\\Distinction} & \makecell{Live\\Time (d)} & \makecell{Exposure\\(kg~yr)} \\
    \hline\hline
    DS1 & 12/31/15 & Inner Cu Shield added         & 74.8 & 3.11 \\
    DS2 & 5/24/16  & Pre-summing                   & 40.1 & 1.67 \\
    DS3 & 8/25/16  & M1 and M2 installed & 29.9 & 1.25 \\
    DS4 & 8/25/16  & M1 and M2 installed & 19.2 & 0.62 \\
    DS5a & 10/13/16& Integrated DAQ (noise)        & 81.6 & 6.02 \\
    DS5b & 1/27/17 & Optimized Grounding           & 39.5 & 2.92 \\
    DS5c & 3/17/17 & Blind                         & 46.2 & 3.40 \\
    DS6a & 5/11/17 & Pre-summing, blind            & 309.8& 22.95 \\
    \hline
  \end{tabular}
\end{center}

\end{table}

Detector calibration is performed using line sources that can be inserted along calibration tracks that wrap around each cryostat.
Once per week, $^{228}$Th sources are deployed into each track, one at a time, for 90~min.
In addition, in January~2019, a $^{56}$Co line source with a nominal activity of 6~kBq was deployed in each track for one week at a time.
This source emits many $\gamma$ rays with an energy of $>1.5$~MeV, which produce inherently multi-site pair production events that are useful for systematic checks.

\section{Detection Signature}
\begin{figure}[tb]
  \centering
  \includegraphics[width=\linewidth]{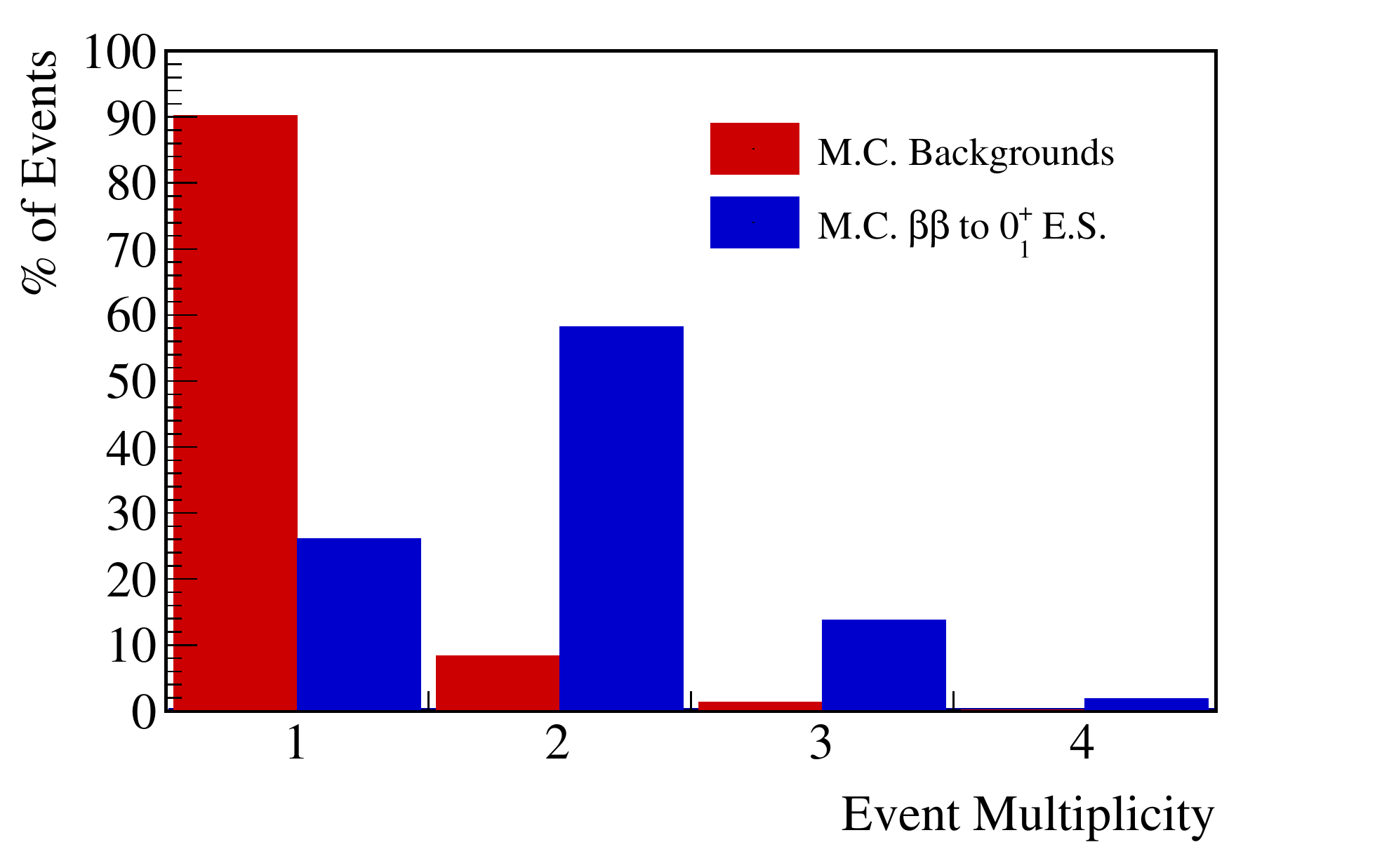}
  \caption{\label{fig:mult_hist}Comparison of fraction of events with different detector multiplicities from monte carlo simulations of backgrounds (red) and $2\nu\beta\beta$ to the $0^+_1$ e.s.\ (blue). Sensitivity can be significantly improved by focussing on events with multiplicity $>1$.}
\end{figure}
The \textsc{Majorana Demonstrator} is searching for the $\beta\beta$-decays of $^{76}$Ge to the $0^+_1$, $2^+_1$ and $2^+_2$ states of $^{76}$Se, in both $0\nu$ and $2\nu$ decay modes (for a total of 6 distinct decay modes).
The Q-value and $\gamma$-ray energies are shown in Fig.~\ref{leveldiagram}.
The $0^+_1$ e.s.\ decay mode has a Q-value of 917~keV and two $\gamma$s, with energies 559~keV and 563~keV.
Due to angular momentum conservation, the $\gamma$s are emitted with an angular correlation of~\cite{SINGH1995}
\begin{equation}
  P(\theta)=1-3\cos^2(\theta)+4\cos^4(\theta)
\end{equation}
where $\theta$ is the angle between the emitted $\gamma$s.
The $2^+_1$ e.s.\ decay mode has a Q-value of 1480~keV and a single $\gamma$, with energy 559~keV.
The $2^+_2$ e.s.\ decay mode has a Q-value of 823~keV and will release a single 1216~keV $\gamma$ 36\% of the time, or two $\gamma$s at 657~keV and 559~keV 64\% of the time.
The 657~keV and 559~keV $\gamma$s are emitted with an angular correlation of~\cite{SINGH1995}
\begin{equation}
  P(\theta)=1-1.218\cos^2(\theta)+1.1005\cos^4(\theta)
\end{equation}
These $\gamma$s are emitted promptly after the $\beta\beta$-decay, and will frequently be absorbed in an active germanium region, resulting in multi-site events.
Thus, the \textsc{Demonstrator} can significantly reduce its backgrounds by searching only for events that involve multiple detector hits, as shown in Fig.~\ref{fig:mult_hist}.
In particular, we search for events with detector multiplicity of 2 or greater, in which one detector hit falls in a peak at the energy of one of the $\gamma$s.
The data from the detector hit in coincidence with a candidate for a given peak can be used to further reduce backgrounds, as will be described in Section~\ref{sec:cuts}.
A peak-sideband analysis will ultimately be performed in Section~\ref{sec:results}, using simulations to estimate the detection efficiency for each decay mode and to study various sources of systematic error in the detection efficiency.
The most likely decay mode to be observed is the $2\nu\beta\beta$ to $0^+_1$ e.s., so the figures and values cited in this section will focus on this decay; however, the same techniques were applied for all decay modes.
\begin{figure*}[tb]
  \centering
  \includegraphics[width=0.8\linewidth]{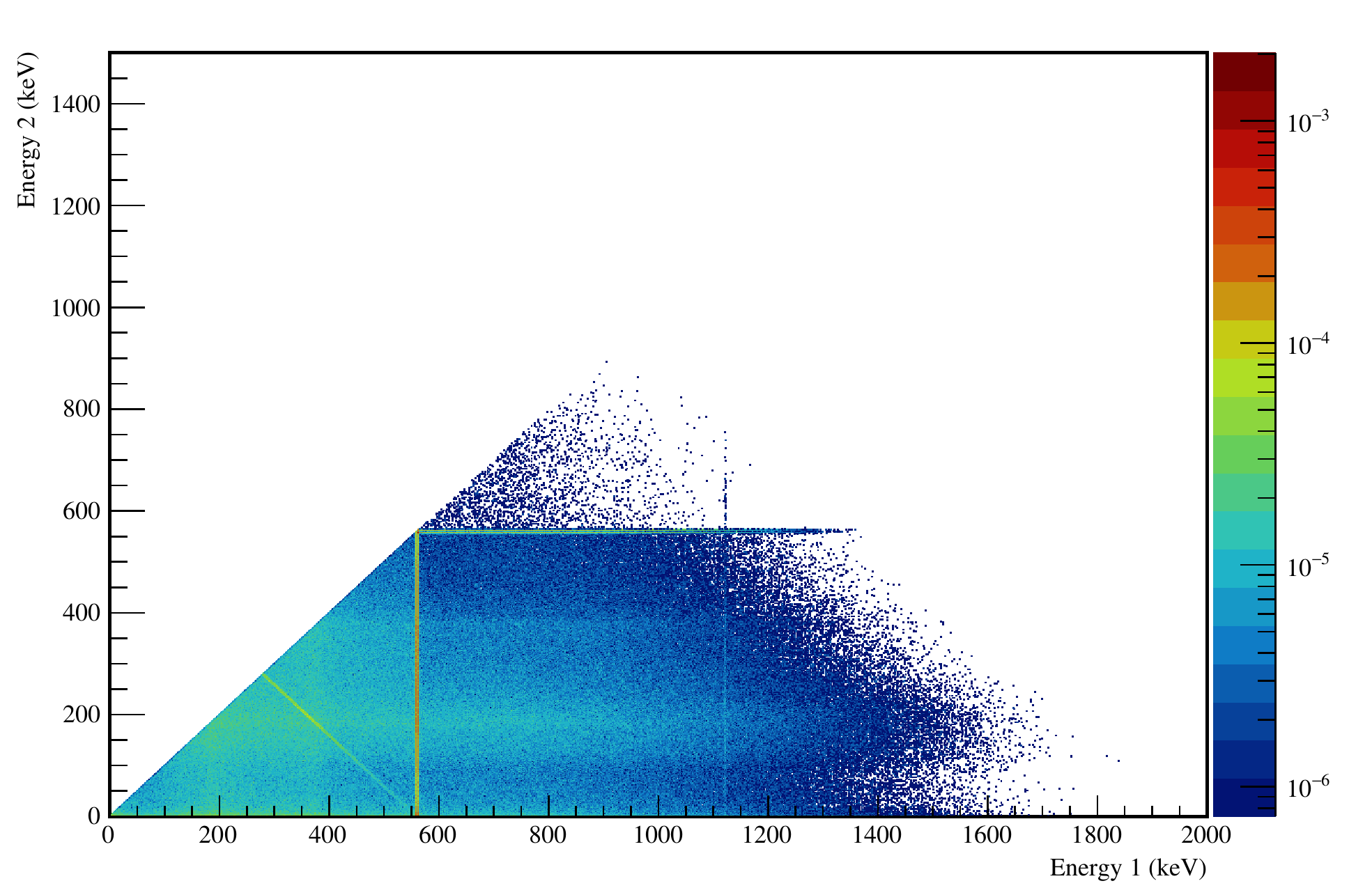}
  \caption{\label{es_sim}Multiplicity~2 energy spectrum produced by a simulation of $2\nu\beta\beta$-decay to the $0^+_1$ e.s.\ of $^{76}$Se. The vertical and horizontal lines at energies 559 and 563~keV act as a clear detection signature for the decay and are used as a region of interest for this search}
\end{figure*}

\subsection{Region of Interest Selection}
Events are selected in a signal region of interest (ROI) around the expected $\gamma$ line energies.
The ROI is determined by optimizing the discovery potential for the peak, based on a parametrized peak shape model.
The peak shape function is described by a Gaussian component with a low energy tail provided by an exponentially modified Gaussian component~\cite{mjd2019}.
The variation of the peak shape parameters with energy is measured using a simultaneous fit of 26 $^{228}$Th calibration $\gamma$ peaks between 215~keV and 2614~keV.
These peak shape parameters and optimal ROIs are measured separately for each dataset listed in Table~\ref{tab:datasets}.

These peak shapes are futher adjusted for gain drift over time and for energy nonlinearities, as described in~\cite{mjd2019}.
In addition to the factors accounted for in the $0\nu\beta\beta$ analysis, detector crosstalk must be accounted for in events involving multiple detector triggers.
The effect of crosstalk was measured by comparing the width of the 583~keV peak in multiplicity~1 events, in which no crosstalk is expected, to multiplicity~2 events, in which the 583~keV $\gamma$ is in coincidence with a $\gamma$ hit in a second detector that may induce an energy shift due crosstalk.
The observed shift in both peak center and FWHM was found to be $<0.01$~keV.

The signal ROI is then optimized for $3\sigma$-discovery potential based on the peak-shape and background index.
For the $2\nu\beta\beta$ to $0^+_1$ e.s.\ decay mode, the ROIs for the 559 and 563~keV $\gamma$s were $1.6-2$~keV wide, depending on the dataset (due to increased noise, DS5a has a wider ROI).
The ROI peak containment efficiency was estimated to be $87-89\%$.

For each decay mode, a background region of interest (BG ROI) was selected to estimate a background index.
The total width of the BG ROIs varied from $50-100$~keV depending on the $\gamma$ energy.
These BG ROIs were asymmetric on either side of the peak, and included discontinuities to exclude $>99.9\%$ of the peak shape of known background peaks.
The signal and BG ROIs selected for each $\gamma$ peak can be seen in Fig.~\ref{fig:alldata_roi}.

\subsection{Simulation of $\beta\beta$-decay to e.s.} \label{sec:sims}
\mage~\cite{mage2011}, a \textsc{Geant4}~\cite{geant2003} based simulation package containing a detailed simulated geometry of the \textsc{Majorana Demonstrator}, was used to simulate $^{76}$Ge $\beta\beta$-decay events and background events.
The $\beta\beta$-decay event generator DECAY0~\cite{Ponkratenko2000} was used in combination with \mage\ to produce simulations of each e.s.\ decay mode for $^{76}$Ge, with several modifications.
First, DECAY0 was modified for this analysis to include angular correlations in the $2^+_2$ e.s.\ decay mode (the other angular correlation for the $0^+_1$ e.s.\ was already included).
Second, the precision of the $\gamma$ energies was increased from 559 to 559.101~keV, from 563 to 563.178~keV, from 657 to 657.041~keV, and from 1216 to 1216.104~keV~\cite{SINGH1995}.
Finally, the seeding for the RANLUX random number generator (RNG) was updated.
Previously, a fixed RNG seed was supplied, but a large number of numbers were thrown out to achieve independance; instead, for these simulations an RNG seed based on the job number was supplied.
Simulations of each excited state mode were produced, with 5,000,000 events in enriched detectors and 213,993 events in natural detectors, in proportion with the fraction of isotopic mass in each detector group.
The multiplicity~2 events from these simulations are shown in Fig.~\ref{es_sim}.

For each decay mode, multiple sets of simulations were produced for systematic studies.
Energy depositions in the lithiated dead layers that extend $\sim1.1$~mm from the surfaces of the crystals will be observed with degraded charge collection, impacting detection efficiency.
The dead layer thicknesses for each detector were measured by the vendor prior to insertion into a module, and using the weekly $^{228}$Th calibration run data; a combination of these measurements is used by \mage. 
Simulations were produced with and without application of dead layers, and the fractional uncertainty from the dead layer measurements was applied to the difference in detection efficiency between both sets of simulations in order to estimate the systematic error from the dead layer thickness.

\begin{figure}
  \centering
  \includegraphics[width=\linewidth]{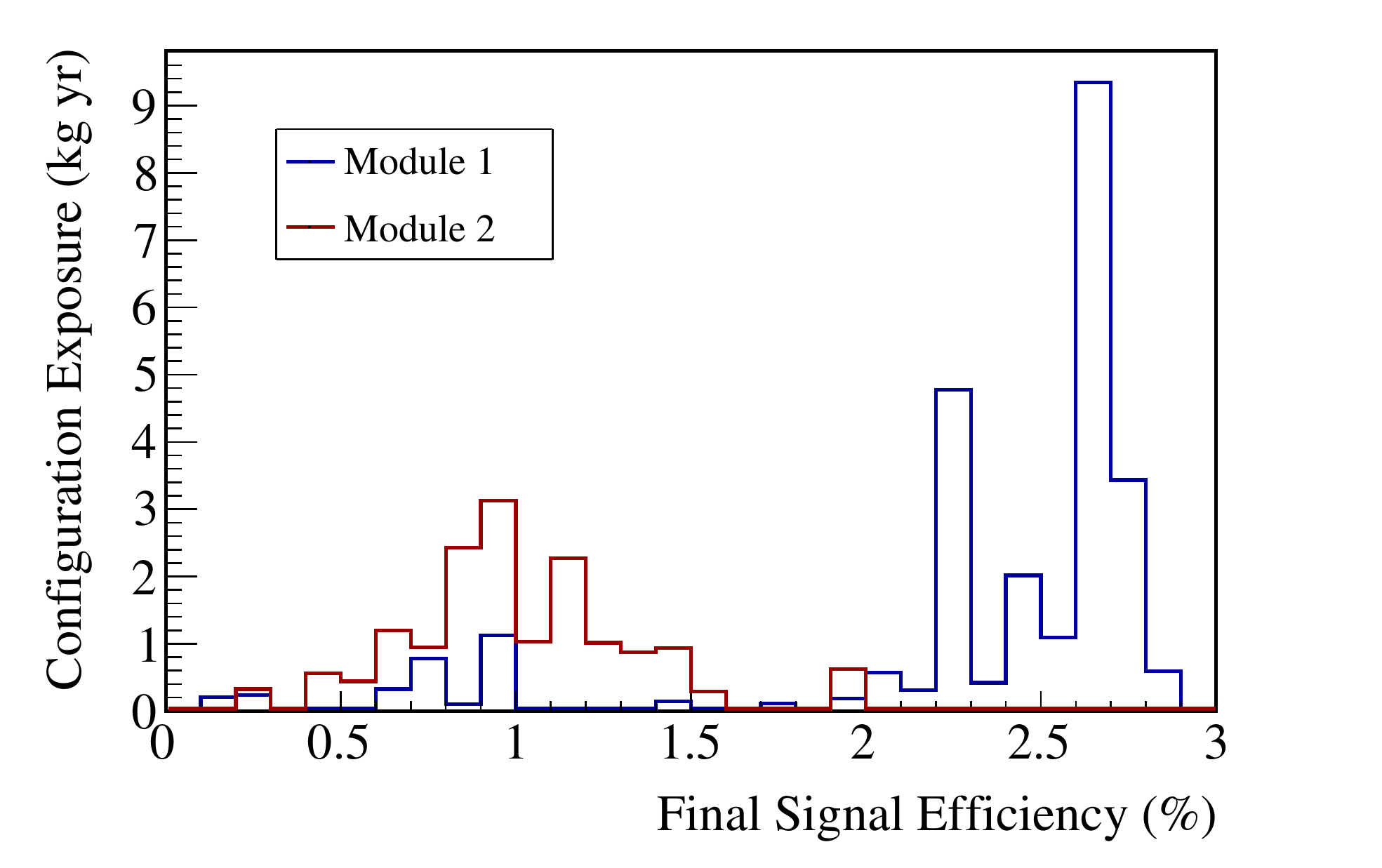}
  \caption{\label{fig:sdseff}
    The datasets were further divided into 80 sub-datasets based on which detectors were active, and the isotopic exposure and detection efficiency for each $\beta\beta$ to e.s.\ mode were separately calculated for each one. Shown above is the distribution of detection efficiencies for the decay to the $0^+_1$ e.s.
  }
\end{figure}

Dead time was included in two different ways.
First, dead time from random, uncorrelated sources such as retrigger dead time, was measured using pulsers injected into the front end electronics for each detector.
Additional simulations were produced in which detector hits were dropped with probability equal to the dead fraction in each detector; similarly to the dead layers, the fractional uncertainty from the dead fraction measurements were applied to the differences between these simulations to determine the systematic error.
In addition, many detectors were disabled for periods of time; as a result, each dataset is further divided into sub-datasets corresponding to the set of detectors enabled.
Sets of simulations were produced for each sub-dataset, and the exposure-weighted average of the simulated detection efficiency was used to obtain a limit.
The detection efficiency for various sub-datasets can vary by a factor of $>2$ from the mean in some sub-datasets, as shown in Fig.~\ref{fig:sdseff}; this approach entirely corrects for this variation.

\subsection{Background cuts} \label{sec:cuts}
To further improve detection sensitivity, a variety of background cuts are used.
Many of these cuts utilize information from the detectors in coincidence with a $\beta\beta$-decay to e.s.\ candidate.
\begin{itemize}
\item The same run selection and data cleaning cuts applied in~\cite{mjd2018, mjd2019} were applied here, excluding runs taken prior to the installation of the inner copper shield. Data cleaning routines remove waveforms caused by non-physical processes such as transient noise, and events that cannot be reliably analyzed such as pileup and saturated waveforms. The combined effect of data cleaning cuts is to remove $<0.1\%$ of physical background events.
\item As already dicussed, events with a detector multiplicity of one are removed. For e.s.\ decay modes with a single $\gamma$, events with detector multiplicity $>2$ are also cut. Detector multiplicity is determined by grouping together waveforms in a 4~$\mu$s rolling window into events. This window is conservative, as $<0.1$\% of true coincidence events are expected to trigger $>1~\mu$s apart. Events were divided based on which module they occured in; events with energy deposition in both modules were cut. This enables an independent analysis to be performed for each module.
\item Events associated with cosmic ray muons are cut by vetoing events near in time to triggers of the muon veto system. The muon veto consists of scintillating panels with $4\pi$ coverage of the modules, and triggers when at least two panels on different surfaces simultaneously surpass an energy threshold. Events are cut within 20~ms before and 1~s after a muon event; this window is expected to remove $>99.9\%$ of muon-associated events. The effect of this cut on the $\beta\beta$ decay to e.s.\ half-life measurement is evaluated by subtracting the veto time windows from the exposure, rather than by simulating its effect on the detection efficiency. This cut removes $<0.1\%$ of livetime.
\item Events are cut in which no hit in coincidence with a hit in the BG or signal ROI is enriched in $^{76}$Ge. One of the coincident detectors is assumed to contain the site of the $\beta\beta$-decay; since $\sim95\%$ of the isotopic mass of $^{76}$Ge is contained in enriched detectors, this cut sacrifices $<5\%$ of $\beta\beta$ to e.s.\ events, while cutting a significantly higher fraction of backgrounds due to the relatively higher fraction of total mass in these detectors, and since they were preferentially placed closer to the outside of the detector arrays.
\item Events where any coincident detector or the sum energy over all detectors have energy within a set of energy ranges are cut. The motivation for this cut is to remove background $\gamma$s with known energies that either compton scatter (for the sum energy cut) or are emitted in a $\gamma$ cascade (for the coincident energy cut). In addition, the energy spectrum produced by Compton scattering of $\gamma$ rays has finite amplitude at low energies, while the $\beta\beta$ spectral amplitude runs to 0 at low energies. For this reason, a low energy threshold is also set by this cut. Because the energy spectrum in coincident enriched and natural detectors is expected to differ due to the different isotopic abundances of $^{76}$Ge, a separate set of energy ranges is used for each. The energy ranges are selected using an optimization process described below. For $0\nu\beta\beta$ to e.s.\ decay modes with a single $\gamma$, a coincident energy window equal to the Q-value of the decay is used in lieu of the optimization process.
\end{itemize}

\begin{figure*}[tb]
  \centering
  \includegraphics[width=0.45\textwidth]{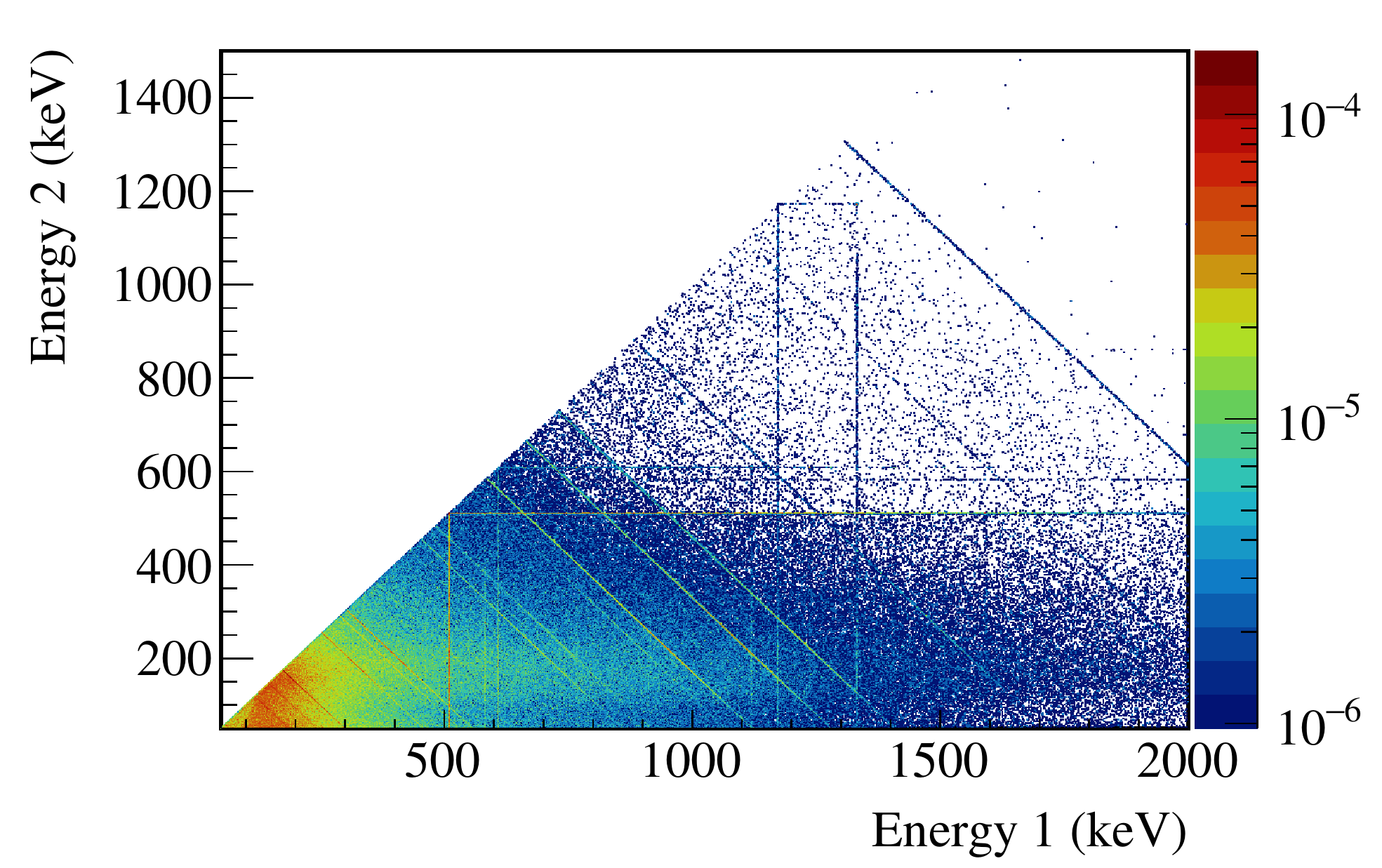}
  \includegraphics[width=0.45\textwidth]{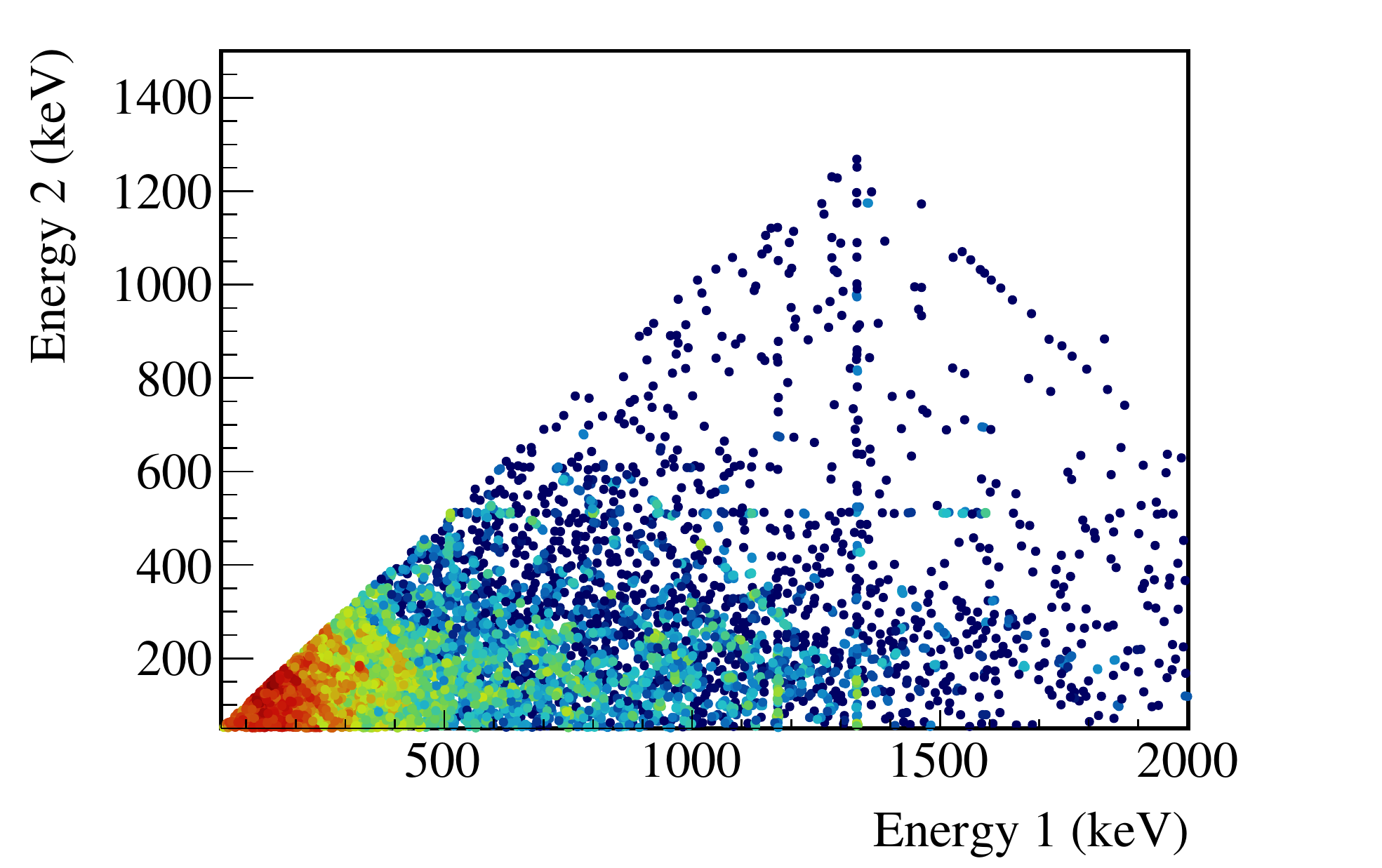}
  \caption{\label{fig:bg_sim_data} 
    Left: simulated multiplicity~2 background 2-dimensional energy spectrum histogram. Verticle and horizontal lines correspond to events in which one hit has a fixed energy, while digonal lines correspond to events with a fixed sum energy.
    Right: Scatter plot of all multiplicity~2 events, including all events in datasets used in this analysis. A Gaussian kernel density estimate with width 5~keV was used to achieve a similar color scale to the simulated spectrum.}
\end{figure*}
A simulation of the background spectrum measured by the \textsc{Demonstrator} was used to optimize sensitivity to $\beta\beta$-decay to e.s.\ using these cuts.
\mage\ simulations of known backgrounds from a variety of decay chains generated in the physical components of the experiment as defined within \mage\ were used.
The activities of the components included were determined using an unpublished fit of a linear combination of background spectra to data that is part of a background study that is still in progress; the early model used is known to be incomplete~\cite{buuck2019}.
Fig.~\ref{fig:bg_sim_data} shows the 2-dimensional energy spectrum for events with multiplicity~$>2$, along with a comparison to data.
This background model includes as background sources a limited number of components with $\gamma$ rays from $^{40}$K, $^{60}$Co, $^{222}$Rn, $^{232}$Th, $^{238}$U, and $^{68}$Ge.
One missing component of this model is cosmogenically activated $^{60}$Co inside of the natural HPGe detectors, which produces multi-detector events with high likelihood, and contribute to the 1173 and 1333~keV peaks in the hit energy spectrum.
Fortunately, the accuracy of the result presented in Section~\ref{sec:results} is not impacted by deficits in the background model used, as it assumes a flat background; still, improving this background model would help in optimizing the result.

An algorithm was written that selects a set of coincident and sum energy ranges in order to optimize the discovery potential as predicted by the background model simulation.
The algorithm begins by identifying candidate events in the BG and signal ROIs in both the background model and $\beta\beta$-decay to e.s.\ decay simulations.
These events are then sorted into energy bins for each coincident hit and for the event sum energy (a single event will fall into multiple bins).
We want to cut an energy bin if doing so improves our discovery potential, meaning, for bin k:
\begin{equation} \label{eq:dp_improvement}
  \frac{DP\big(s\cdot N_{BG}\big)}{N_{sig}} < \frac{DP\big(s\cdot (N_{BG} - n_{BG,k})\big)}{N_{sig} - n_{sig,k}}
\end{equation}
DP is the $3\sigma$ discovery potential, defined as the signal strength for which we have a 50\% chance of claiming $3\sigma$ discovery, based on the Poisson counting statistics of the signal and background events; $N_{sig}$ and $N_{BG}$ are the number of counts in the simulated ROIs for the e.s.\ decay and the background model; $n_{sig,k}$ and $n_{BG,k}$ are the number of simulated e.s.\ and background counts removed by cutting events in an energy bin; $s$ is a scaling factor for the background to reflect the expected measured background counts.
The scaling factor is determined using the ratio of events in the BG ROI in unblinded data to those found in simulations, and then increased to reflect the additional exposure from blinded data; for the $\beta\beta$ to $0^+_1$ e.s.\ decay mode, $s$ had a value of 0.001.
For each bin, we will calculate the probability that Eq.~\ref{eq:dp_improvement} holds based on Poisson counting statistics for $n_{BG,k}$ and $n_{sig,k}$.
The bin with the highest probability of improving discovery potential is then added to the cut.
After this, cut events are removed from other bins, the probability of improvement in discovery probability is recalculated, and the process is repeated until no bin has a $>50\%$ chance of improving the cut.

\begin{figure*}[tb]
  \centering
  \centering
  \subfloat[\scriptsize Simulated background coincident energy spectrum]{\includegraphics[width=0.45\textwidth]{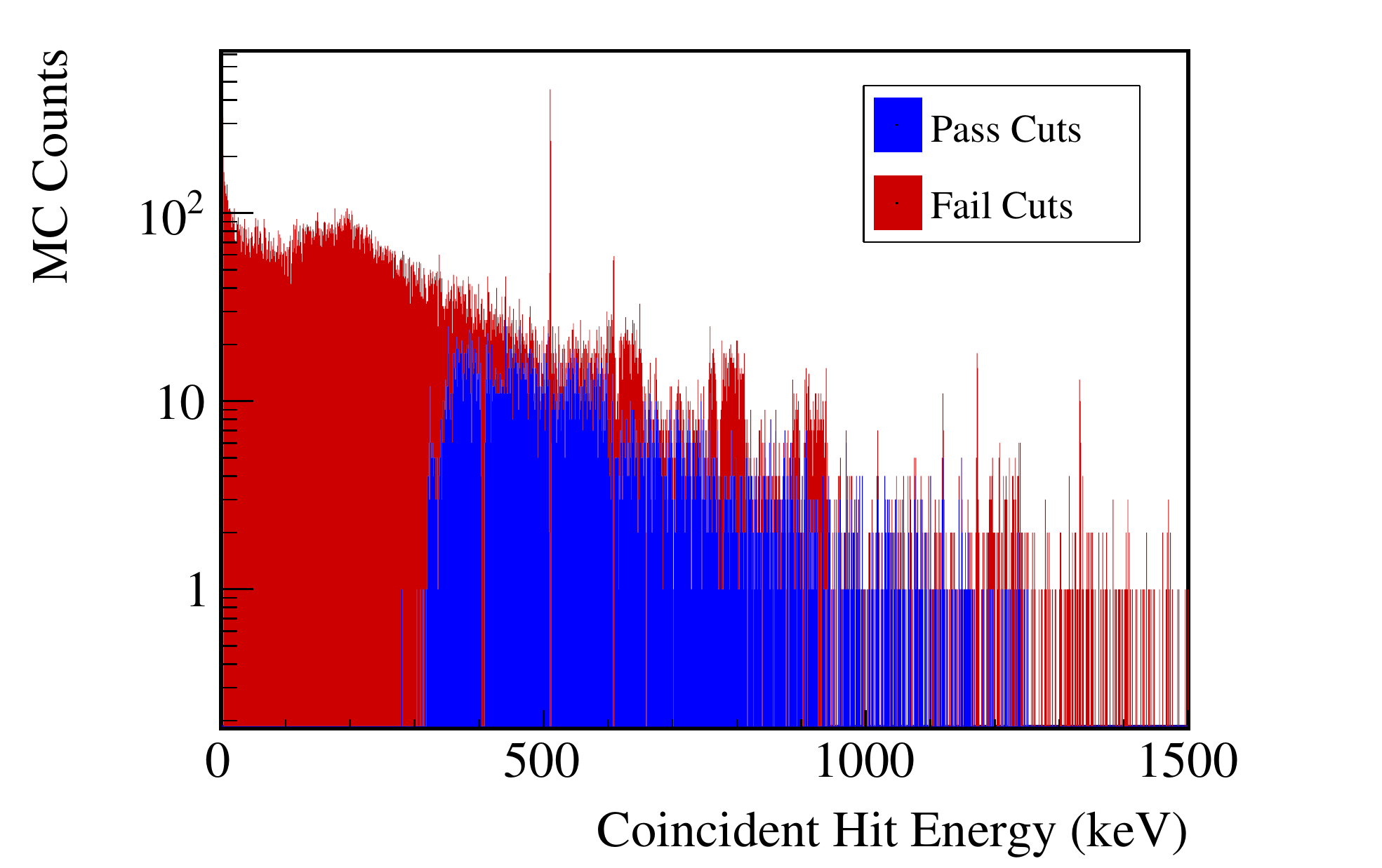}}
  \subfloat[\scriptsize Simulated background sum energy spectrum]{\includegraphics[width=0.45\textwidth]{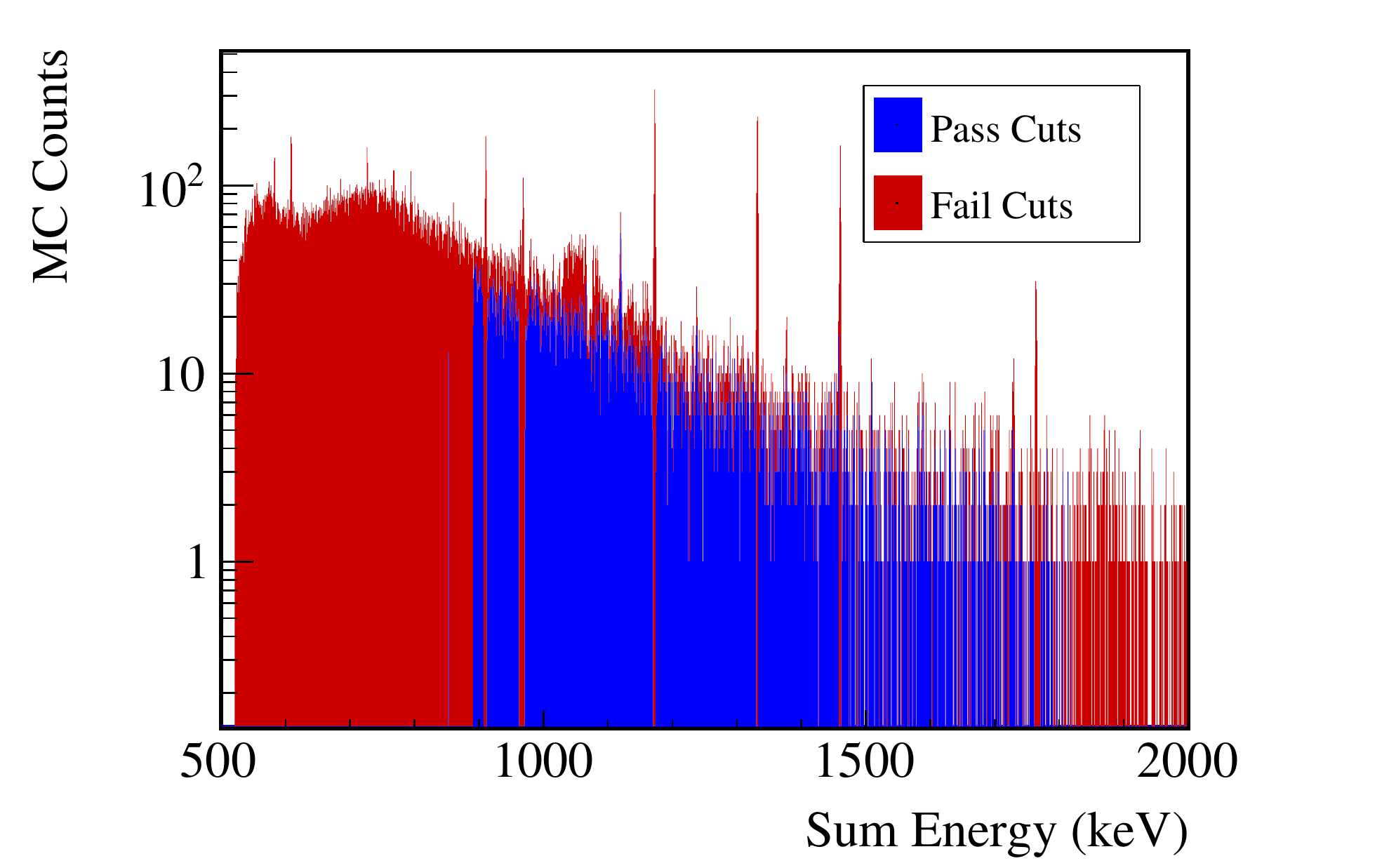}}
  \\
  \subfloat[\scriptsize Simulated $2\nu\beta\beta$ to $0^+_1$!E.S. coincident energy spectrum]{\includegraphics[width=0.45\textwidth]{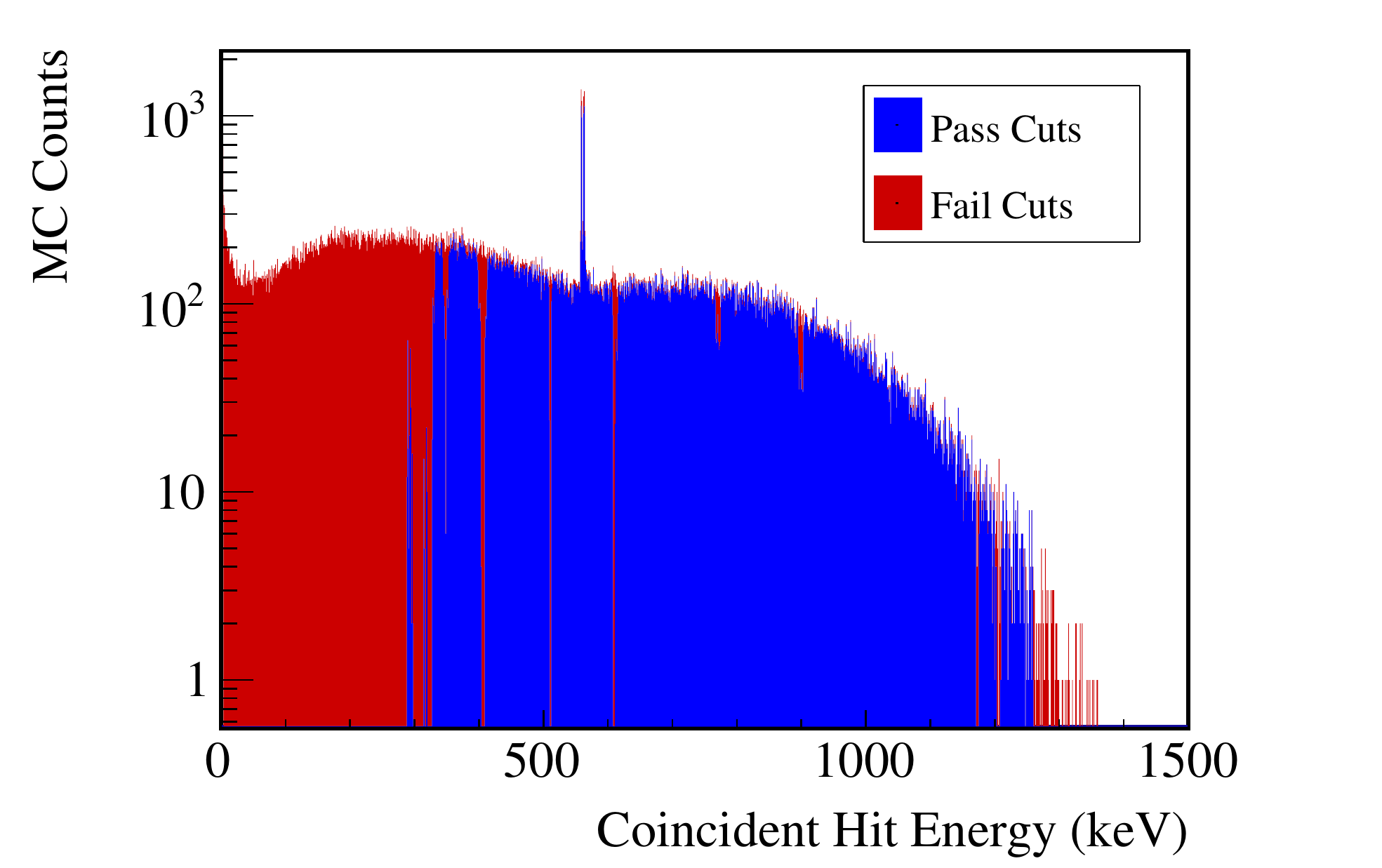}}
    \subfloat[\scriptsize Simulated $2\nu\beta\beta$ to $0^+_1$!E.S. sum energy spectrum]{\includegraphics[width=0.45\textwidth]{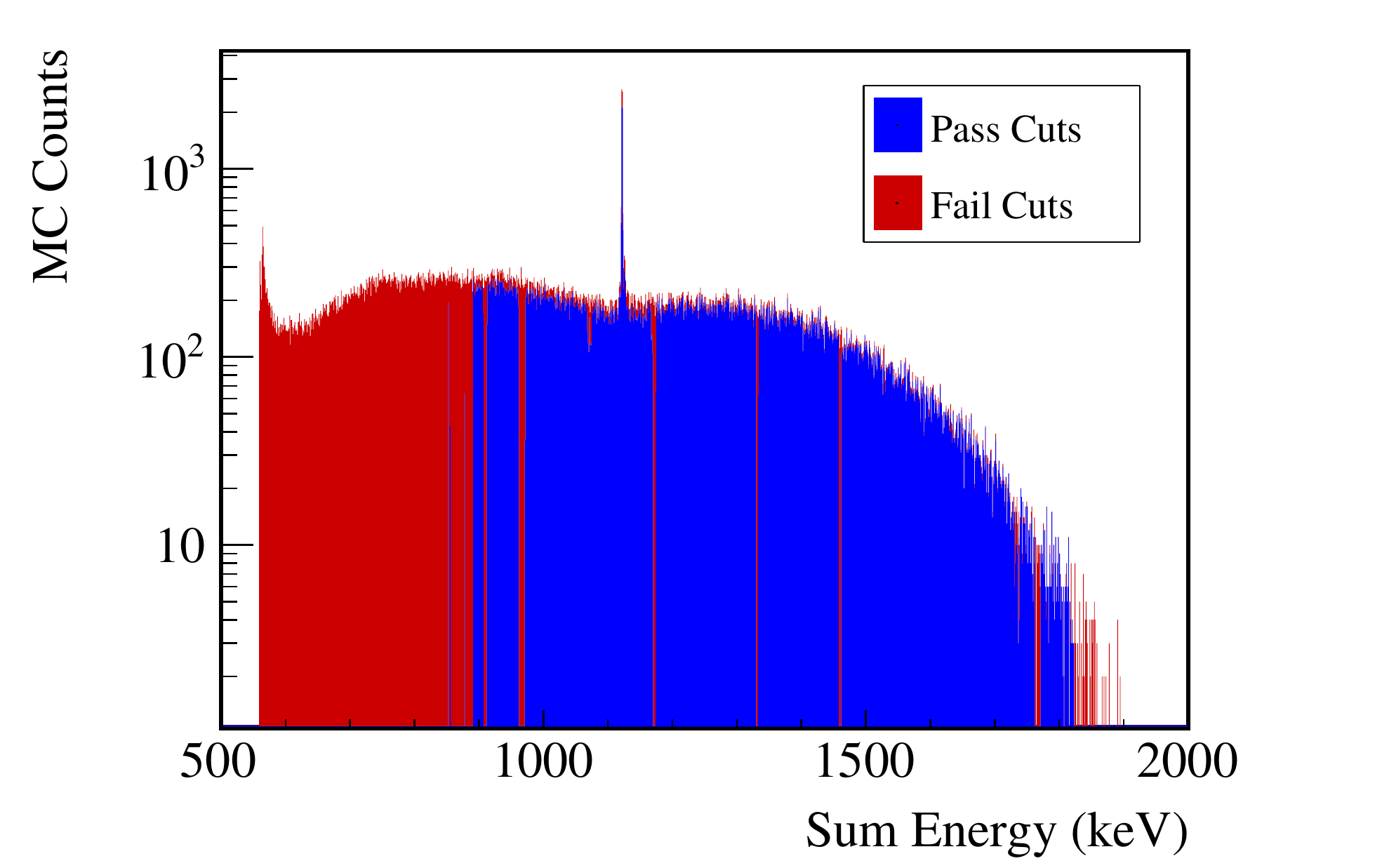}}
  \caption{\label{fig:coinandsumEcuts}
    Energy spectra demonstrating the effects of applying the sum- and coincident-energy cuts for the $2\nu\beta\beta$-decay to $0^+_1$ decay mode. All multiplicity~2 or greater events are included, with passing events in blue and cut events stacked atop these in red. Left: Energy spectrum for all BG (top) and e.s. (bottom) simulated hits in coincidence with a hit in the signal or BG ROI. Right: Sum energy spectrum for BG (top) and e.s. (bottom) simulated events including a hit in the signal or BG ROI. Fig.~\ref{fig:2DEcuts} shows similar information, including only multiplicity~2 events and plotted over 2~dimensions; this includes the cyan and green events in that figure, projected along the single-hit or sum-energies. Note in the BG spectra that narrow ranges around prominent peaks are cut, as intended.}
\end{figure*}

\begin{figure*}
  \includegraphics[width=0.8\textwidth]{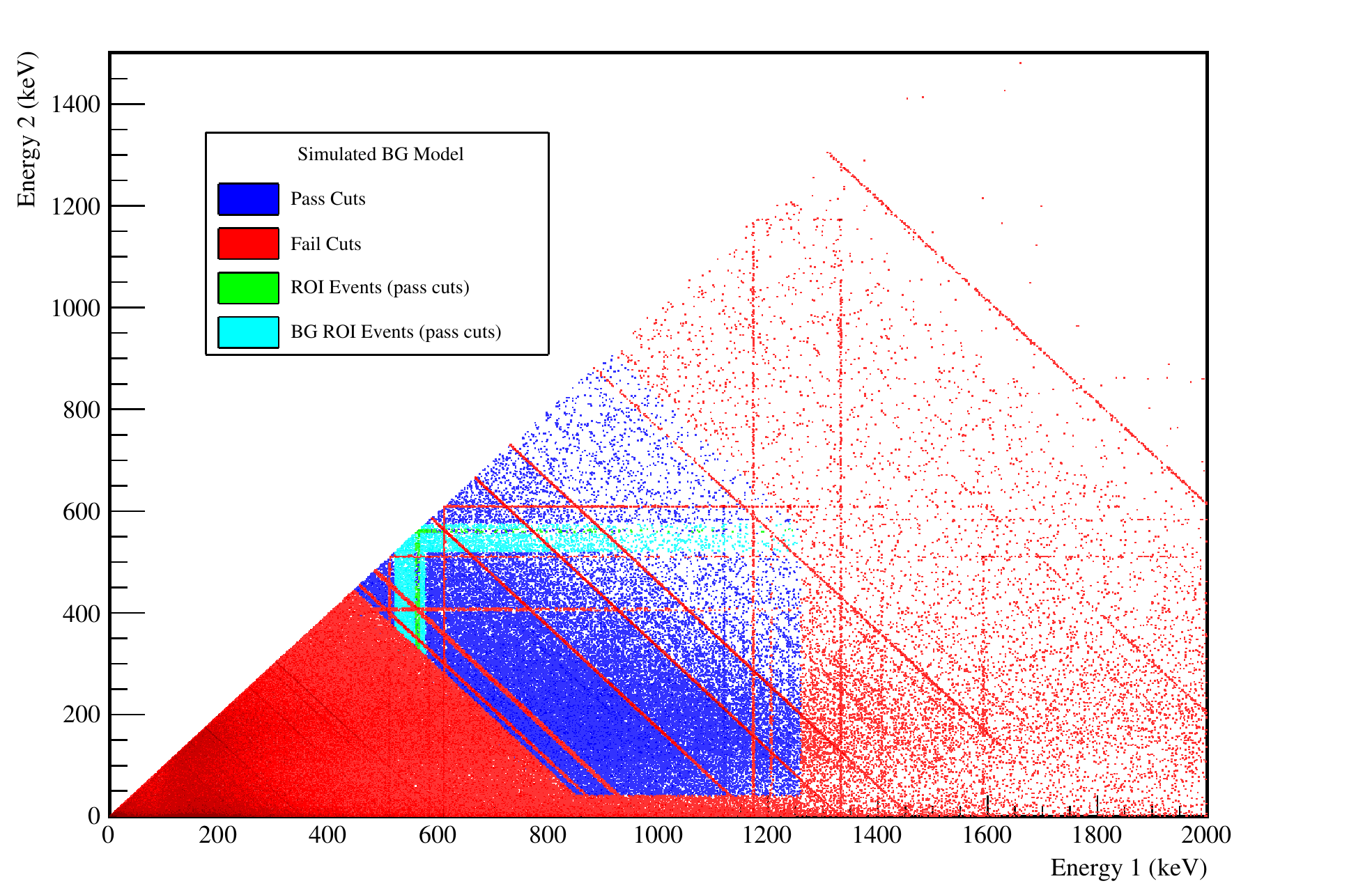} \\
  \includegraphics[width=0.8\textwidth]{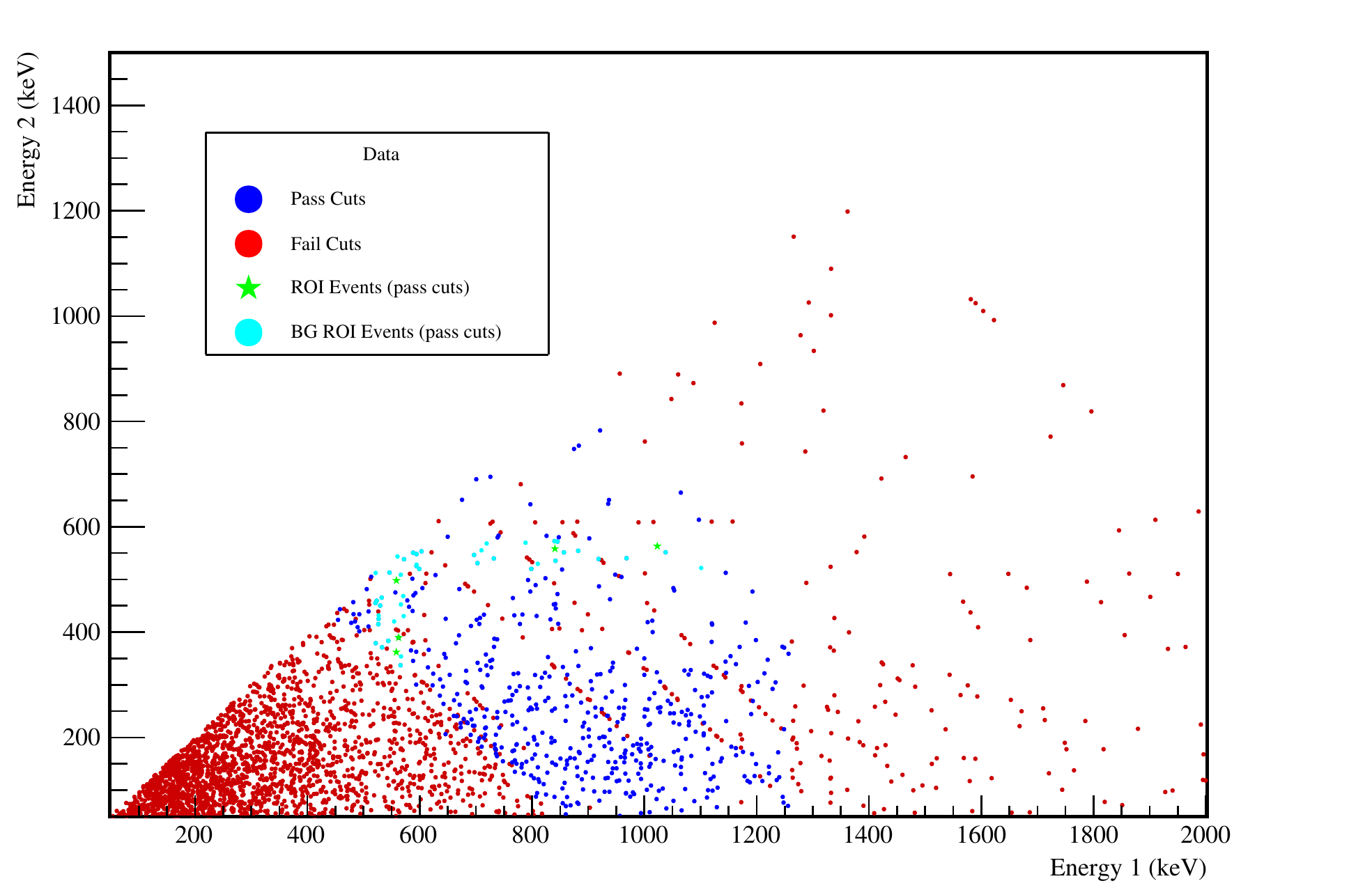}
  \caption{\label{fig:2DEcuts}
Top: Simulated multiplicity~2 energy spectrum for the background model. /edit{The colors represent cut events (red), surviving events (blue), and events that pass all cuts in the signal ROI (green) and BG ROI (cyan). Fig.~\ref{fig:coinandsumEcuts} shows similar information, with the events in the green and cyan ROIs projected onto the single hit (vertical and horizontal) and sum energy (diagonal) axes}. Bottom: scatter plot of multiplicity~2 events in data used in this analysis.}
\end{figure*}
The sampling statistics for the background modeling simulation are limited due to the very low probability of an event primary producing a multi-dector event.
As a result, this process is biased to cut energy bins with an upward fluctuation in background counts, causing it to cut more events than optimal.
In order to counter this bias, a penalty term is added to the probability above so that a new energy range will only be introduced to the cut if there is a $99.8\%$ chance that cutting it will improve discovery potential.
This penalty term is similar to the one applied when using the Akaike Information Criterion (AIC)~\cite{Akaike1974}.
To ensure that the bin contents are large enough to overcome this penalty, a bin width of 6.4~keV is initially used in determining this cut.
The energy ranges selected are then modified by reducing the bin width by a factor of 2 at a time in order to improve the energy resolution of the cut, with a final binning of width 0.2~keV.

The end result of this cut optimization routine can be seen in Figs.~\ref{fig:coinandsumEcuts}, \ref{fig:2DEcuts} and \ref{fig:bg_data_cuts}.
The combination of the energy and enriched source detector cuts is expected to remove 82\% of Module~1 background events and 87\% of Module~2, while sacrificing 41\% and 49\% of $\beta\beta$ to $0^+_1$ e.s.\ events in Modules~1 and~2, respectively.
The systematic error on the sacrifice for the enriched source detector cut is determined based on the uncretainty in the total isotopic mass in the enriched and natural detectors, and was found to be $<0.1\%$.
Systematic error in the sum and coincident energy cuts can originate from spectral distortions that may shift events in and out of cut regions.
Various spectral distortions were investigated, including energy nonlinearities and error in the phase space factor for $\beta\beta$ decay.
Ultimately, the largest possible source of error was determined to be the phase space integral, for which Kotila and Iachello reported an fractional uncertainty of $0.5\%$~\cite{Kotila2012}.
Here we take a conservative approach and apply this as the uncertainty for this cut efficiency; this is equivalent to assuming that all of the error in phase space calculations is concentrated in either events that are cut or uncut.
Even so, the systematic uncertainties applied to these cut efficiencies are subdominant to other sources.

\subsection{Simulation validation}
To validate the simulated detection efficiency for $\beta\beta$-decay to e.s., measurements of pair-production peaks were compared between simulations and calibration data.
Pair-production events involve the production of an $e^+-e^-$ pair in the bulk of a detector, and the prompt emission of two 511~keV $\gamma$s from the $e^+$ annihilation.
Because these events involve a single pair production site and the prompt emission of $\gamma$s which may be absorbed in a separate detector, they make a good proxy for $\beta\beta$-decay to e.s.\ events.
In single-escape peak (SEP) events, one gamma is absorbed in the detector containing the pair-production, while the other escapes, resulting in a source detector hit with energy equal to the $\gamma$ energy minus 511~keV.
In double-escape peak (DEP) events, both gammas escape the detector, resulting in a source detector hit with energy equal to the $\gamma$ energy minus 1022~keV.
Both SEP and DEP events present the possibility for a second 511~keV detector hit.
By comparing the rate of multiplicity-1 events in the SEPs and DEPs to the rate of multiplicity-2 events in which one hit falls into one of these peaks and the other falls into the 511~keV peak, we can measure a proxy for the detection efficiency of our multi-detector event signature.
$^{56}$Co produces a large number of $\gamma$s at energies high enough to cause pair production, which allows for a comparison of many peaks to our simulation.
This comparison was performed using 168.1~h and 167.1~h of data with the $^{56}$Co line source inserted into the Module~1 and Module~2 calibration tracks, respectively, and a simulation of 3~billion event primaries generated by \mage.

\begin{figure*}[tb]
  \centering
  \includegraphics[width=0.45\textwidth]{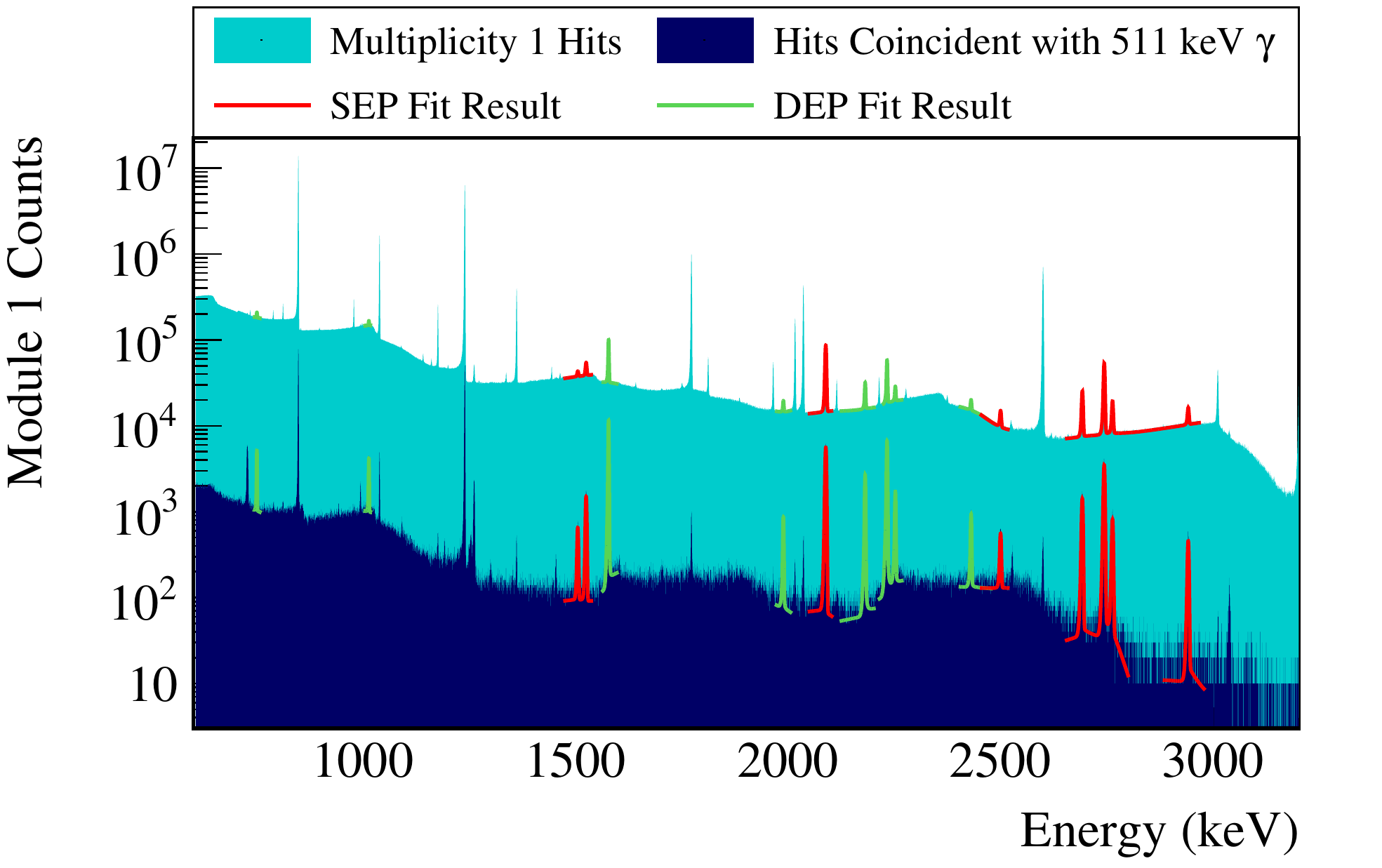}
  \includegraphics[width=0.45\textwidth]{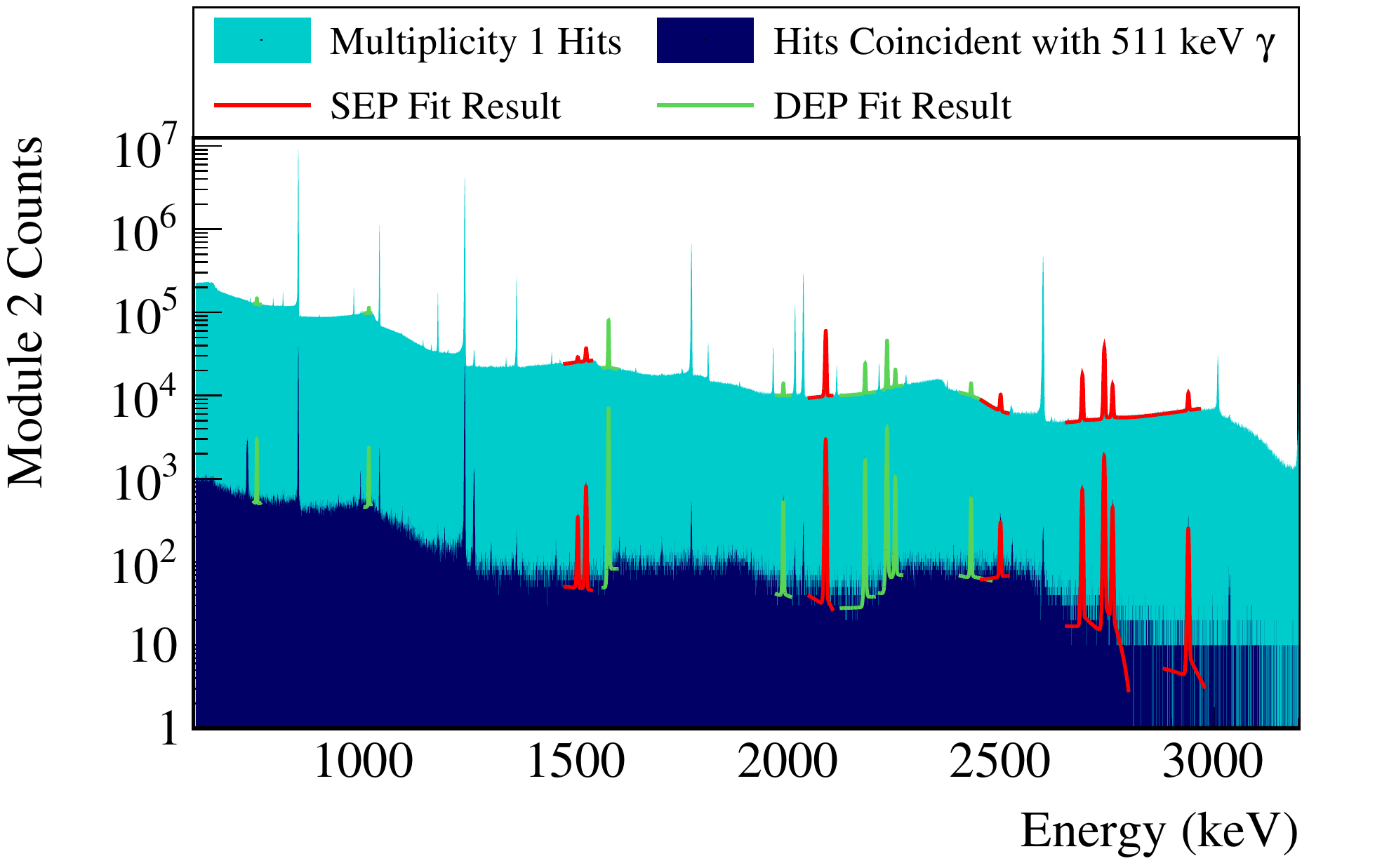}\\
  \subfloat[Module 1]{\includegraphics[width=0.45\textwidth]{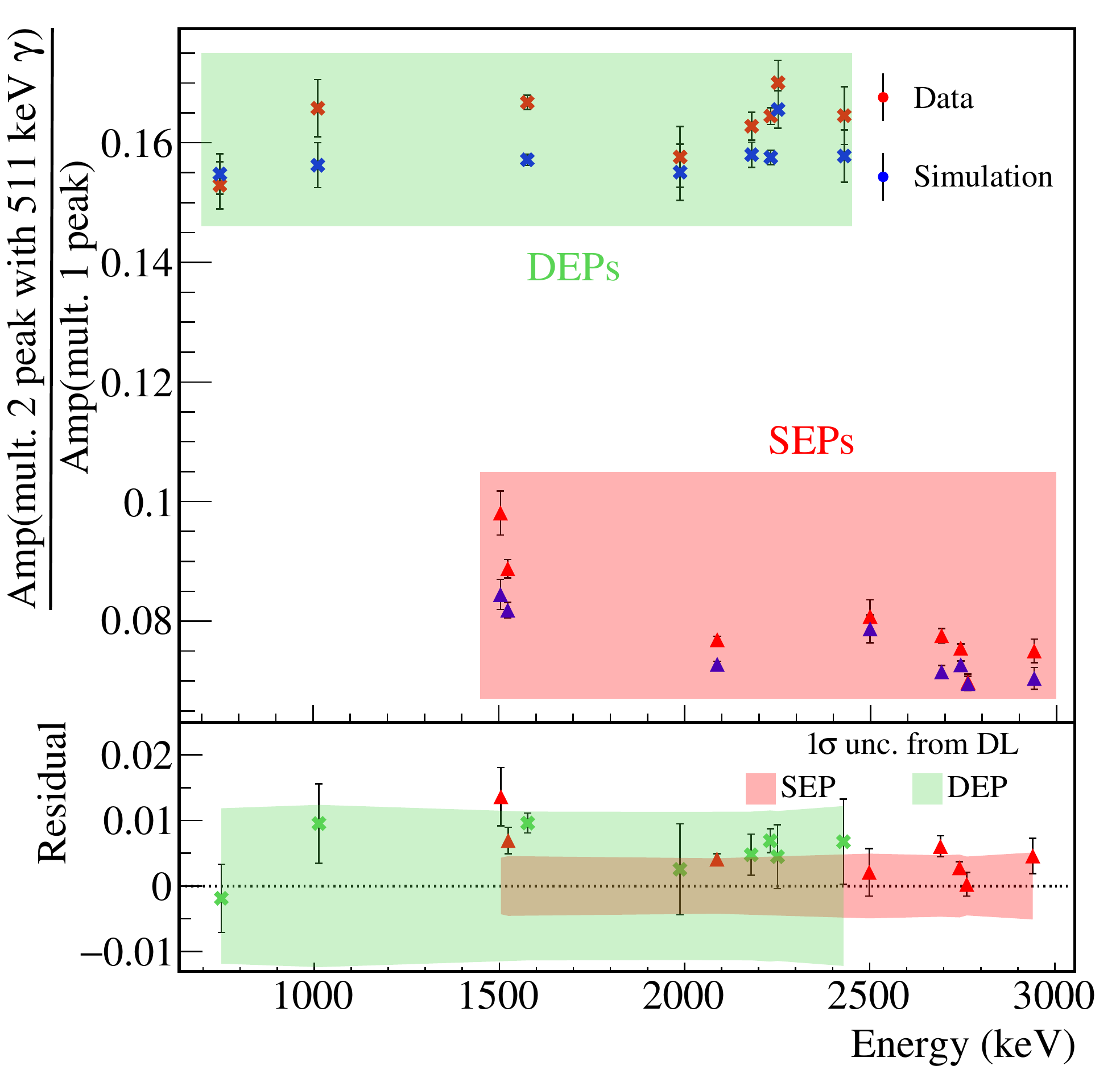}}
  \subfloat[Module 2]{\includegraphics[width=0.45\textwidth]{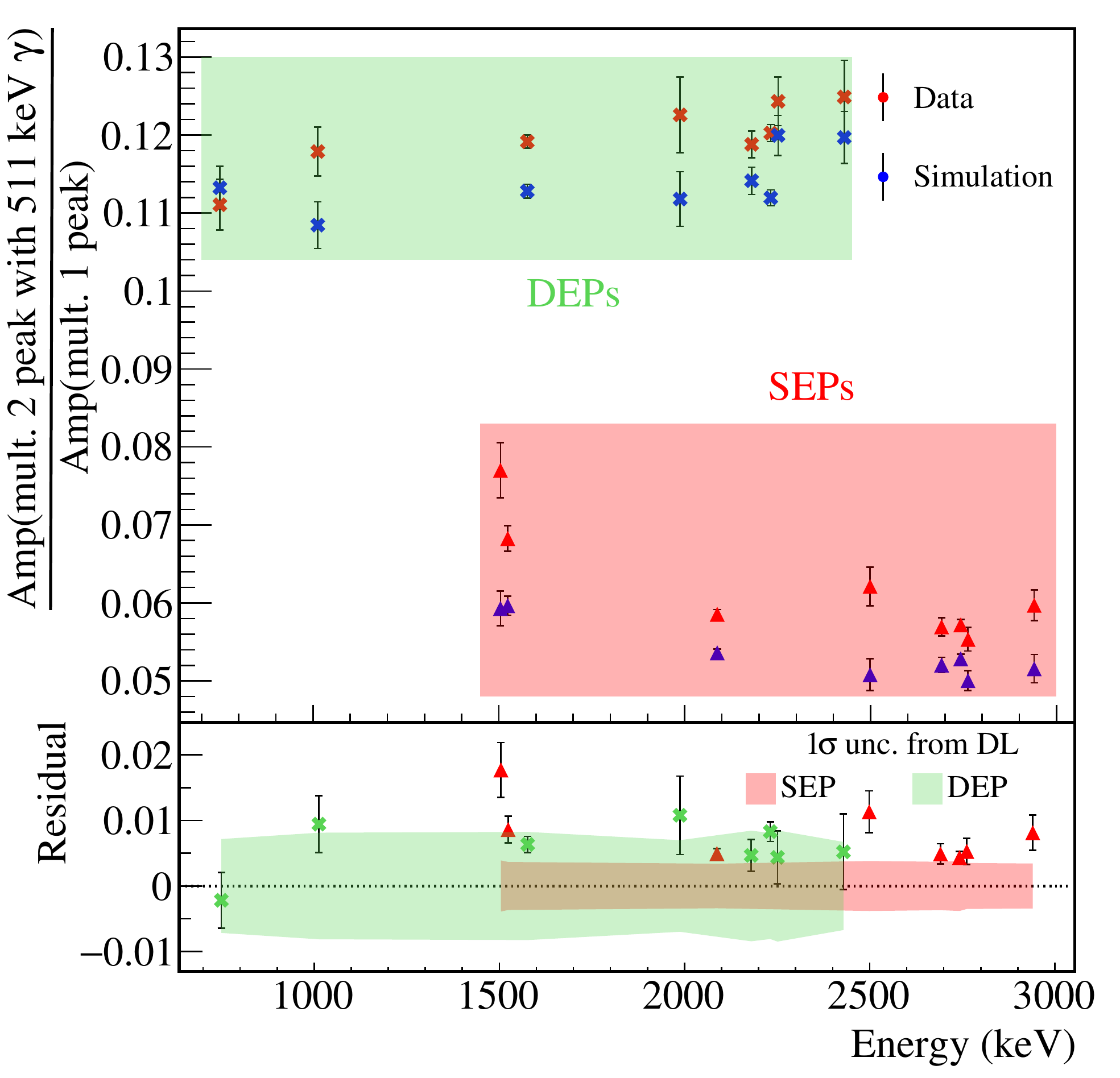}}
  \caption{\label{fig:co56results}Top: Energy spectra recorded while the $^{56}$Co line source was inserted into the calibration track for Module~1 (left) and Module~2 (right). Spectra are shown for multiplicity~1 events and multiplicity~2 events in which the other detector hit fell within the 511~keV peak. The SEPs (red triangle) and DEPs (green~x) that were fit and used for simulation validation are shown. Bottom: the ratio of peak amplitudes from the selected SEPs and DEPs for Module~1 (left) and Module~2 (right). The expected systematic error from the dead layer thickness for SEPs and DEPs is shown on top of the residuals.}
\end{figure*}
The result of this comparison is shown in Fig.~\ref{fig:co56results}, with an overall offset that cannot be fully explained by statistical error.
Some of this discrepency can be explained by uncertainty in the dead layer thickness; the remaining difference represents systematic error from an unknown source such as errors in the \mage\ geometry.
A systematic error term is added to the detection efficiencies until these results are consistent.
The error terms are 0.20\% for Module~1 and 0.47\% for Module~2, which are added to the detection efficiencies for $\beta\beta$ to e.s.
The measured error in Module~1 was found to be consistent with the expected error from dead layers, which was not the case for Module~2; for this reason, Module~2 has a much larger error term which will be the dominant error in the $\beta\beta$ to e.s.\ efficiencies.

\subsection{Simulated Detection Efficiency}
The detection efficiency for $\beta\beta$-decay to e.s.\ is calculated by applying the previously mentioned cuts to the simulations of each $\beta\beta$-decay mode.
Table~\ref{tab:efficiency} lists the effect of each source of signal loss described in this section, determined using simulations.
For each source of signal loss, the systematic uncertainty estimations described in the previous sub-sections are listed.
These uncertainties are assumed to be Gaussian and uncorrellated, and they are combined by adding the fractional uncertainties in quadrature.
The domininant sources of uncertainty come from the dead layer thickness and the error found in the $^{56}$Co spectral comparisons.
Using these simulations, the detection efficiency for the $2\nu\beta\beta$ to $0^+_1$ e.s.\ decay mode was estimated to be $2.3\pm0.2$\% for Module~1 and $1.0\pm0.2$\% for Module~2.
The primary reason Module~1's detection efficiency is greater than Module~2's is that Module~2 has more disabled detectors than Module~1.
\begin{table*}
  \centering
  \caption{\label{tab:efficiency}
    Detection efficiency for the $2\nu\beta\beta$ to $0^+_1$ e.s.\ mode. The first entry is the efficiency prior to applying cuts and other effects. The effeciency for individual effects is listed as if it was the last effect applied; as a result, since these effects are not statistically independent, their product as listed will not be the final efficiency. Uncertainties shown are determined from systematics measurements described in the text.
  }
  \begin{tabular}{|c|c|c|l|}
  \hline
  Source & \makecell{Module 1\\efficiency} & \makecell{Module 2\\efficiency} & Dominant Source of Uncertainty \\
\hline
  \makecell{Multi-Detector with\\Full Energy $\gamma$} & $5.6 \pm 0.2\%$ & $3.1 \pm 0.5\%$ & $^{56}$Co validation measurement  \\
  ROI Containment & $86.8 \pm 1.5\%$ & $86.8 \pm 1.5\%$ & Energy calibration and resolution systematics \\
  Dead Layer & $74.7 \pm 4.3\%$ & $63.8 \pm 6.3\%$ & Dead layer thickness \\
  Detector Dead Times & $98.3 \pm 0.8\%$ & $98.4 \pm 0.8\%$ & Detector dead time \\
  Enriched Source Detector Cut & $96.9 \pm{}<\!0.1\%$ & $90.6 \pm{}<\!0.1\%$ & Detector mass and isotopic abundance \\
  Coincident Energy Cut & $91.4 \pm 0.5\%$ & $89.7 \pm 0.5\%$ & Phase space integral \\
  Sum Energy Cut & $62.8 \pm 0.5\%$ & $56.4 \pm 0.5\%$ & Phase space integral \\
  \hline Final Efficiency & $2.33 \pm 0.17\%$ & $1.01 \pm 0.18\%$ & \\
\hline
\end{tabular}

\end{table*}

\section{Results} \label{sec:results}
Data collected between January 12, 2016 and April 17, 2018 were used for this analysis.
The total isotopic exposure was $25.819\pm0.037~\mathrm{kg\cdot yr}$ for Module~1 and $16.104\pm0.024~\mathrm{kg\cdot yr}$ for Module~2.
Isotopic exposure is defined here as the total mass of $^{76}$Ge in a module times the run time for the module.
This stands in contrast with the \textsc{Demonstrator}'s $0\nu\beta\beta$ result~\cite{mjd2019}, which subtracts dead layers and inactive detectors from the mass used to calculate active isotopic exposure; instead, as previously described, inactive isotopic material is instead accounted for in the detection efficiency.

Of this exposure, $12.463~\mathrm{kg\cdot yr}$ in Module~1 and $8.232~\mathrm{kg\cdot yr}$ in Module~2 consisted of blinded data.
Data were unblinded in a staged fashion; first, multiplicity~1 data excluding hits below 200~keV and hits in the $0\nu\beta\beta$ region of interest were unblinded.
Events with multiplicity~2 or greater were unblinded for this analysis after review of the unblinded results, which have been publicly presented~\cite{Guinn_2020} but not published in a peer reviewed journal.
Immediately after unblinding, an error in the application of detector selection was detected based on irregularities in the rates of high-multiplicity events; this was fixed prior to analyzing the data for $\beta\beta$-decay to e.s.
Additionally, after performing an unblinded analysis, two errors were uncovered in DECAY0.
First, the incorrect $\gamma$ correlation factors were used for the $2^+_2$ e.s.\ decay mode; second, for correlated $\gamma$ emissions, the RNG used was not using the seed provided to the program.
These errors were rectified in the simulations without any changes to data selection; the detection efficiency was recalculated and agreed with the old values within uncertainty.

\begin{figure*}[tb]
  \centering
  \includegraphics[width=0.45\textwidth]{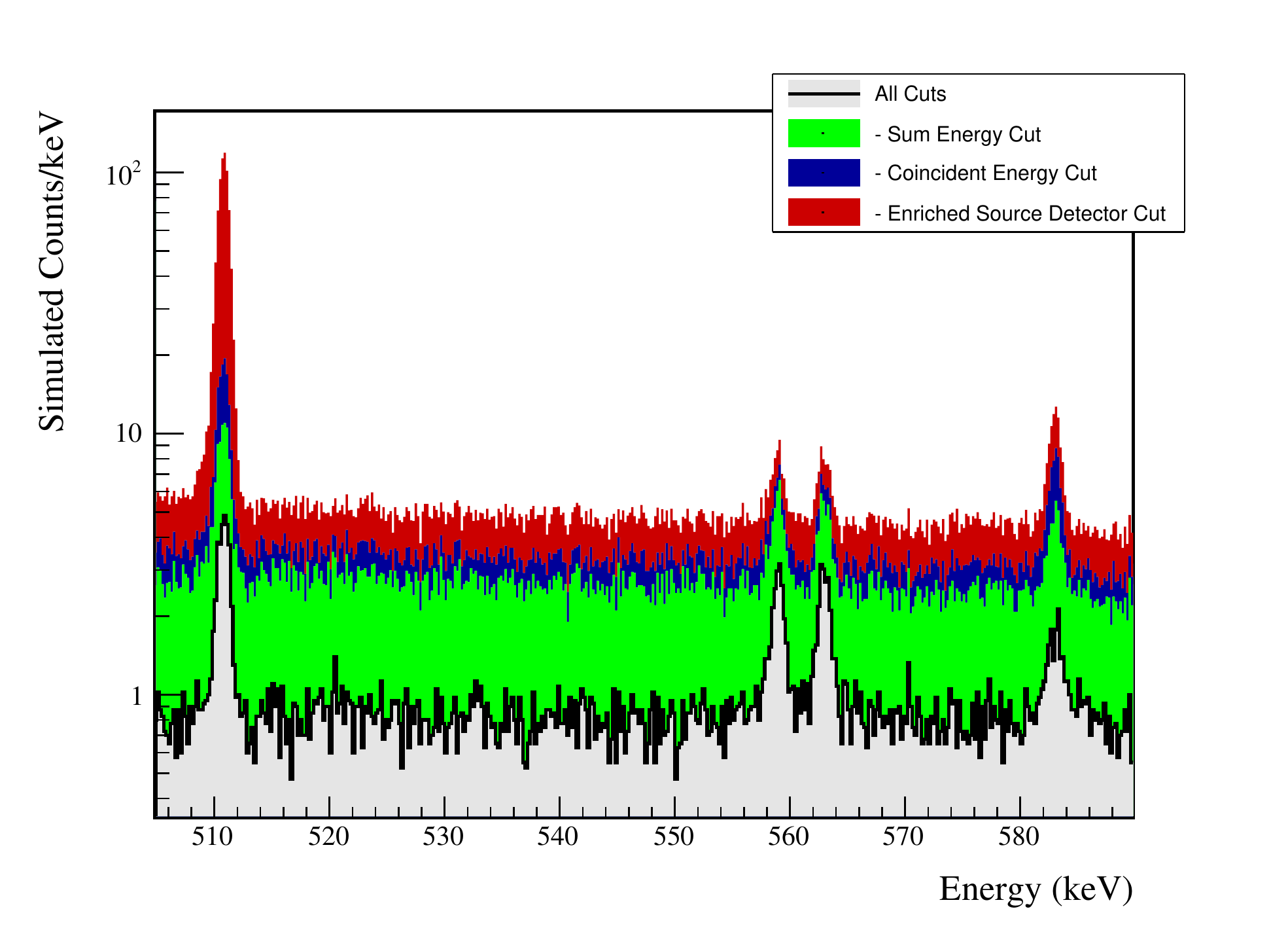}
  \includegraphics[width=0.45\textwidth]{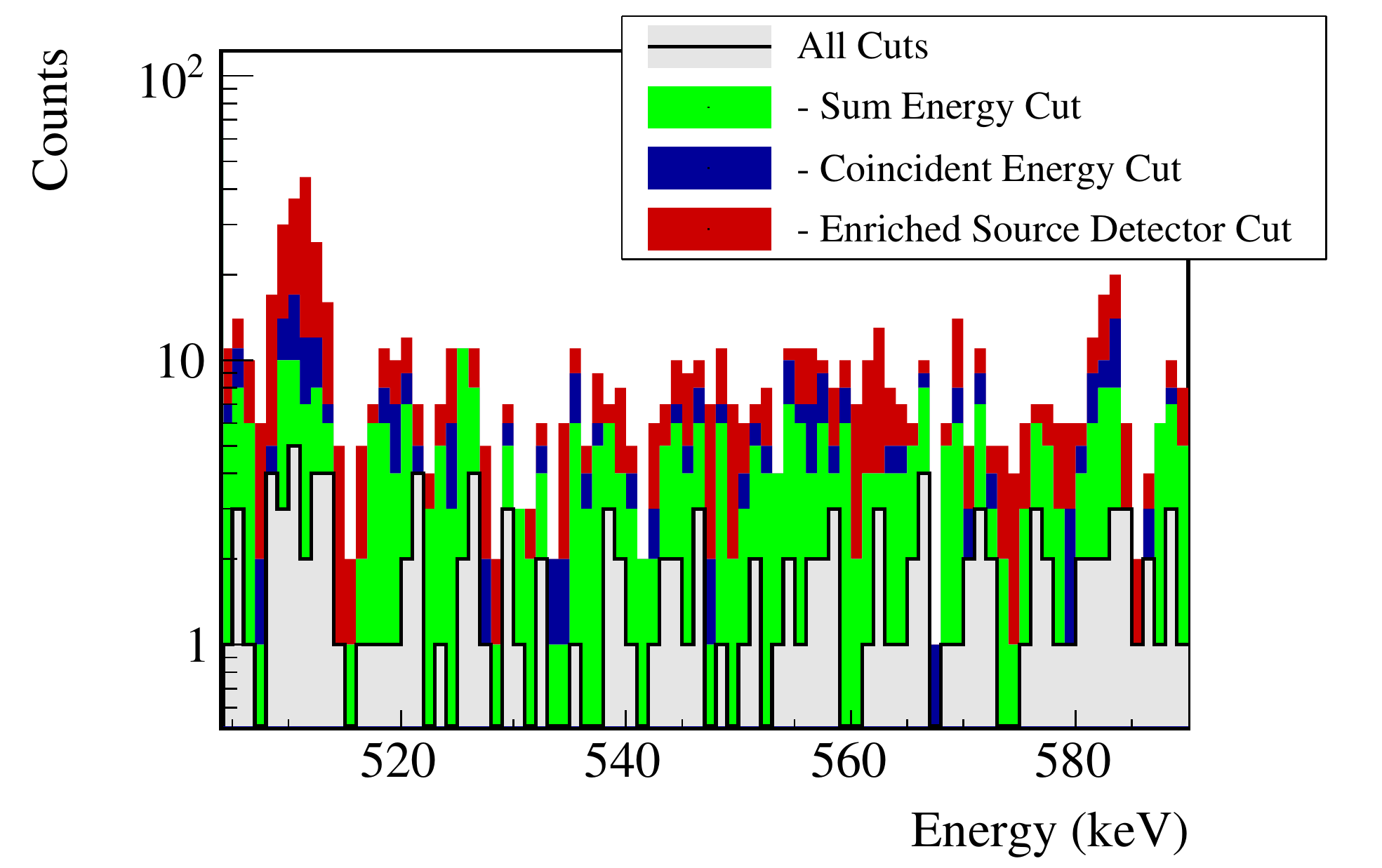}
  \caption{\label{fig:bg_data_cuts}Left: Simulated energy spectrum including $42~\mathrm{kg\cdot yr}$ of exposure from the background model and the $2\nu\beta\beta$-decay to $0^+_1$ e.s.\ peaks at 559 and 563~keV, assuming a half-life of $10^{24}$~yr. The  cuts optimized for this decay mode are lincluded. Right: Measured data energy spectrum with the same cuts applied.}
\end{figure*}
Fig.~\ref{fig:bg_data_cuts} shows data around the 559~ and 563~keV ROIs for the $2\nu\beta\beta$-decay to $0^+_1$ e.s.\ decay mode including the effect of cuts, compared with the background model simulation.
The 511~keV peak from $e^+$ annihilation is notably wider in data than in the simulation; this can be explained by doppler broadening, which is not included in the \mage\ simulation.

A frequentist analysis was performed to calculate Neyman confidence intervals for the half-life of each $\beta\beta$-decay to e.s.\ decay mode.
For a given decay peak $k$, we can calculate the expected number of counts in the signal ROI using
\begin{equation}
  \langle s_k\rangle = \ln2 \frac{N_A}{m_{76}}\epsilon_k\frac{M_{iso}T_{live}}{T_{1/2}}
\end{equation}
where $N_A$ is Avogadro's number, $m_{76}$ is the isotopic mass of $^{76}$Ge, $\epsilon_k$ is the detection efficiency for the peak, $M_{iso}T_{live}$ is the isotopic exposure, and $T_{1/2}$ is the half-life of the decay mode.
For convenience, we will define
\begin{equation}\label{eq:tstar}
  T^*_k=\ln2 \frac{N_A}{m_{76}}\epsilon_kM_{iso}T_{live}
\end{equation}
which is the decay half-life that would produce on average one count in signal ROI $k$.
The following likelihood function is used:
\begin{equation}
  \label{eq:rolke}
  \begin{aligned}
    \mathcal{L}_k&(T_{1/2},T^*_k,b_k|n_k,m_k,\langle T^*_k\rangle, \sigma_{T^*_k},\tau) =
    \\ &\frac{\mu_k^{n_k}e^{-\mu_k}}{n_k!} \cdot \frac{(b_k/\tau)^{m_k}e^{-b_k/\tau}}{m_k!} \cdot
    \frac{1}{\sigma_{T^*_k}\sqrt{2\pi}}e^{-\frac{(T^*_k-\langle T^*_k\rangle)^2}{2\sigma_{T^*_k}^2}} \\
    \mu_k &= s_k+b_k = \frac{T^*_k}{T_{1/2}} + b_k
  \end{aligned}
\end{equation}
In this likelihood function, $T_{1/2}$ is the half-life of the decay, $T^*_k$ is defined in Eq.~\ref{eq:tstar} and assumes Gaussian statistics, and $b_k$, $s_k$ and $\mu_k$ are the expected number of background, signal and total counts in the signal ROI assuming Poisson statistics.
As input parameters, $n_k$ and $m_k$ are the observed counts in the signal and BG ROIs, respectively, $\langle T^*_k\rangle$ and$\sigma_{T^*_k}$ are the measured value and uncertainty on $T^*_k$ calculated from the signal efficiency and exposure, and $\tau$ is a ratio used to compute the expected number of backgrounds in the signal ROI ($b_k\tau$).
In order avoid double counting, events with multiple hits in either the signal or BG ROI only add one count to $n_k$ and $m_k$, respectively.
If an event has a hit in both the signal  and BG ROIs, then it is only counted as in $n_k$; the small excess this creates in the signal ROI is accounted for in $\tau$.

The likelihood function is maximized over $T^*_k$ and $b_k$ as prescribed by Rolke~\cite{Rolke2005} in order to produce a 90\% confidence interval for each individual peak-module combination.
A combined result is then calculated for each decay mode by constructing a likelilhood from the product of Eq.~\ref{eq:rolke}.
Confidence intervals were calculated by profiling the negative log likelihood along $T_{1/2}^{-1}$ until it increased by 2.7.
For all modes, the 90\% confidence intervals were bounded at $T_{1/2}^{-1}=0$, meaning that lower half-life limits are presented.

The detection sensitivity is computed by constructing a toy Monte Carlo for each decay mode, assuming that each $T_{1/2}^{-1}=0$.
For each sample $i$, a random $n_i$ and $m_i$ is drawn from a Poisson distribution with mean $b_k$ and $m_k$.
The confidence interval for a measurement with these values is computed.
The median sensitivity is extracted by taking the median lower half-life limit over 100001 samples.

\begin{figure*}[tb]
  \subfloat[\scriptsize$2\nu\beta\beta$ to $0^+_1$ e.s.\ 559+563 keV $\gamma$ peaks]{\includegraphics[width=0.33\textwidth]{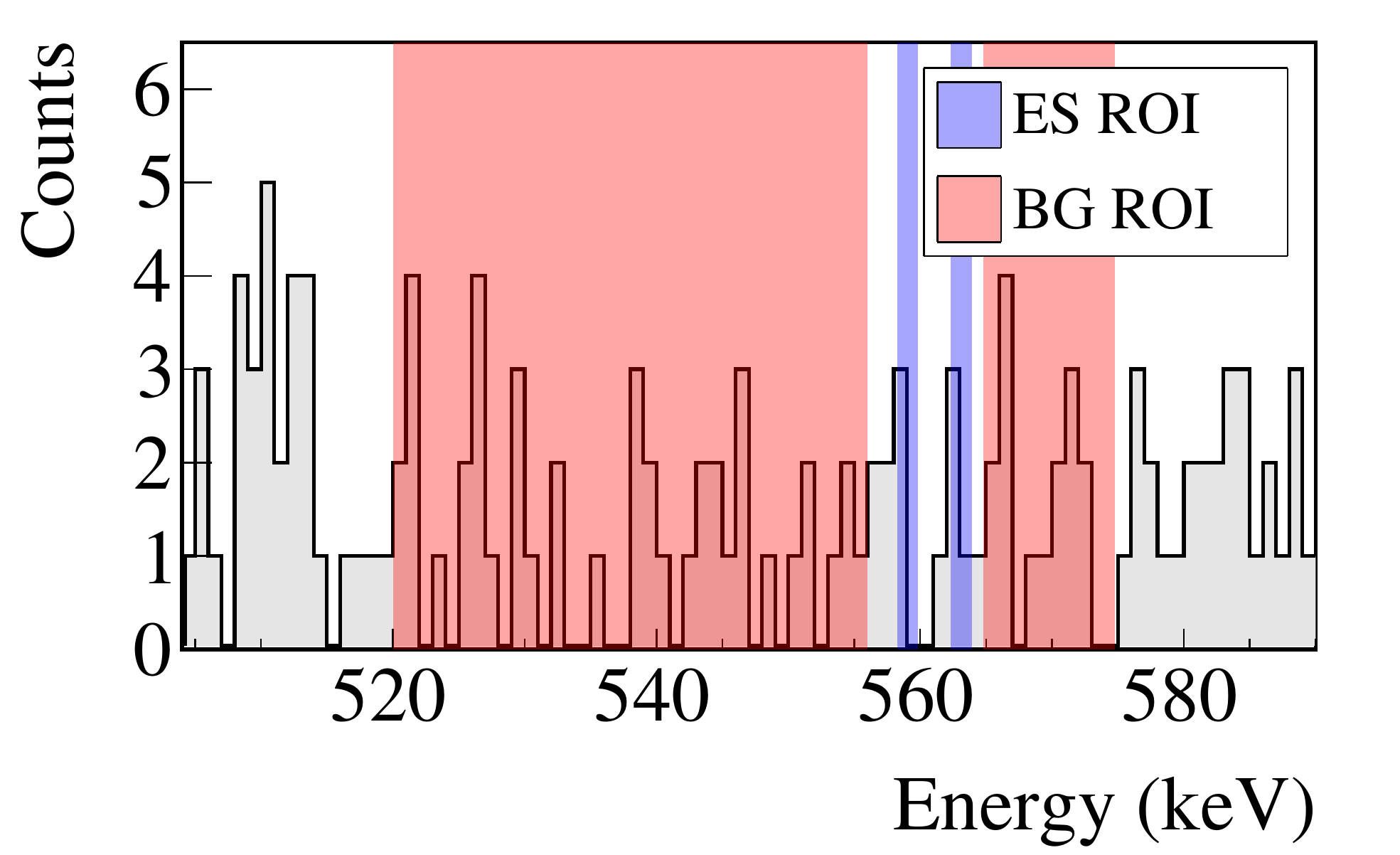}}
  \subfloat[\scriptsize$2\nu\beta\beta$ to $2^+_1$ e.s.\ 559 keV $\gamma$ peak]{\includegraphics[width=0.33\textwidth]{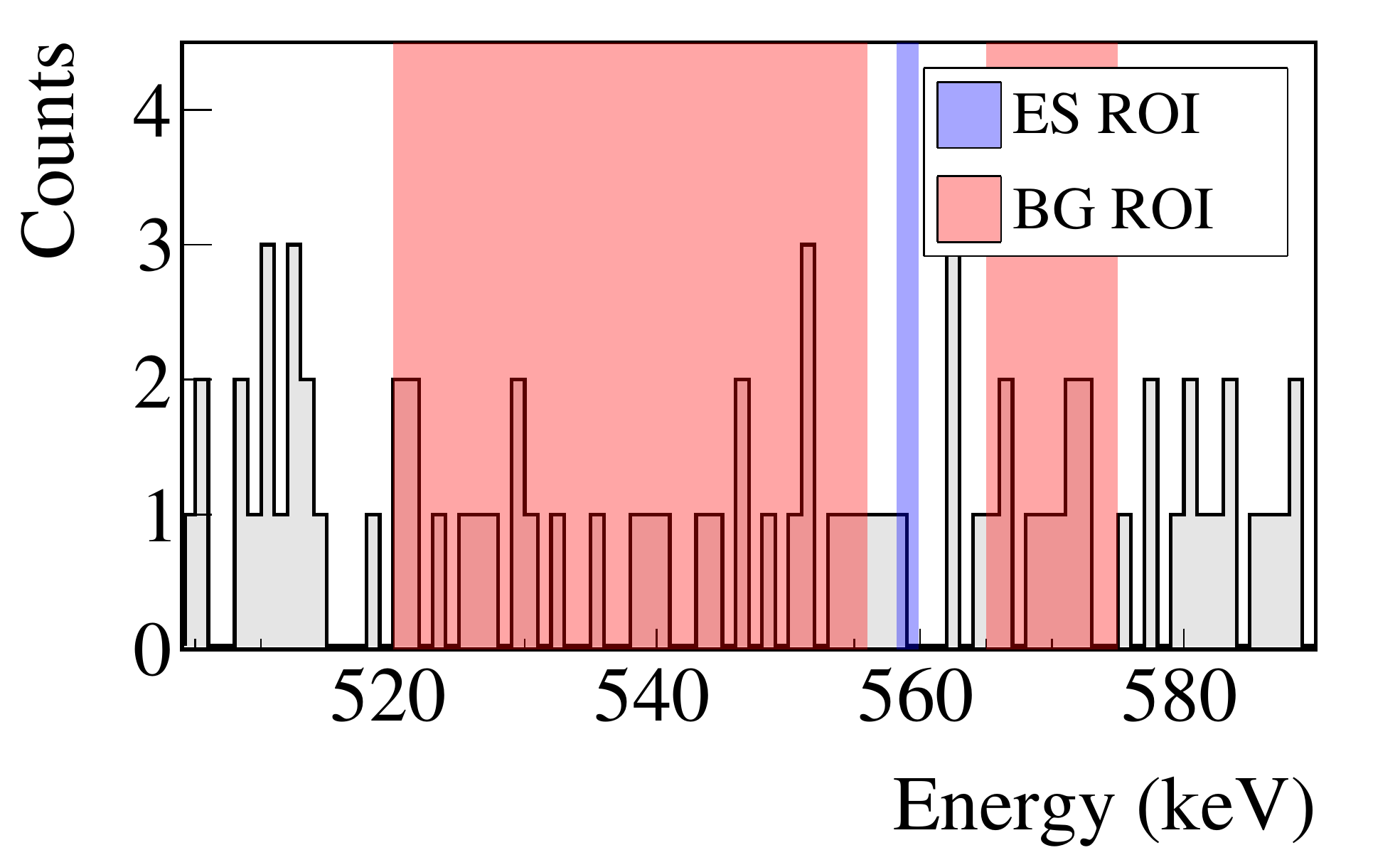}} \\
  \subfloat[\scriptsize$2\nu\beta\beta$ to $2^+_2$ e.s.\ 559 keV $\gamma$ peak]{\includegraphics[width=0.33\textwidth]{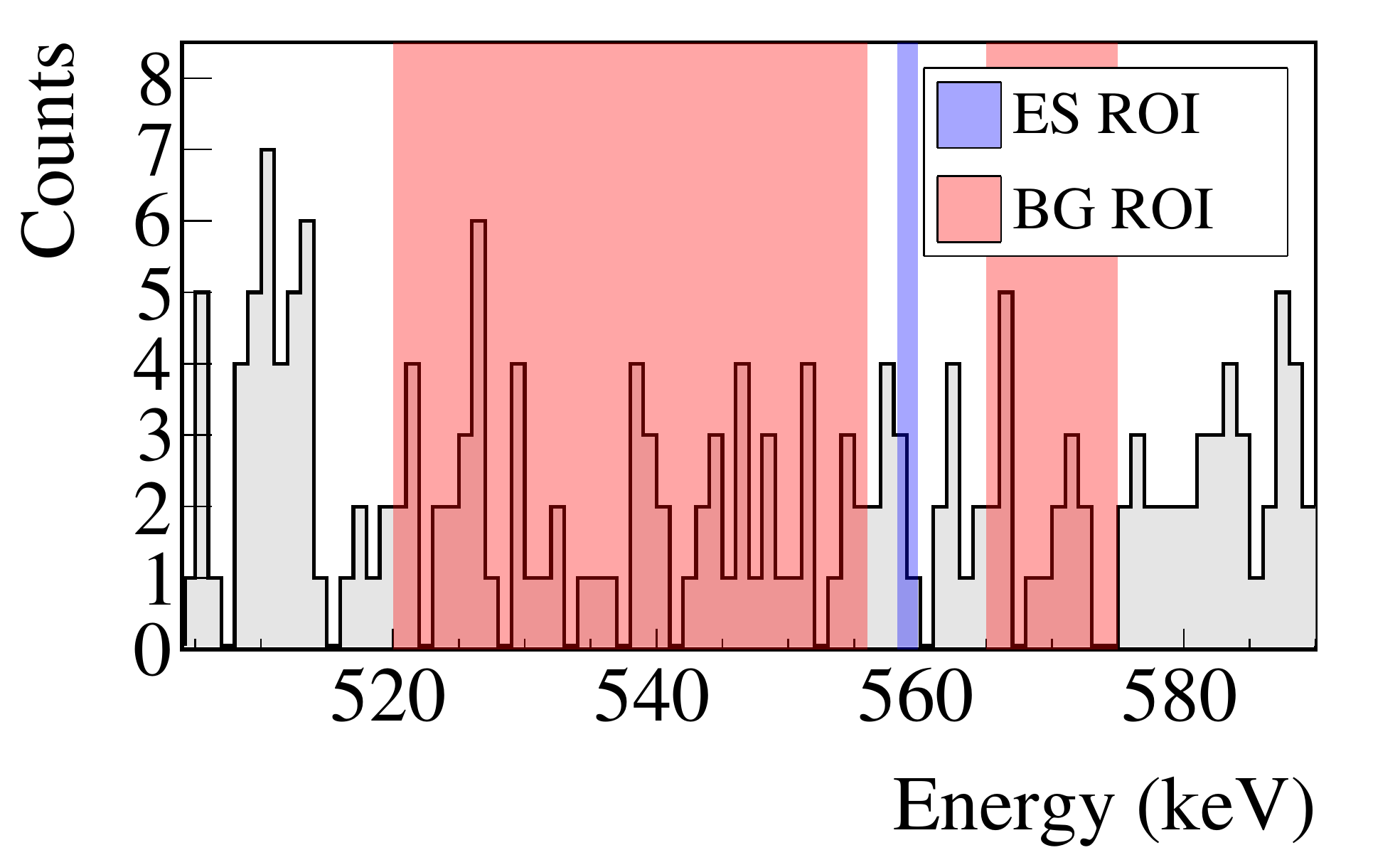}} 
  \subfloat[\scriptsize$2\nu\beta\beta$ to $2^+_2$ e.s.\ 657 keV $\gamma$ peak]{\includegraphics[width=0.33\textwidth]{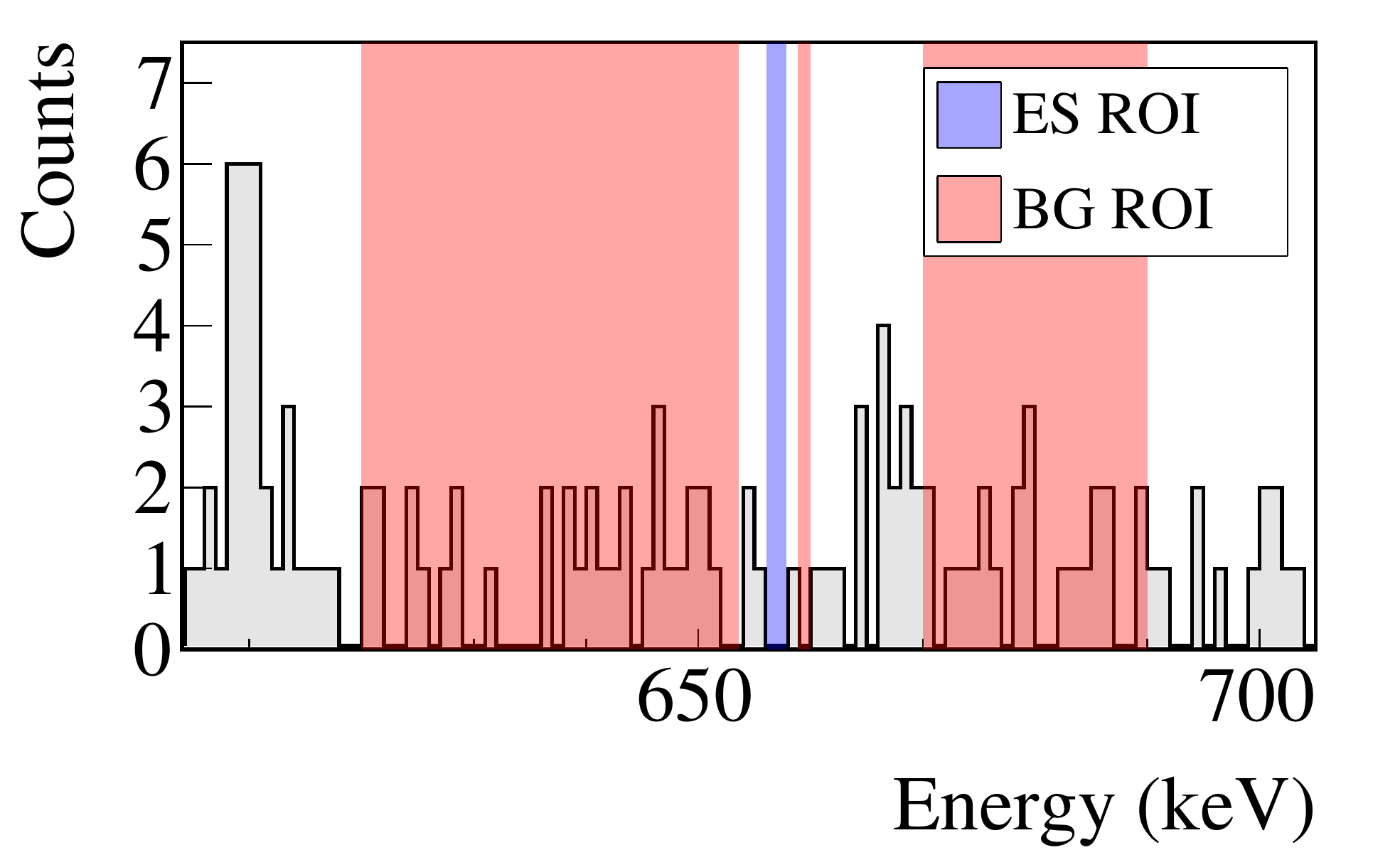}} 
  \subfloat[\scriptsize$2\nu\beta\beta$ to $2^+_2$ e.s.\ 1216 keV $\gamma$ peak]{\includegraphics[width=0.33\textwidth]{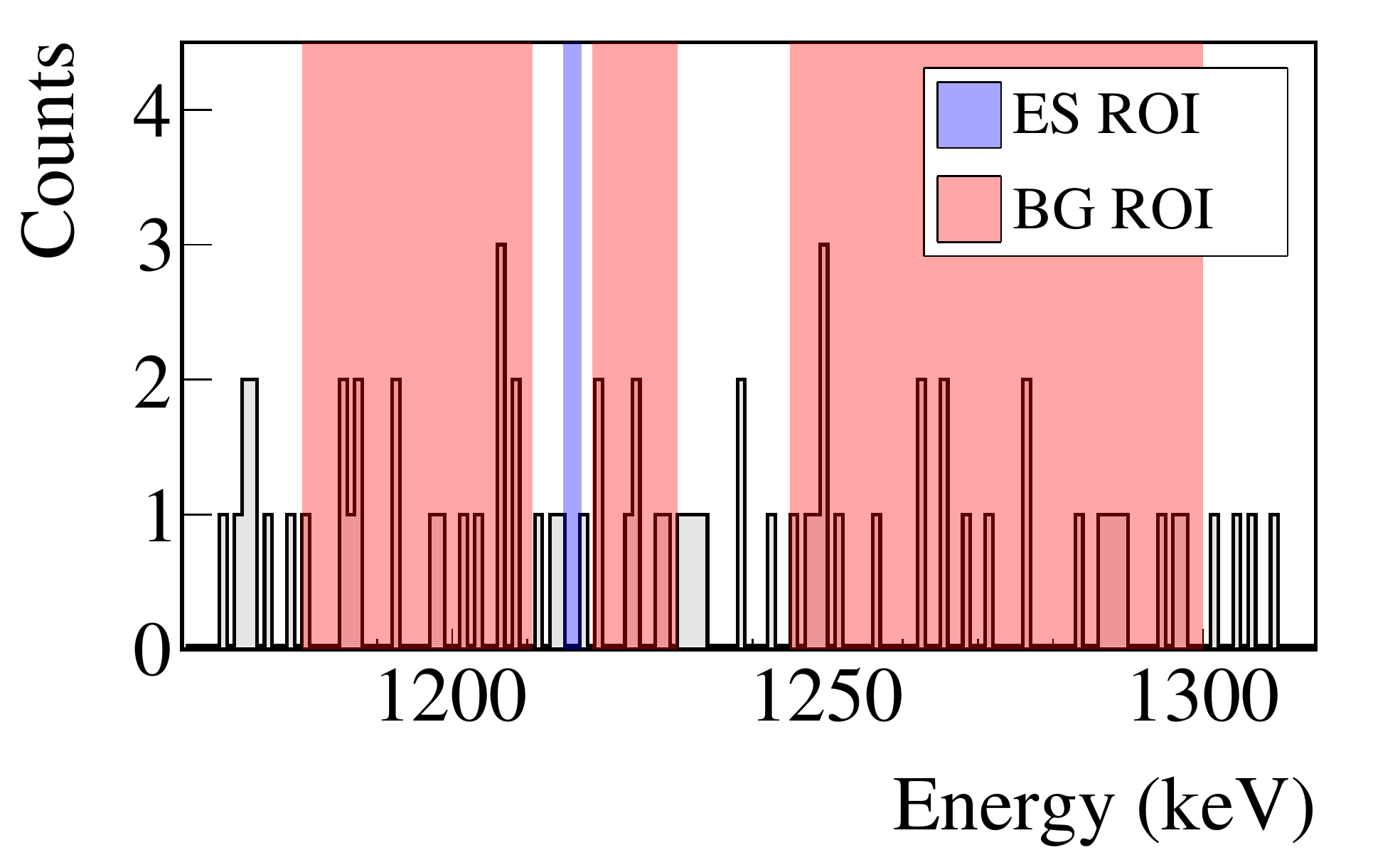}} \\
  \subfloat[\scriptsize$0\nu\beta\beta$ to $0^+_1$ e.s.\ 559+563 keV $\gamma$ peaks]{\includegraphics[width=0.33\textwidth]{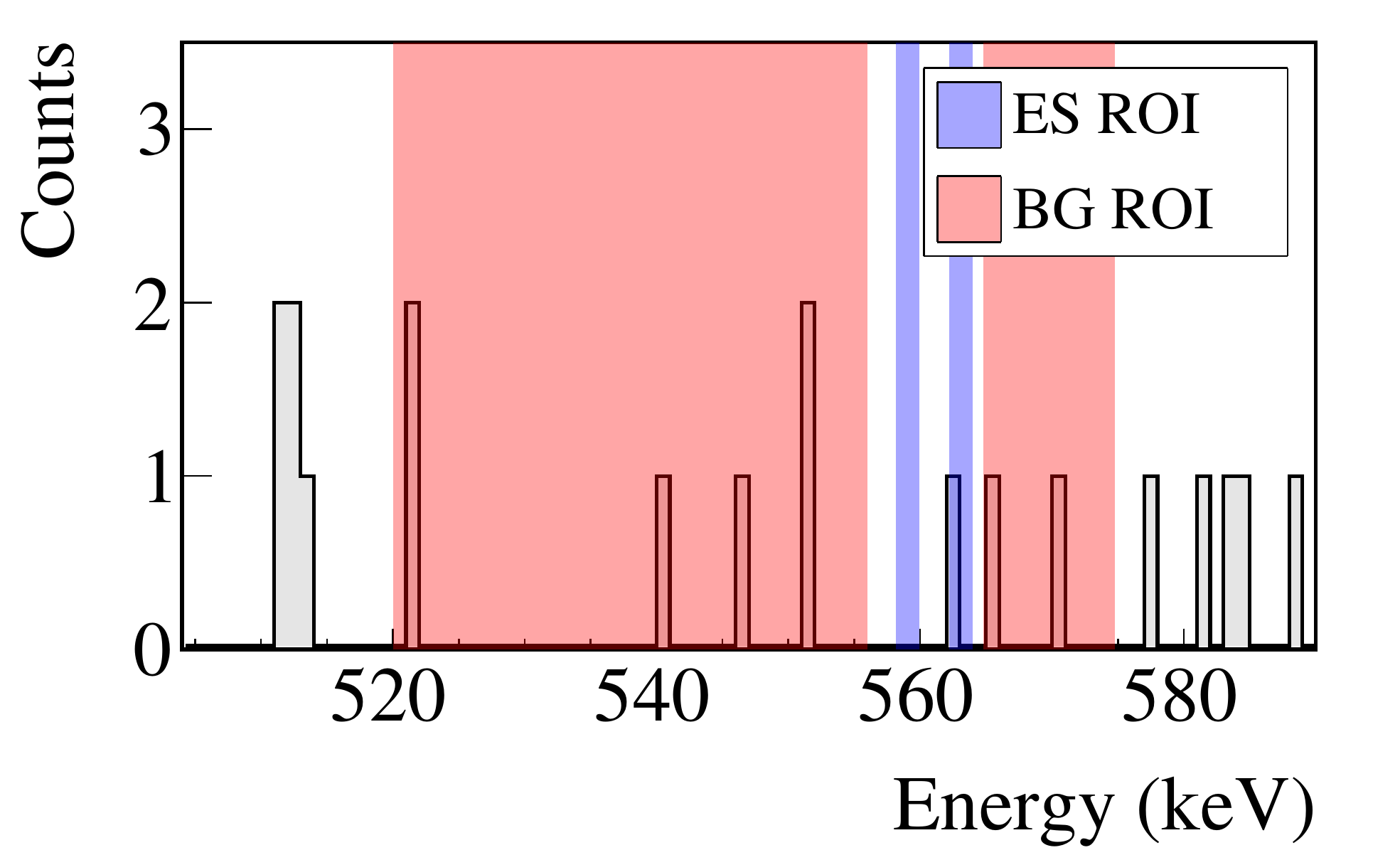}} 
  \subfloat[\scriptsize$0\nu\beta\beta$ to $2^+_1$ e.s.\ 559 keV $\gamma$ peak]{\includegraphics[width=0.33\textwidth]{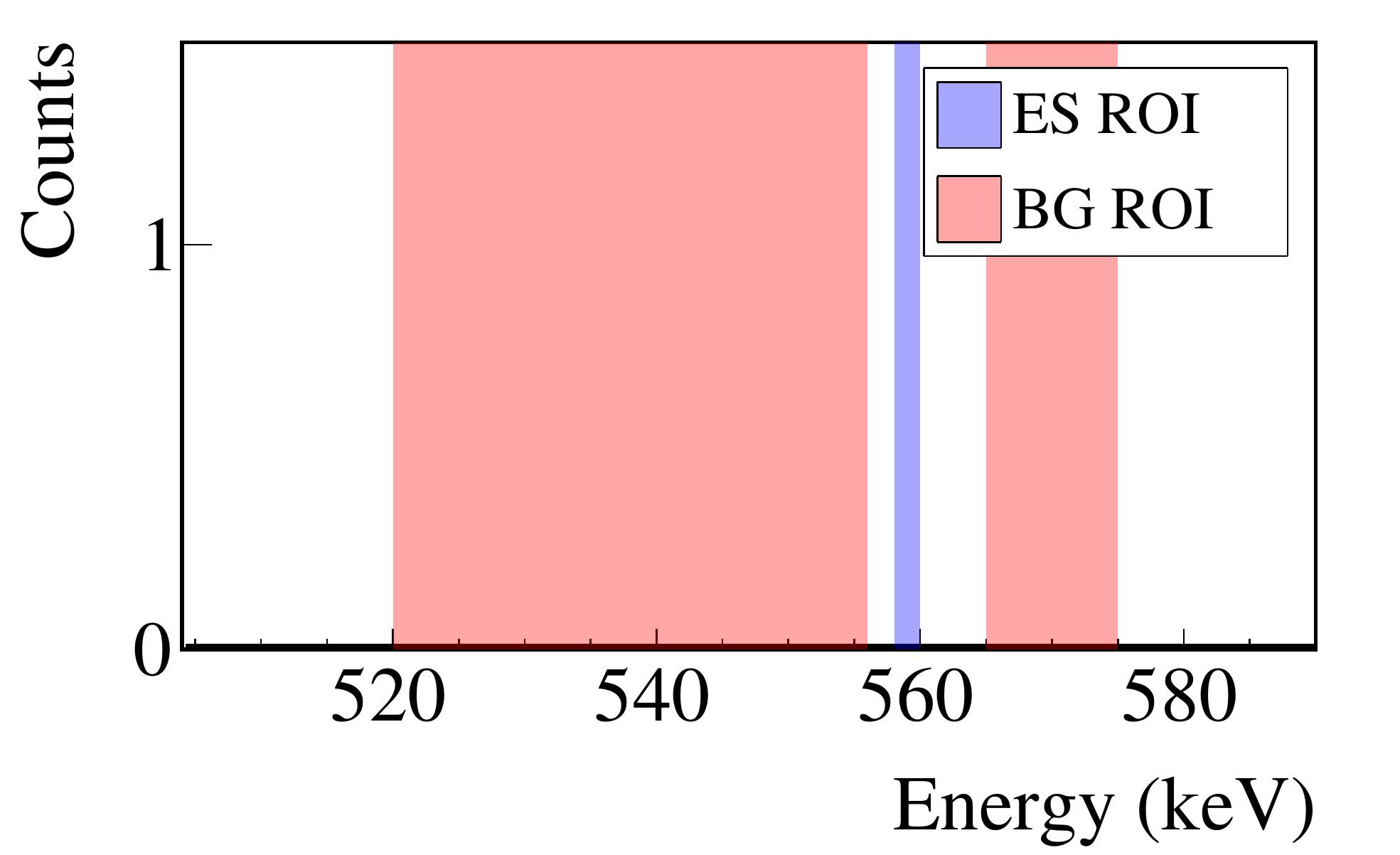}} \\
  \subfloat[\scriptsize$0\nu\beta\beta$ to $2^+_2$ e.s.\ 559 keV $\gamma$ peak]{\includegraphics[width=0.33\textwidth]{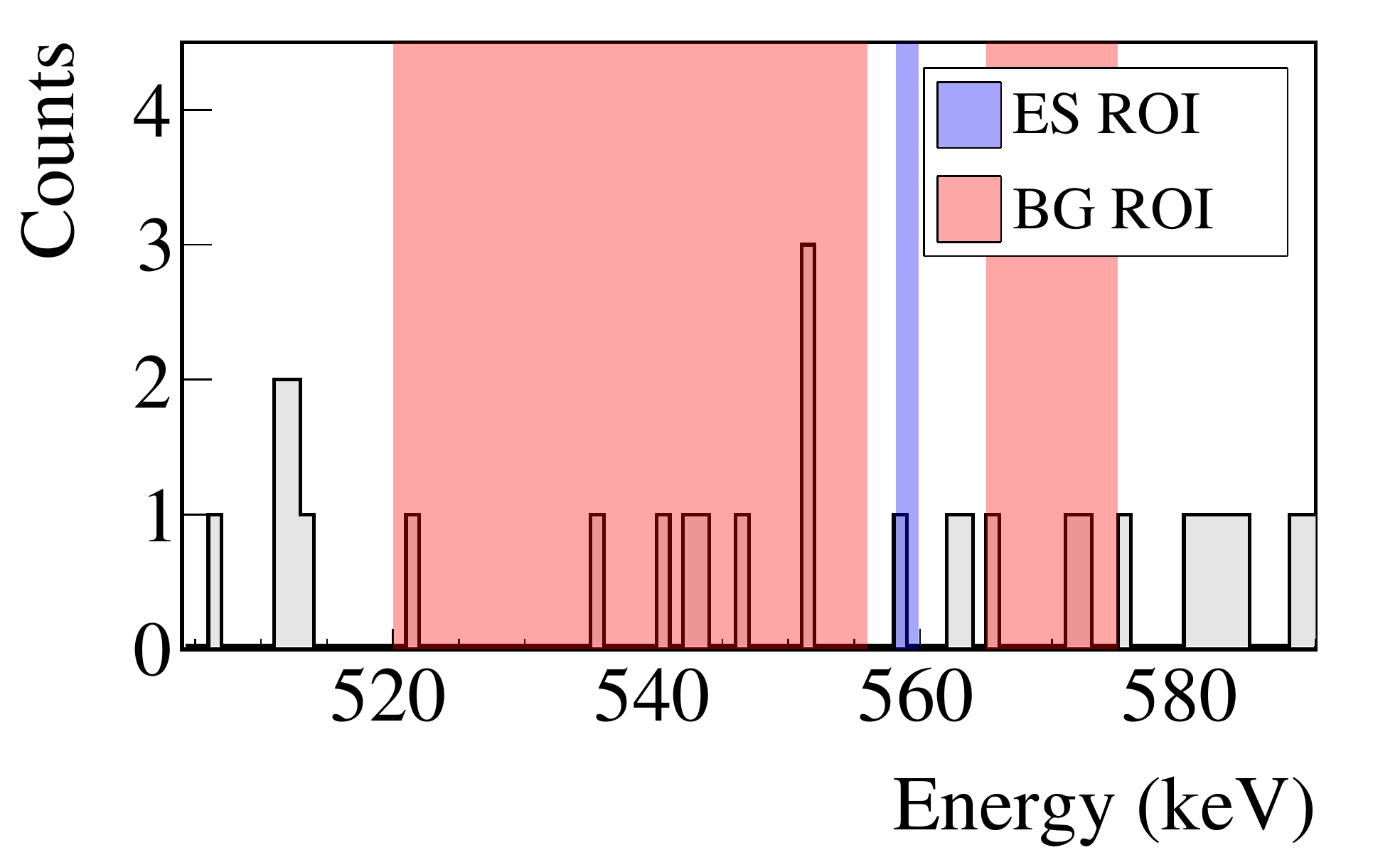}} 
  \subfloat[\scriptsize$0\nu\beta\beta$ to $2^+_2$ e.s.\ 657 keV $\gamma$ peak]{\includegraphics[width=0.33\textwidth]{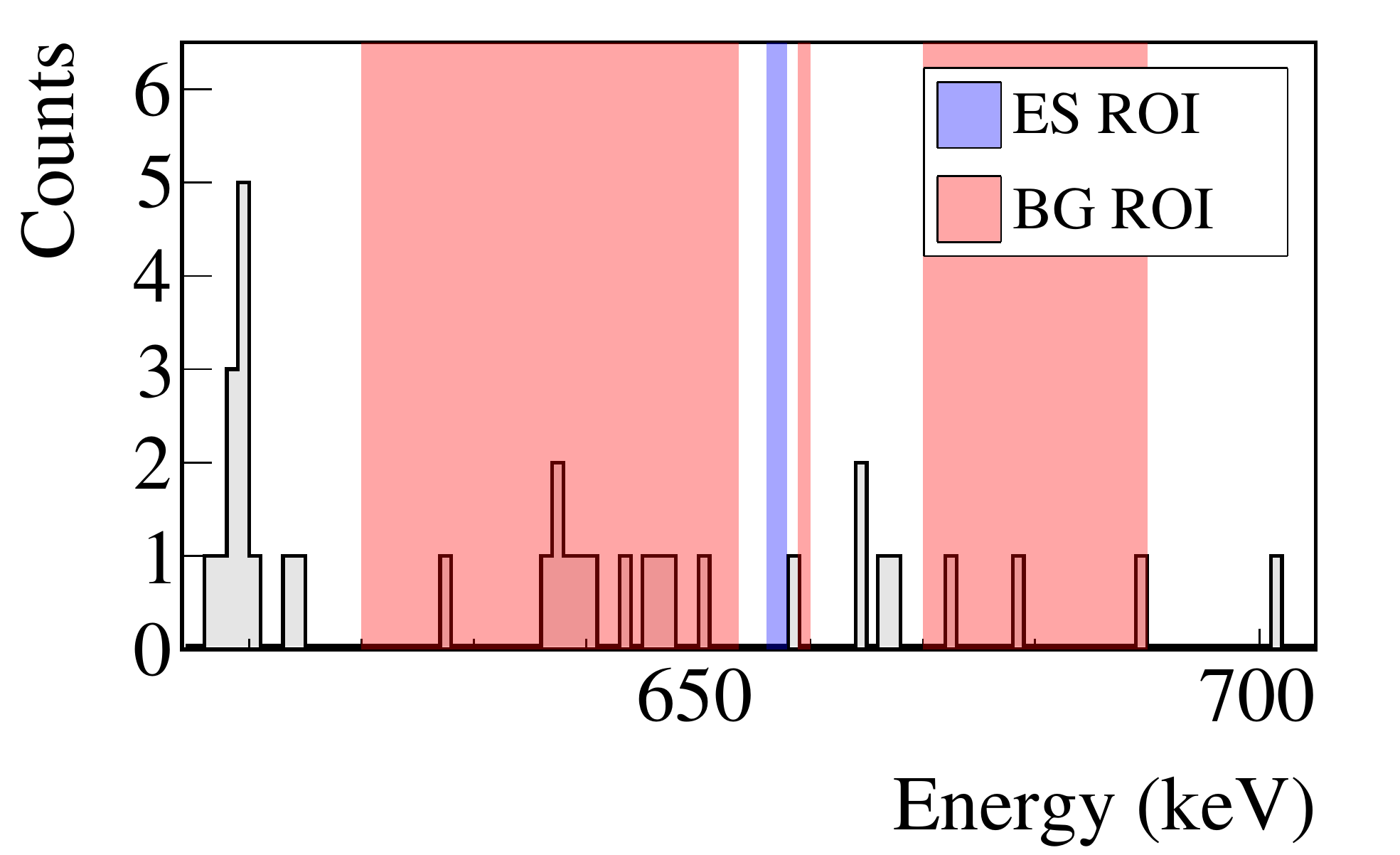}}
  \subfloat[\scriptsize$0\nu\beta\beta$ to $2^+_2$ e.s.\ 1216 keV $\gamma$ peak]{\includegraphics[width=0.33\textwidth]{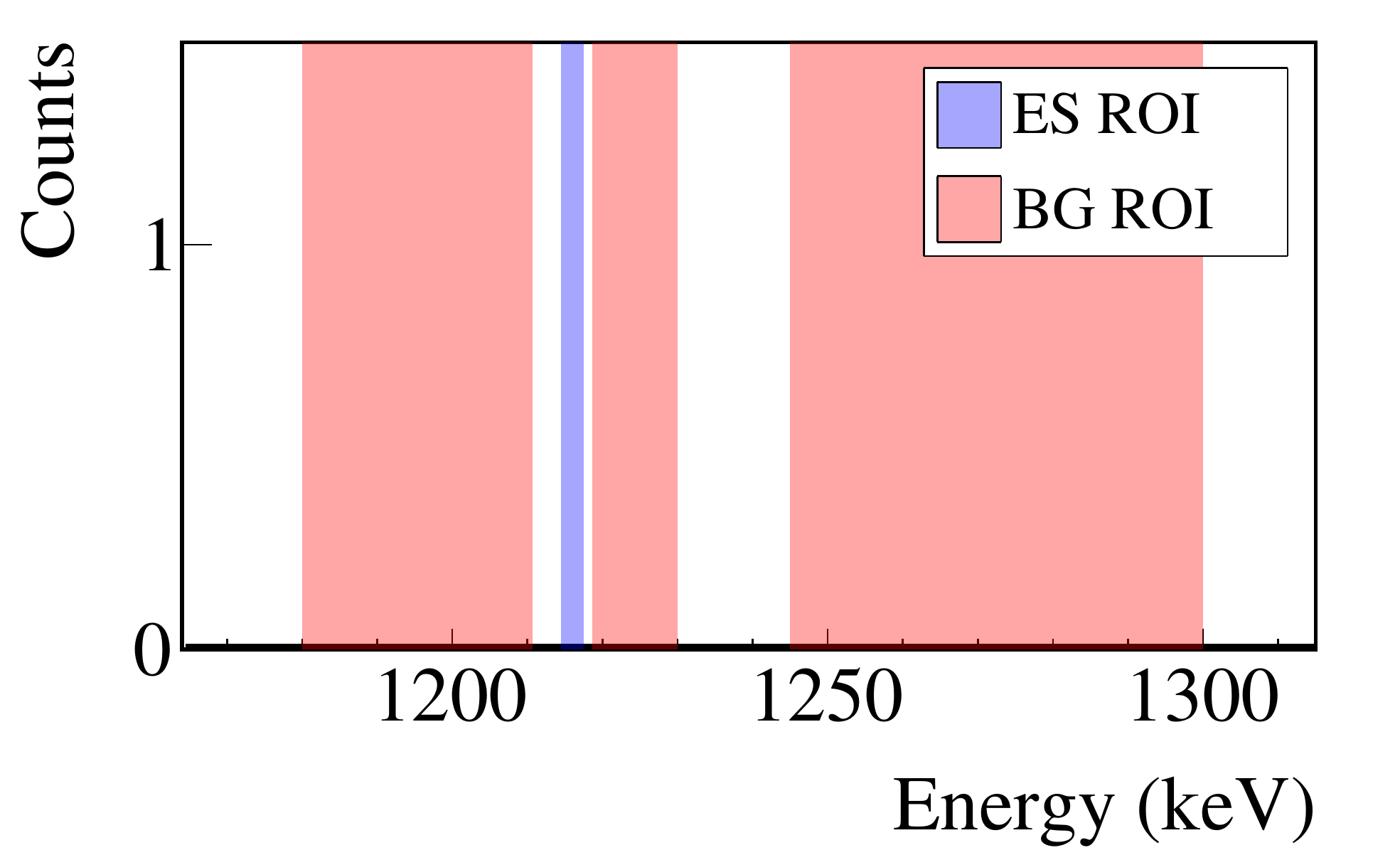}}

  \caption{\label{fig:alldata_roi}Energy spectra for each $\beta\beta$-decay to e.s.\ decay mode after applying optimized cuts. The signal and BG ROIs are highlighted in blue and red, respectively.}
\end{figure*}

\begin{table*}
  \centering
  \caption[Final results for all decay modes]{ \label{tab:alllimits}
  Results for all decay modes. $n_k$ and $m_k$ are the observed counts in the signal and BG ROI, respectively, and can be seen, combining modules, in Fig.~\ref{fig:alldata_roi}. The expected ROI BGs corresponds to $\tau\cdot m_k$ from Eq.~\ref{eq:rolke}. $\epsilon$ is the final signal detection efficiency measured from simulations as shown in Table~\ref{tab:efficiency}. $\langle T^*_k\rangle$ is the value calculated from exposure and efficiency estimations, as defined in Eq.~\ref{eq:tstar}.
  }
  %\scriptsize
  \newcommand{\decaySP}[3]{$0^+_{g.s.}\xrightarrow{#1\nu\beta\beta}#2^+_{#3}$}
\begin{tabular}{|c|c|c c c c c r|c|c|}
\hline  Decay Mode & Peak & Module & $n_k$ & $m_k$ & \makecell{Expected\\ROI BGs} & $\epsilon~(\%)$ & $\langle T^*_k\rangle\,(\times 10^{23} \mathrm{yr})$ & \makecell{$T_{1/2}\,(\times 10^{23} \mathrm{yr})$ \\ 90\% Limit} & \makecell{$T_{1/2}\,(\times 10^{23} \mathrm{yr})$ \\ 90\% Sensitivity} \\
\hline
\multirow{5}{*}{\decaySP{2}{0}{1}} & \multirow{2}{*}{559 keV} & M1 & 2 & 51 & 1.77 & 1.2 & $16.5 \pm 1.2$ & $>4.6$ & $>5.1$ \\
     &      & M2 & 1 & 6 & 0.25 & 0.5 & $4.5 \pm 0.8$ & $>1.3$ & $>3.2$ \\
     & \multirow{2}{*}{563 keV} & M1 & 2 & 51 & 1.95 & 1.2 & $16.5 \pm 1.2$ & $>4.9$ & $>5.1$ \\
     &      & M2 & 0 & 6 & 0.22 & 0.5 & $4.4 \pm 0.8$ & $>3.2$ & $>3.2$ \\
     & Combined &  &  &  &  &  &  & $>7.5$ & $>10.5$ \\
\hline\multirow{3}{*}{\decaySP{2}{2}{1}} & \multirow{2}{*}{559 keV} & M1 & 0 & 35 & 1.43 & 1.4 & $19.2 \pm 2.0$ & $>14.1$ & $>7.8$ \\
     &      & M2 & 1 & 2 & 0.10 & 0.6 & $5.2 \pm 1.7$ & $>1.2$ & $>3.3$ \\
     & Combined &  &  &  &  &  &  & $>7.7$ & $>10.2$ \\
\hline\multirow{7}{*}{\decaySP{2}{2}{2}} & \multirow{2}{*}{559 keV} & M1 & 3 & 74 & 2.57 & 1.0 & $13.8 \pm 1.7$ & $>3.2$ & $>4.3$ \\
     &      & M2 & 1 & 8 & 0.32 & 0.4 & $3.8 \pm 1.7$ & $>0.7$ & $>2.3$ \\
     & \multirow{2}{*}{657 keV} & M1 & 0 & 46 & 1.48 & 0.8 & $11.2 \pm 1.5$ & $>8.2$ & $>4.6$ \\
     &      & M2 & 0 & 6 & 0.19 & 0.4 & $3.1 \pm 1.6$ & $>1.8$ & $>1.8$ \\
     & \multirow{2}{*}{1216 keV} & M1 & 0 & 41 & 1.07 & 0.4 & $5.8 \pm 1.6$ & $>4.0$ & $>2.1$ \\
     &      & M2 & 0 & 7 & 0.24 & 0.2 & $1.5 \pm 1.8$ & $>2.2$ & $>2.2$ \\
     & Combined &  &  &  &  &  &  & $>12.8$ & $>8.2$ \\
\hline\multirow{5}{*}{\decaySP{0}{0}{1}} & \multirow{2}{*}{559 keV} & M1 & 0 & 6 & 0.24 & 1.5 & $21.5 \pm 1.8$ & $>15.8$ & $>15.8$ \\
     &      & M2 & 0 & 1 & 0.06 & 0.6 & $5.7 \pm 1.1$ & $>4.1$ & $>4.1$ \\
     & \multirow{2}{*}{563 keV} & M1 & 0 & 6 & 0.25 & 1.5 & $21.2 \pm 1.8$ & $>15.6$ & $>15.6$ \\
     &      & M2 & 0 & 1 & 0.06 & 0.6 & $5.7 \pm 1.1$ & $>4.1$ & $>4.1$ \\
     & Combined &  &  &  &  &  &  & $>39.9$ & $>39.9$ \\
\hline\multirow{3}{*}{\decaySP{0}{2}{1}} & \multirow{2}{*}{559 keV} & M1 & 0 & 0 & 0.00 & 1.6 & $22.9 \pm 2.5$ & $>16.8$ & $>16.8$ \\
     &      & M2 & 0 & 0 & 0.00 & 0.7 & $6.0 \pm 2.1$ & $>4.0$ & $>4.0$ \\
     & Combined &  &  &  &  &  &  & $>21.2$ & $>21.2$ \\
\hline\multirow{7}{*}{\decaySP{0}{2}{2}} & \multirow{2}{*}{559 keV} & M1 & 0 & 11 & 0.40 & 1.0 & $13.8 \pm 1.8$ & $>10.0$ & $>10.0$ \\
     &      & M2 & 1 & 1 & 0.07 & 0.4 & $3.7 \pm 1.8$ & $>0.6$ & $>2.2$ \\
     & \multirow{2}{*}{657 keV} & M1 & 1 & 10 & 0.41 & 0.9 & $13.5 \pm 1.9$ & $>4.1$ & $>9.8$ \\
     &      & M2 & 0 & 1 & 0.01 & 0.4 & $3.5 \pm 1.8$ & $>2.0$ & $>2.0$ \\
     & \multirow{2}{*}{1216 keV} & M1 & 0 & 0 & 0.00 & 0.4 & $6.2 \pm 1.7$ & $>4.3$ & $>4.3$ \\
     &      & M2 & 0 & 0 & 0.00 & 0.2 & $1.6 \pm 1.9$ & $>0.3$ & $>0.3$ \\
     & Combined &  &  &  &  &  &  & $>9.7$ & $>18.6$ \\
\hline\end{tabular}

\end{table*}
Table~\ref{tab:alllimits} contains a summary of the results for each decay mode.
For the $2\nu\beta\beta$-decay to $0^+_1$ e.s.\ mode, 5 events passed all cuts in the combined 559 and 563~keV signal ROIs, with 4.2 events expected from backgrounds.
This set a 90\% CI limit on the half-life of $T_{1/2}>7.5\times10^{23}$~yr, compared with a 90\% median sensitivity of $T_{1/2}>1.05\times10^{24}$~yr.
Fig.~\ref{fig:alldata_roi} shows the events that passed all cuts for all $\beta\beta$-decay to e.s.\ $\gamma$ peaks, with the signal and BG ROIs highlighted.

\subsection{Discussion of Results}
This result sets the most stringent limits and has the greatest sensitivity to date for $\beta\beta$-decay of $^{76}$Ge to all excited states of $^{76}$Se.
Table~\ref{limits} lists the previous best limits along with those set by this work.
The \textsc{Majorana Demonstrator} derives its increased sensitivity relative to the results from GERDA Phase~I~\cite{gerda:ES} from several factors.
First, the \textsc{Demonstrator} had higher detection efficiency due to the lack of liquid argon surrounding the detectors, which shielded the dexcitation $\gamma$s.
Second, the dominant background in GERDA's search for excited state decays came from cosmogenic $^{42}$K in its liquid argon shield, which does not exist in the \textsc{Demonstrator}.
Finally, the \textsc{Demonstrator} had significantly better energy resolution due to the lack of cross-talk between detectors, which worsened GERDA's resolution for multi-detector events.
\begin{table}[h]
  \centering
  \caption{\label{limits} Table of limits at 90\% CL for each $\beta\beta$-decay to e.s.\ decay mode}
  \begin{tabular}{|c|c|c|}
  \hline
  Decay Mode & Previous Limit & MJD Limit \\
  \hline\hline
  $0^+_{g.s.} \xrightarrow{2\nu\beta\beta} 0^+_1$ & $3.7\times10^{23}$ yr \cite{gerda:ES} & $7.5\times10^{23}$ yr \\
  $0^+_{g.s.} \xrightarrow{2\nu\beta\beta} 2^+_1$ & $1.6\times10^{23}$ yr \cite{gerda:ES} & $7.7\times10^{23}$ yr \\
  $0^+_{g.s.} \xrightarrow{2\nu\beta\beta} 2^+_2$ & $2.3\times10^{23}$ yr \cite{gerda:ES} & $1.3\times10^{24}$ yr \\
  $0^+_{g.s.} \xrightarrow{0\nu\beta\beta} 0^+_1$ & $1.3\times10^{22}$ yr \cite{morales1988} & $4.0\times10^{24}$ yr \\
  $0^+_{g.s.} \xrightarrow{0\nu\beta\beta} 2^+_1$ & $1.3\times10^{23}$ yr \cite{maier1994} & $2.1\times10^{24}$ yr \\
  $0^+_{g.s.} \xrightarrow{0\nu\beta\beta} 2^+_2$ & $1.4\times10^{21}$ yr \cite{barabash1995:2} & $9.7\times10^{23}$ yr \\
  \hline
\end{tabular}

\end{table}

This result has also begun to probe recent theoretical predictions for the half-life for $2\nu\beta\beta$ to the $0^+_1$ e.s.\ of $^{76}$Se.
A recent half-life prediction using Renormalized proton-neutron Quasi-Random Phase Approximation (RQRPA) of $(1.2-5.8)\times 10^{23}$~yr~\cite{gerda:ES} has been excluded with CL 97\% by this result.
Combining the best measurement of the $2\nu\beta\beta$ to g.s.\ half-life~\cite{gerda:ES} and phase-space integrals~\cite{Stoica2019} with nuclear matrix element calculations applying an effective field theory (EFT) framework~\cite{menendez2018} and interacting boson model~\cite{barea2015}, yield half-life predictions of $1.7\times10^{24}$~yr and $7.1\times10^{24}$~yr, respectively.

More progress will still be required to test half-life predictions for $2\nu\beta\beta$ to the $2^+$ e.s.\ of $^{76}$Se.
The most recent predictions for the $2^+_1$ e.s.\ range using RQRPA~\cite{Toivanen1997, Schwieger1998, Raduta2007, Unlu2014} and EFT~\cite{menendez2018} techniques yield half-lives in excess of $1.0\times10^{26}$~yr, well beyond the reach of the \textsc{Majorana Demonstrator}.
Similarly, a calculation using RQRPA for the $2^+_2$ e.s.\ yielded a half-life in excess of $7\times10^{27}$~yr~\cite{Toivanen1997}.
Without knowledge of $|m_{\beta\beta}|$ or other BSM physics parameters involved in generating $0\nu\beta\beta$, it is impossible to generate a specific half-life prediction for neutrinoless decay modes to e.s.
By applying nuclear matrix elements calculated for $0\nu\beta\beta$ to the $0^+_1$ e.s.\ under the assumption of light neutrino exchange, we can calculate upper limits on $|m_{\beta\beta}|$ of $3.2-7.7$~eV~\cite{Hyvarinen2016, menendez2009}.

This analysis leaves some room for improvment; for example, by taking advantage of the PPC detectors' sensitivity to events that are multi-site within a single detector\cite{mjd:avse}, we could refine the search to achieve greater signal acceptance.
In fact, one could potentially use pulse-shape information to change the signal selection criterion to include single-detector events in which a deexcitation $\gamma$ is absorbed in the same detector as the $\beta\beta$ site, which could greatly improve signal acceptance.
Furthermore, the \textsc{Majorana Demonstrator} is continuing to collect data.
A future analysis, with increased exposure and improved signal sensitivity, may be able to test the effective field theory prediction.

\section{Acknowledgments}
This material is based upon work supported by the U.S.~Department of Energy, Office of Science, Office of Nuclear Physics under contract / award numbers DE-AC02-05CH11231,  DE-AC05-00OR22725, DE-AC05-76RL0130, DE-FG02-97ER41020, DE-FG02-97ER41033, DE-FG02-97ER41041, DE-SC0012612, DE-SC0014445, DE-SC0018060, and LANLEM77. We acknowledge support from the Particle Astrophysics Program and Nuclear Physics Program of the National Science Foundation through grant numbers MRI-0923142, PHY-1003399, PHY-1102292, PHY-1206314, PHY-1614611, PHY-1812409, and PHY-1812356.  We gratefully acknowledge the support of the Laboratory Directed Research \& Development (LDRD) program at Lawrence Berkeley National Laboratory for this work. We gratefully acknowledge the support of the U.S.~Department of Energy through the Los Alamos National Laboratory LDRD Program and through the Pacific Northwest National Laboratory LDRD Program for this work.  We acknowledge support from the Russian Foundation for Basic Research, grant No.~15-02-02919. We acknowledge the support of the Natural Sciences and Engineering Research Council of Canada, funding reference number SAPIN-2017-00023, and from the Canada Foundation for Innovation John R.~Evans Leaders Fund.  This research used resources provided by the Oak Ridge Leadership Computing Facility at Oak Ridge National Laboratory and by the National Energy Research Scientific Computing Center, a U.S.~Department of Energy Office of Science User Facility. We thank our hosts and colleagues at the Sanford Underground Research Facility for their support.

\bibliography{mjd_excited_states}

\begin{thebibliography}{58}%
\makeatletter
\providecommand \@ifxundefined [1]{%
 \@ifx{#1\undefined}
}%
\providecommand \@ifnum [1]{%
 \ifnum #1\expandafter \@firstoftwo
 \else \expandafter \@secondoftwo
 \fi
}%
\providecommand \@ifx [1]{%
 \ifx #1\expandafter \@firstoftwo
 \else \expandafter \@secondoftwo
 \fi
}%
\providecommand \natexlab [1]{#1}%
\providecommand \enquote  [1]{``#1''}%
\providecommand \bibnamefont  [1]{#1}%
\providecommand \bibfnamefont [1]{#1}%
\providecommand \citenamefont [1]{#1}%
\providecommand \href@noop [0]{\@secondoftwo}%
\providecommand \href [0]{\begingroup \@sanitize@url \@href}%
\providecommand \@href[1]{\@@startlink{#1}\@@href}%
\providecommand \@@href[1]{\endgroup#1\@@endlink}%
\providecommand \@sanitize@url [0]{\catcode `\\12\catcode `\$12\catcode
  `\&12\catcode `\#12\catcode `\^12\catcode `\_12\catcode `\%12\relax}%
\providecommand \@@startlink[1]{}%
\providecommand \@@endlink[0]{}%
\providecommand \url  [0]{\begingroup\@sanitize@url \@url }%
\providecommand \@url [1]{\endgroup\@href {#1}{\urlprefix }}%
\providecommand \urlprefix  [0]{URL }%
\providecommand \Eprint [0]{\href }%
\providecommand \doibase [0]{http://dx.doi.org/}%
\providecommand \selectlanguage [0]{\@gobble}%
\providecommand \bibinfo  [0]{\@secondoftwo}%
\providecommand \bibfield  [0]{\@secondoftwo}%
\providecommand \translation [1]{[#1]}%
\providecommand \BibitemOpen [0]{}%
\providecommand \bibitemStop [0]{}%
\providecommand \bibitemNoStop [0]{.\EOS\space}%
\providecommand \EOS [0]{\spacefactor3000\relax}%
\providecommand \BibitemShut  [1]{\csname bibitem#1\endcsname}%
\let\auto@bib@innerbib\@empty
%</preamble>
\bibitem [{\citenamefont {Goeppert-Mayer}(1935)}]{GoeppertMayer1935}%
  \BibitemOpen
  \bibfield  {author} {\bibinfo {author} {\bibfnamefont {M.}~\bibnamefont
  {Goeppert-Mayer}},\ }\href {\doibase 10.1103/PhysRev.48.512} {\bibfield
  {journal} {\bibinfo  {journal} {Phys. Rev.}\ }\textbf {\bibinfo {volume}
  {48}},\ \bibinfo {pages} {512} (\bibinfo {year} {1935})}\BibitemShut
  {NoStop}%
\bibitem [{\citenamefont {Furry}(1939)}]{Furry1939}%
  \BibitemOpen
  \bibfield  {author} {\bibinfo {author} {\bibfnamefont {W.~H.}\ \bibnamefont
  {Furry}},\ }\href {\doibase 10.1103/PhysRev.56.1184} {\bibfield  {journal}
  {\bibinfo  {journal} {Phys. Rev.}\ }\textbf {\bibinfo {volume} {56}},\
  \bibinfo {pages} {1184} (\bibinfo {year} {1939})}\BibitemShut {NoStop}%
\bibitem [{\citenamefont {Avignone}\ \emph {et~al.}(2008)\citenamefont
  {Avignone}, \citenamefont {Elliott},\ and\ \citenamefont
  {Engel}}]{Avignone2008}%
  \BibitemOpen
  \bibfield  {author} {\bibinfo {author} {\bibfnamefont {F.~T.}\ \bibnamefont
  {Avignone}}, \bibinfo {author} {\bibfnamefont {S.~R.}\ \bibnamefont
  {Elliott}}, \ and\ \bibinfo {author} {\bibfnamefont {J.}~\bibnamefont
  {Engel}},\ }\href {\doibase 10.1103/RevModPhys.80.481} {\bibfield  {journal}
  {\bibinfo  {journal} {Rev. Mod. Phys.}\ }\textbf {\bibinfo {volume} {80}},\
  \bibinfo {pages} {481} (\bibinfo {year} {2008})}\BibitemShut {NoStop}%
\bibitem [{\citenamefont {Vergados}\ \emph {et~al.}(2012)\citenamefont
  {Vergados}, \citenamefont {Ejiri},\ and\ \citenamefont
  {{\v{S}}imkovic}}]{Vergados2012}%
  \BibitemOpen
  \bibfield  {author} {\bibinfo {author} {\bibfnamefont {J.~D.}\ \bibnamefont
  {Vergados}}, \bibinfo {author} {\bibfnamefont {H.}~\bibnamefont {Ejiri}}, \
  and\ \bibinfo {author} {\bibfnamefont {F.}~\bibnamefont {{\v{S}}imkovic}},\
  }\href {\doibase 10.1088/0034-4885/75/10/106301} {\bibfield  {journal}
  {\bibinfo  {journal} {Reports on Progress in Physics}\ }\textbf {\bibinfo
  {volume} {75}},\ \bibinfo {pages} {106301} (\bibinfo {year}
  {2012})}\BibitemShut {NoStop}%
\bibitem [{\citenamefont {Päs}\ and\ \citenamefont
  {Rodejohann}(2015)}]{Rodejohann2015}%
  \BibitemOpen
  \bibfield  {author} {\bibinfo {author} {\bibfnamefont {H.}~\bibnamefont
  {Päs}}\ and\ \bibinfo {author} {\bibfnamefont {W.}~\bibnamefont
  {Rodejohann}},\ }\href {\doibase 10.1088/1367-2630/17/11/115010} {\bibfield
  {journal} {\bibinfo  {journal} {New Journal of Physics}\ }\textbf {\bibinfo
  {volume} {17}},\ \bibinfo {pages} {115010} (\bibinfo {year}
  {2015})}\BibitemShut {NoStop}%
\bibitem [{\citenamefont {Dell’Oro}\ \emph {et~al.}(2016)\citenamefont
  {Dell’Oro}, \citenamefont {Marcocci}, \citenamefont {Viel},\ and\
  \citenamefont {Vissani}}]{DelloOro2016}%
  \BibitemOpen
  \bibfield  {author} {\bibinfo {author} {\bibfnamefont {S.}~\bibnamefont
  {Dell’Oro}}, \bibinfo {author} {\bibfnamefont {S.}~\bibnamefont
  {Marcocci}}, \bibinfo {author} {\bibfnamefont {M.}~\bibnamefont {Viel}}, \
  and\ \bibinfo {author} {\bibfnamefont {F.}~\bibnamefont {Vissani}},\ }\href
  {\doibase 10.1155/2016/2162659} {\bibfield  {journal} {\bibinfo  {journal}
  {Advances in High Energy Physics}\ }\textbf {\bibinfo {volume} {2016}},\
  \bibinfo {pages} {2162659} (\bibinfo {year} {2016})}\BibitemShut {NoStop}%
\bibitem [{\citenamefont {Dolinski}\ \emph {et~al.}(2019)\citenamefont
  {Dolinski}, \citenamefont {Poon},\ and\ \citenamefont
  {Rodejohann}}]{Dolinski2019}%
  \BibitemOpen
  \bibfield  {author} {\bibinfo {author} {\bibfnamefont {M.~J.}\ \bibnamefont
  {Dolinski}}, \bibinfo {author} {\bibfnamefont {A.~W.}\ \bibnamefont {Poon}},
  \ and\ \bibinfo {author} {\bibfnamefont {W.}~\bibnamefont {Rodejohann}},\
  }\href {\doibase 10.1146/annurev-nucl-101918-023407} {\bibfield  {journal}
  {\bibinfo  {journal} {Annual Review of Nuclear and Particle Science}\
  }\textbf {\bibinfo {volume} {69}},\ \bibinfo {pages} {219} (\bibinfo {year}
  {2019})},\ \Eprint
  {http://arxiv.org/abs/https://doi.org/10.1146/annurev-nucl-101918-023407}
  {https://doi.org/10.1146/annurev-nucl-101918-023407} \BibitemShut {NoStop}%
\bibitem [{\citenamefont {Majorana}(2008)}]{Majorana1937}%
  \BibitemOpen
  \bibfield  {author} {\bibinfo {author} {\bibfnamefont {E.}~\bibnamefont
  {Majorana}},\ }\href {\doibase 10.1007/BF02961314} {\bibfield  {journal}
  {\bibinfo  {journal} {Il Nuovo Cimento (1924-1942)}\ }\textbf {\bibinfo
  {volume} {14}},\ \bibinfo {pages} {171} (\bibinfo {year} {2008})}\BibitemShut
  {NoStop}%
\bibitem [{\citenamefont {Schechter}\ and\ \citenamefont
  {Valle}(1982)}]{Schechter1982}%
  \BibitemOpen
  \bibfield  {author} {\bibinfo {author} {\bibfnamefont {J.}~\bibnamefont
  {Schechter}}\ and\ \bibinfo {author} {\bibfnamefont {J.~W.~F.}\ \bibnamefont
  {Valle}},\ }\href {\doibase 10.1103/PhysRevD.25.2951} {\bibfield  {journal}
  {\bibinfo  {journal} {Phys. Rev. D}\ }\textbf {\bibinfo {volume} {25}},\
  \bibinfo {pages} {2951} (\bibinfo {year} {1982})}\BibitemShut {NoStop}%
\bibitem [{\citenamefont {{Sakharov}}(1967)}]{Sakharov1967}%
  \BibitemOpen
  \bibfield  {author} {\bibinfo {author} {\bibfnamefont {A.~D.}\ \bibnamefont
  {{Sakharov}}},\ }\href {\doibase 10.1070/PU1991v034n05ABEH002497} {\bibfield
  {journal} {\bibinfo  {journal} {Pisma Zh. Eksp. Teor. Fiz.}\ }\textbf
  {\bibinfo {volume} {5}},\ \bibinfo {pages} {32} (\bibinfo {year}
  {1967})}\BibitemShut {NoStop}%
\bibitem [{\citenamefont {Fukugita}\ and\ \citenamefont
  {Yanagida}(1986)}]{FUKUGITA1986}%
  \BibitemOpen
  \bibfield  {author} {\bibinfo {author} {\bibfnamefont {M.}~\bibnamefont
  {Fukugita}}\ and\ \bibinfo {author} {\bibfnamefont {T.}~\bibnamefont
  {Yanagida}},\ }\href {\doibase https://doi.org/10.1016/0370-2693(86)91126-3}
  {\bibfield  {journal} {\bibinfo  {journal} {Physics Letters B}\ }\textbf
  {\bibinfo {volume} {174}},\ \bibinfo {pages} {45 } (\bibinfo {year}
  {1986})}\BibitemShut {NoStop}%
\bibitem [{\citenamefont {Suhonen}(2017)}]{Suhonen2017}%
  \BibitemOpen
  \bibfield  {author} {\bibinfo {author} {\bibfnamefont {J.}~\bibnamefont
  {Suhonen}},\ }\href {\doibase 10.1103/PhysRevC.96.055501} {\bibfield
  {journal} {\bibinfo  {journal} {Phys. Rev. C}\ }\textbf {\bibinfo {volume}
  {96}},\ \bibinfo {pages} {055501} (\bibinfo {year} {2017})}\BibitemShut
  {NoStop}%
\bibitem [{\citenamefont {Kotila}\ and\ \citenamefont
  {Iachello}(2012)}]{Kotila2012}%
  \BibitemOpen
  \bibfield  {author} {\bibinfo {author} {\bibfnamefont {J.}~\bibnamefont
  {Kotila}}\ and\ \bibinfo {author} {\bibfnamefont {F.}~\bibnamefont
  {Iachello}},\ }\href {\doibase 10.1103/PhysRevC.85.034316} {\bibfield
  {journal} {\bibinfo  {journal} {Phys. Rev. C}\ }\textbf {\bibinfo {volume}
  {85}},\ \bibinfo {pages} {034316} (\bibinfo {year} {2012})}\BibitemShut
  {NoStop}%
\bibitem [{\citenamefont {Stoica}\ and\ \citenamefont
  {Mirea}(2019)}]{Stoica2019}%
  \BibitemOpen
  \bibfield  {author} {\bibinfo {author} {\bibfnamefont {S.}~\bibnamefont
  {Stoica}}\ and\ \bibinfo {author} {\bibfnamefont {M.}~\bibnamefont {Mirea}},\
  }\href {\doibase 10.3389/fphy.2019.00012} {\bibfield  {journal} {\bibinfo
  {journal} {Frontiers in Physics}\ }\textbf {\bibinfo {volume} {7}},\ \bibinfo
  {pages} {12} (\bibinfo {year} {2019})}\BibitemShut {NoStop}%
\bibitem [{\citenamefont {Engel}\ and\ \citenamefont
  {Men{\'{e}}ndez}(2017)}]{Engel2017}%
  \BibitemOpen
  \bibfield  {author} {\bibinfo {author} {\bibfnamefont {J.}~\bibnamefont
  {Engel}}\ and\ \bibinfo {author} {\bibfnamefont {J.}~\bibnamefont
  {Men{\'{e}}ndez}},\ }\href {\doibase 10.1088/1361-6633/aa5bc5} {\bibfield
  {journal} {\bibinfo  {journal} {Reports on Progress in Physics}\ }\textbf
  {\bibinfo {volume} {80}},\ \bibinfo {pages} {046301} (\bibinfo {year}
  {2017})}\BibitemShut {NoStop}%
\bibitem [{\citenamefont {Barabash}(2015)}]{barabash2015}%
  \BibitemOpen
  \bibfield  {author} {\bibinfo {author} {\bibfnamefont {A.}~\bibnamefont
  {Barabash}},\ }\href {\doibase
  https://doi.org/10.1016/j.nuclphysa.2015.01.001} {\bibfield  {journal}
  {\bibinfo  {journal} {Nuclear Physics A}\ }\textbf {\bibinfo {volume}
  {935}},\ \bibinfo {pages} {52 } (\bibinfo {year} {2015})}\BibitemShut
  {NoStop}%
\bibitem [{\citenamefont {Barabash}\ \emph
  {et~al.}(1995{\natexlab{a}})\citenamefont {Barabash}, \citenamefont
  {Avignone}, \citenamefont {Collar}, \citenamefont {Guerard}, \citenamefont
  {Arthur}, \citenamefont {Brodzinski}, \citenamefont {Miley}, \citenamefont
  {Reeves}, \citenamefont {Meier}, \citenamefont {Ruddick},\ and\ \citenamefont
  {Umatov}}]{BARABASH1995}%
  \BibitemOpen
  \bibfield  {author} {\bibinfo {author} {\bibfnamefont {A.}~\bibnamefont
  {Barabash}}, \bibinfo {author} {\bibfnamefont {F.}~\bibnamefont {Avignone}},
  \bibinfo {author} {\bibfnamefont {J.}~\bibnamefont {Collar}}, \bibinfo
  {author} {\bibfnamefont {C.}~\bibnamefont {Guerard}}, \bibinfo {author}
  {\bibfnamefont {R.}~\bibnamefont {Arthur}}, \bibinfo {author} {\bibfnamefont
  {R.}~\bibnamefont {Brodzinski}}, \bibinfo {author} {\bibfnamefont
  {H.}~\bibnamefont {Miley}}, \bibinfo {author} {\bibfnamefont
  {J.}~\bibnamefont {Reeves}}, \bibinfo {author} {\bibfnamefont
  {J.}~\bibnamefont {Meier}}, \bibinfo {author} {\bibfnamefont
  {K.}~\bibnamefont {Ruddick}}, \ and\ \bibinfo {author} {\bibfnamefont
  {V.}~\bibnamefont {Umatov}},\ }\href {\doibase
  https://doi.org/10.1016/0370-2693(94)01657-X} {\bibfield  {journal} {\bibinfo
   {journal} {Physics Letters B}\ }\textbf {\bibinfo {volume} {345}},\ \bibinfo
  {pages} {408 } (\bibinfo {year} {1995}{\natexlab{a}})}\BibitemShut {NoStop}%
\bibitem [{\citenamefont {Barabash}\ \emph {et~al.}(2004)\citenamefont
  {Barabash}, \citenamefont {Hubert}, \citenamefont {Hubert},\ and\
  \citenamefont {Umatov}}]{Barabash2004}%
  \BibitemOpen
  \bibfield  {author} {\bibinfo {author} {\bibfnamefont {A.~S.}\ \bibnamefont
  {Barabash}}, \bibinfo {author} {\bibfnamefont {F.}~\bibnamefont {Hubert}},
  \bibinfo {author} {\bibfnamefont {P.}~\bibnamefont {Hubert}}, \ and\ \bibinfo
  {author} {\bibfnamefont {V.~I.}\ \bibnamefont {Umatov}},\ }\href {\doibase
  https://doi.org/10.1134/1.1675911} {\bibfield  {journal} {\bibinfo  {journal}
  {JETP Lett.}\ }\textbf {\bibinfo {volume} {79}},\ \bibinfo {pages} {10}
  (\bibinfo {year} {2004})}\BibitemShut {NoStop}%
\bibitem [{\citenamefont {Barabash}\ \emph {et~al.}(2009)\citenamefont
  {Barabash}, \citenamefont {Hubert}, \citenamefont {Nachab},\ and\
  \citenamefont {Umatov}}]{Barabash2009}%
  \BibitemOpen
  \bibfield  {author} {\bibinfo {author} {\bibfnamefont {A.~S.}\ \bibnamefont
  {Barabash}}, \bibinfo {author} {\bibfnamefont {P.}~\bibnamefont {Hubert}},
  \bibinfo {author} {\bibfnamefont {A.}~\bibnamefont {Nachab}}, \ and\ \bibinfo
  {author} {\bibfnamefont {V.~I.}\ \bibnamefont {Umatov}},\ }\href {\doibase
  10.1103/PhysRevC.79.045501} {\bibfield  {journal} {\bibinfo  {journal} {Phys.
  Rev. C}\ }\textbf {\bibinfo {volume} {79}},\ \bibinfo {pages} {045501}
  (\bibinfo {year} {2009})}\BibitemShut {NoStop}%
\bibitem [{\citenamefont {Barabash}(2017)}]{Barabash2017}%
  \BibitemOpen
  \bibfield  {author} {\bibinfo {author} {\bibfnamefont {A.~S.}\ \bibnamefont
  {Barabash}},\ }\href {\doibase 10.1063/1.5007627} {\bibfield  {journal}
  {\bibinfo  {journal} {AIP Conference Proceedings}\ }\textbf {\bibinfo
  {volume} {1894}},\ \bibinfo {pages} {020002} (\bibinfo {year} {2017})},\
  \Eprint
  {http://arxiv.org/abs/https://aip.scitation.org/doi/pdf/10.1063/1.5007627}
  {https://aip.scitation.org/doi/pdf/10.1063/1.5007627} \BibitemShut {NoStop}%
\bibitem [{\citenamefont {Abgrall}\ \emph {et~al.}(2014)\citenamefont {Abgrall}
  \emph {et~al.}}]{mjd2014}%
  \BibitemOpen
  \bibfield  {author} {\bibinfo {author} {\bibfnamefont {N.}~\bibnamefont
  {Abgrall}} \emph {et~al.} (\bibinfo {collaboration} {Majorana
  Collaboration}),\ }\href {\doibase 10.1155/2014/365432} {\bibfield  {journal}
  {\bibinfo  {journal} {Advances in High Energy Physics}\ }\textbf {\bibinfo
  {volume} {2014}},\ \bibinfo {pages} {365432} (\bibinfo {year}
  {2014})}\BibitemShut {NoStop}%
\bibitem [{\citenamefont {Alvis}\ \emph
  {et~al.}(2019{\natexlab{a}})\citenamefont {Alvis} \emph {et~al.}}]{mjd2019}%
  \BibitemOpen
  \bibfield  {author} {\bibinfo {author} {\bibfnamefont {S.~I.}\ \bibnamefont
  {Alvis}} \emph {et~al.} (\bibinfo {collaboration} {Majorana Collaboration}),\
  }\href {\doibase 10.1103/PhysRevC.100.025501} {\bibfield  {journal} {\bibinfo
   {journal} {Phys. Rev. C}\ }\textbf {\bibinfo {volume} {100}},\ \bibinfo
  {pages} {025501} (\bibinfo {year} {2019}{\natexlab{a}})}\BibitemShut
  {NoStop}%
\bibitem [{\citenamefont {Agostini}\ \emph {et~al.}(2019)\citenamefont
  {Agostini} \emph {et~al.}}]{gerda:2019}%
  \BibitemOpen
  \bibfield  {author} {\bibinfo {author} {\bibfnamefont {M.}~\bibnamefont
  {Agostini}} \emph {et~al.} (\bibinfo {collaboration} {GERDA Collaboration}),\
  }\href {\doibase 10.1126/science.aav8613} {\bibfield  {journal} {\bibinfo
  {journal} {Science}\ }\textbf {\bibinfo {volume} {365}},\ \bibinfo {pages}
  {1445} (\bibinfo {year} {2019})}\BibitemShut {NoStop}%
\bibitem [{\citenamefont {Agostini}\ \emph {et~al.}(2020)\citenamefont
  {Agostini} \emph {et~al.}}]{gerda:2020}%
  \BibitemOpen
  \bibfield  {author} {\bibinfo {author} {\bibfnamefont {M.}~\bibnamefont
  {Agostini}} \emph {et~al.} (\bibinfo {collaboration} {GERDA Collaboration}),\
  }\href {\doibase 10.1103/PhysRevLett.125.252502} {\bibfield  {journal}
  {\bibinfo  {journal} {Phys. Rev. Lett.}\ }\textbf {\bibinfo {volume} {125}},\
  \bibinfo {pages} {252502} (\bibinfo {year} {2020})}\BibitemShut {NoStop}%
\bibitem [{\citenamefont {Abgrall}\ \emph {et~al.}(2017)\citenamefont {Abgrall}
  \emph {et~al.}}]{LEGEND200}%
  \BibitemOpen
  \bibfield  {author} {\bibinfo {author} {\bibfnamefont {N.}~\bibnamefont
  {Abgrall}} \emph {et~al.},\ }\href {\doibase 10.1063/1.5007652} {\bibfield
  {journal} {\bibinfo  {journal} {AIP Conference Proceedings}\ }\textbf
  {\bibinfo {volume} {1894}},\ \bibinfo {pages} {020027} (\bibinfo {year}
  {2017})}\BibitemShut {NoStop}%
\bibitem [{\citenamefont {Fiorini}(1978)}]{Fiorini1977}%
  \BibitemOpen
  \bibfield  {author} {\bibinfo {author} {\bibfnamefont {E.}~\bibnamefont
  {Fiorini}},\ }in\ \href@noop {} {\emph {\bibinfo {booktitle} {{Proc. Int.
  Conf. Neutrino '77}}}},\ Vol.~\bibinfo {volume} {2}\ (\bibinfo  {publisher}
  {{Akademy of Sciences of USSR}},\ \bibinfo {address} {Moscow},\ \bibinfo
  {year} {1978})\ pp.\ \bibinfo {pages} {315--320}\BibitemShut {NoStop}%
\bibitem [{\citenamefont {Agostini}\ \emph {et~al.}(2015)\citenamefont
  {Agostini} \emph {et~al.}}]{gerda:ES}%
  \BibitemOpen
  \bibfield  {author} {\bibinfo {author} {\bibfnamefont {M.}~\bibnamefont
  {Agostini}} \emph {et~al.} (\bibinfo {collaboration} {GERDA Collaboration}),\
  }\href {\doibase 10.1088/0954-3899/42/11/115201} {\bibfield  {journal}
  {\bibinfo  {journal} {Journal of Physics G: Nuclear and Particle Physics}\
  }\textbf {\bibinfo {volume} {42}},\ \bibinfo {pages} {115201} (\bibinfo
  {year} {2015})}\BibitemShut {NoStop}%
\bibitem [{\citenamefont {Dolgov}\ and\ \citenamefont
  {Smirnov}(2005)}]{Dolgov2005}%
  \BibitemOpen
  \bibfield  {author} {\bibinfo {author} {\bibfnamefont {A.}~\bibnamefont
  {Dolgov}}\ and\ \bibinfo {author} {\bibfnamefont {A.}~\bibnamefont
  {Smirnov}},\ }\href {\doibase https://doi.org/10.1016/j.physletb.2005.06.035}
  {\bibfield  {journal} {\bibinfo  {journal} {Physics Letters B}\ }\textbf
  {\bibinfo {volume} {621}},\ \bibinfo {pages} {1 } (\bibinfo {year}
  {2005})}\BibitemShut {NoStop}%
\bibitem [{\citenamefont {Barabash}\ \emph {et~al.}(2007)\citenamefont
  {Barabash}, \citenamefont {Dolgov}, \citenamefont {Dvornický}, \citenamefont
  {Šimkovic},\ and\ \citenamefont {Smirnov}}]{Barabash2007}%
  \BibitemOpen
  \bibfield  {author} {\bibinfo {author} {\bibfnamefont {A.}~\bibnamefont
  {Barabash}}, \bibinfo {author} {\bibfnamefont {A.}~\bibnamefont {Dolgov}},
  \bibinfo {author} {\bibfnamefont {R.}~\bibnamefont {Dvornický}}, \bibinfo
  {author} {\bibfnamefont {F.}~\bibnamefont {Šimkovic}}, \ and\ \bibinfo
  {author} {\bibfnamefont {A.}~\bibnamefont {Smirnov}},\ }\href {\doibase
  https://doi.org/10.1016/j.nuclphysb.2007.05.033} {\bibfield  {journal}
  {\bibinfo  {journal} {Nuclear Physics B}\ }\textbf {\bibinfo {volume}
  {783}},\ \bibinfo {pages} {90 } (\bibinfo {year} {2007})}\BibitemShut
  {NoStop}%
\bibitem [{\citenamefont {{{\v S}imkovic}}\ and\ \citenamefont
  {{Faessler}}(2002)}]{Simkovic2002}%
  \BibitemOpen
  \bibfield  {author} {\bibinfo {author} {\bibfnamefont {F.}~\bibnamefont {{{\v
  S}imkovic}}}\ and\ \bibinfo {author} {\bibfnamefont {A.}~\bibnamefont
  {{Faessler}}},\ }\href {\doibase 10.1016/S0146-6410(02)00125-4} {\bibfield
  {journal} {\bibinfo  {journal} {Progress in Particle and Nuclear Physics}\
  }\textbf {\bibinfo {volume} {48}},\ \bibinfo {pages} {201} (\bibinfo {year}
  {2002})},\ \Eprint {http://arxiv.org/abs/hep-ph/0112272} {hep-ph/0112272}
  \BibitemShut {NoStop}%
\bibitem [{\citenamefont {Barbeau}\ \emph {et~al.}(2007)\citenamefont
  {Barbeau}, \citenamefont {Collar},\ and\ \citenamefont
  {Tench}}]{Barbeau2007}%
  \BibitemOpen
  \bibfield  {author} {\bibinfo {author} {\bibfnamefont {P.~S.}\ \bibnamefont
  {Barbeau}}, \bibinfo {author} {\bibfnamefont {J.~I.}\ \bibnamefont {Collar}},
  \ and\ \bibinfo {author} {\bibfnamefont {O.}~\bibnamefont {Tench}},\ }\href
  {\doibase 10.1088/1475-7516/2007/09/009} {\bibfield  {journal} {\bibinfo
  {journal} {Journal of Cosmology and Astroparticle Physics}\ }\textbf
  {\bibinfo {volume} {2007}},\ \bibinfo {pages} {009} (\bibinfo {year}
  {2007})}\BibitemShut {NoStop}%
\bibitem [{\citenamefont {Alvis}\ \emph
  {et~al.}(2019{\natexlab{b}})\citenamefont {Alvis} \emph {et~al.}}]{mjd:avse}%
  \BibitemOpen
  \bibfield  {author} {\bibinfo {author} {\bibfnamefont {S.~I.}\ \bibnamefont
  {Alvis}} \emph {et~al.} (\bibinfo {collaboration} {Majorana Collaboration}),\
  }\href {\doibase 10.1103/PhysRevC.99.065501} {\bibfield  {journal} {\bibinfo
  {journal} {Phys. Rev. C}\ }\textbf {\bibinfo {volume} {99}},\ \bibinfo
  {pages} {065501} (\bibinfo {year} {2019}{\natexlab{b}})}\BibitemShut
  {NoStop}%
\bibitem [{\citenamefont {Abgrall}\ \emph {et~al.}(2016)\citenamefont {Abgrall}
  \emph {et~al.}}]{mjd:assay}%
  \BibitemOpen
  \bibfield  {author} {\bibinfo {author} {\bibfnamefont {N.}~\bibnamefont
  {Abgrall}} \emph {et~al.},\ }\href {\doibase
  https://doi.org/10.1016/j.nima.2016.04.070} {\bibfield  {journal} {\bibinfo
  {journal} {Nuclear Instruments and Methods in Physics Research Section A:
  Accelerators, Spectrometers, Detectors and Associated Equipment}\ }\textbf
  {\bibinfo {volume} {828}},\ \bibinfo {pages} {22 } (\bibinfo {year}
  {2016})}\BibitemShut {NoStop}%
\bibitem [{\citenamefont {{Barton}}\ \emph {et~al.}(2011)\citenamefont
  {{Barton}}, \citenamefont {{Luke}}, \citenamefont {{Amman}}, \citenamefont
  {{Chan}}, \citenamefont {{Detwiler}}, \citenamefont {{Loach}}, \citenamefont
  {{Martin}}, \citenamefont {{Poon}}, \citenamefont {{Tindall}},\ and\
  \citenamefont {{Vetter}}}]{mjd:LMFE}%
  \BibitemOpen
  \bibfield  {author} {\bibinfo {author} {\bibfnamefont {P.}~\bibnamefont
  {{Barton}}}, \bibinfo {author} {\bibfnamefont {P.}~\bibnamefont {{Luke}}},
  \bibinfo {author} {\bibfnamefont {M.}~\bibnamefont {{Amman}}}, \bibinfo
  {author} {\bibfnamefont {Y.}~\bibnamefont {{Chan}}}, \bibinfo {author}
  {\bibfnamefont {J.}~\bibnamefont {{Detwiler}}}, \bibinfo {author}
  {\bibfnamefont {J.}~\bibnamefont {{Loach}}}, \bibinfo {author} {\bibfnamefont
  {R.}~\bibnamefont {{Martin}}}, \bibinfo {author} {\bibfnamefont
  {A.}~\bibnamefont {{Poon}}}, \bibinfo {author} {\bibfnamefont
  {C.}~\bibnamefont {{Tindall}}}, \ and\ \bibinfo {author} {\bibfnamefont
  {K.}~\bibnamefont {{Vetter}}},\ }in\ \href {\doibase
  10.1109/NSSMIC.2011.6154397} {\emph {\bibinfo {booktitle} {Proceedings of the
  2011 IEEE Nuclear Science Symposium and Medical Imaging Conference
  (NSS/MIC2011)}}}\ (\bibinfo  {publisher} {IEEE},\ \bibinfo {year} {2011})\
  pp.\ \bibinfo {pages} {1976--1979}\BibitemShut {NoStop}%
\bibitem [{\citenamefont {Bugg}\ \emph {et~al.}(2014)\citenamefont {Bugg},
  \citenamefont {Efremenko},\ and\ \citenamefont {Vasilyev}}]{mjd:muonveto}%
  \BibitemOpen
  \bibfield  {author} {\bibinfo {author} {\bibfnamefont {W.}~\bibnamefont
  {Bugg}}, \bibinfo {author} {\bibfnamefont {Y.}~\bibnamefont {Efremenko}}, \
  and\ \bibinfo {author} {\bibfnamefont {S.}~\bibnamefont {Vasilyev}},\ }\href
  {\doibase https://doi.org/10.1016/j.nima.2014.05.055} {\bibfield  {journal}
  {\bibinfo  {journal} {Nuclear Instruments and Methods in Physics Research
  Section A: Accelerators, Spectrometers, Detectors and Associated Equipment}\
  }\textbf {\bibinfo {volume} {758}},\ \bibinfo {pages} {91 } (\bibinfo {year}
  {2014})}\BibitemShut {NoStop}%
\bibitem [{\citenamefont {Vetter}\ \emph {et~al.}(2000)\citenamefont {Vetter},
  \citenamefont {Kuhn}, \citenamefont {Lee}, \citenamefont {Clark},
  \citenamefont {Cromaz}, \citenamefont {Deleplanque}, \citenamefont {Diamond},
  \citenamefont {Fallon}, \citenamefont {Lane}, \citenamefont {Macchiavelli},
  \citenamefont {Maier}, \citenamefont {Stephens}, \citenamefont {Svensson},\
  and\ \citenamefont {Yaver}}]{Vetter2000}%
  \BibitemOpen
  \bibfield  {author} {\bibinfo {author} {\bibfnamefont {K.}~\bibnamefont
  {Vetter}}, \bibinfo {author} {\bibfnamefont {A.}~\bibnamefont {Kuhn}},
  \bibinfo {author} {\bibfnamefont {I.}~\bibnamefont {Lee}}, \bibinfo {author}
  {\bibfnamefont {R.}~\bibnamefont {Clark}}, \bibinfo {author} {\bibfnamefont
  {M.}~\bibnamefont {Cromaz}}, \bibinfo {author} {\bibfnamefont
  {M.}~\bibnamefont {Deleplanque}}, \bibinfo {author} {\bibfnamefont
  {R.}~\bibnamefont {Diamond}}, \bibinfo {author} {\bibfnamefont
  {P.}~\bibnamefont {Fallon}}, \bibinfo {author} {\bibfnamefont
  {G.}~\bibnamefont {Lane}}, \bibinfo {author} {\bibfnamefont {A.}~\bibnamefont
  {Macchiavelli}}, \bibinfo {author} {\bibfnamefont {M.}~\bibnamefont {Maier}},
  \bibinfo {author} {\bibfnamefont {F.}~\bibnamefont {Stephens}}, \bibinfo
  {author} {\bibfnamefont {C.}~\bibnamefont {Svensson}}, \ and\ \bibinfo
  {author} {\bibfnamefont {H.}~\bibnamefont {Yaver}},\ }\href {\doibase
  https://doi.org/10.1016/S0168-9002(00)00431-9} {\bibfield  {journal}
  {\bibinfo  {journal} {Nuclear Instruments and Methods in Physics Research
  Section A: Accelerators, Spectrometers, Detectors and Associated Equipment}\
  }\textbf {\bibinfo {volume} {452}},\ \bibinfo {pages} {105 } (\bibinfo {year}
  {2000})}\BibitemShut {NoStop}%
\bibitem [{\citenamefont {{Anderson}}\ \emph {et~al.}(2009)\citenamefont
  {{Anderson}}, \citenamefont {{Brito}}, \citenamefont {{Doering}},
  \citenamefont {{Hayden}}, \citenamefont {{Holmes}}, \citenamefont {{Joseph}},
  \citenamefont {{Yaver}},\ and\ \citenamefont {{Zimmermann}}}]{Anderson2009}%
  \BibitemOpen
  \bibfield  {author} {\bibinfo {author} {\bibfnamefont {J.}~\bibnamefont
  {{Anderson}}}, \bibinfo {author} {\bibfnamefont {R.}~\bibnamefont {{Brito}}},
  \bibinfo {author} {\bibfnamefont {D.}~\bibnamefont {{Doering}}}, \bibinfo
  {author} {\bibfnamefont {T.}~\bibnamefont {{Hayden}}}, \bibinfo {author}
  {\bibfnamefont {B.}~\bibnamefont {{Holmes}}}, \bibinfo {author}
  {\bibfnamefont {J.}~\bibnamefont {{Joseph}}}, \bibinfo {author}
  {\bibfnamefont {H.}~\bibnamefont {{Yaver}}}, \ and\ \bibinfo {author}
  {\bibfnamefont {S.}~\bibnamefont {{Zimmermann}}},\ }\href {\doibase
  10.1109/TNS.2008.2009444} {\bibfield  {journal} {\bibinfo  {journal} {IEEE
  Transactions on Nuclear Science}\ }\textbf {\bibinfo {volume} {56}},\
  \bibinfo {pages} {258} (\bibinfo {year} {2009})}\BibitemShut {NoStop}%
\bibitem [{\citenamefont {{Abgrall}}\ \emph {et~al.}(2020)\citenamefont
  {{Abgrall}} \emph {et~al.}}]{mjd:digitizer_nonlinearity}%
  \BibitemOpen
  \bibfield  {author} {\bibinfo {author} {\bibfnamefont {N.}~\bibnamefont
  {{Abgrall}}} \emph {et~al.} (\bibinfo {collaboration} {Majorana
  Collaboration}),\ }\href@noop {} {\bibfield  {journal} {\bibinfo  {journal}
  {IEEE Trans. Nucl. Sci. (submitted)}\ ,\ \bibinfo {eid} {arXiv:2003.04128}}
  (\bibinfo {year} {2020})},\ \Eprint {http://arxiv.org/abs/2003.04128}
  {arXiv:2003.04128 [physics.ins-det]} \BibitemShut {NoStop}%
\bibitem [{\citenamefont {Singh}(1995)}]{SINGH1995}%
  \BibitemOpen
  \bibfield  {author} {\bibinfo {author} {\bibfnamefont {B.}~\bibnamefont
  {Singh}},\ }\href {\doibase https://doi.org/10.1006/ndsh.1995.1005}
  {\bibfield  {journal} {\bibinfo  {journal} {Nuclear Data Sheets}\ }\textbf
  {\bibinfo {volume} {74}},\ \bibinfo {pages} {63 } (\bibinfo {year}
  {1995})}\BibitemShut {NoStop}%
\bibitem [{\citenamefont {{Boswell}}\ \emph {et~al.}(2011)\citenamefont
  {{Boswell}}, \citenamefont {{Chan}}, \citenamefont {{Detwiler}},
  \citenamefont {{Finnerty}}, \citenamefont {{Henning}}, \citenamefont
  {{Gehman}}, \citenamefont {{Johnson}}, \citenamefont {{Jordan}},
  \citenamefont {{Kazkaz}}, \citenamefont {{Knapp}}, \citenamefont
  {{Kroninger}}, \citenamefont {{Lenz}}, \citenamefont {{Leviner}},
  \citenamefont {{Liu}}, \citenamefont {{Liu}}, \citenamefont {{MacMullin}},
  \citenamefont {{Marino}}, \citenamefont {{Mokhtarani}}, \citenamefont
  {{Pandola}}, \citenamefont {{Schubert}}, \citenamefont {{Schubert}},
  \citenamefont {{Tomei}},\ and\ \citenamefont {{Volynets}}}]{mage2011}%
  \BibitemOpen
  \bibfield  {author} {\bibinfo {author} {\bibfnamefont {M.}~\bibnamefont
  {{Boswell}}}, \bibinfo {author} {\bibfnamefont {Y.}~\bibnamefont {{Chan}}},
  \bibinfo {author} {\bibfnamefont {J.~A.}\ \bibnamefont {{Detwiler}}},
  \bibinfo {author} {\bibfnamefont {P.}~\bibnamefont {{Finnerty}}}, \bibinfo
  {author} {\bibfnamefont {R.}~\bibnamefont {{Henning}}}, \bibinfo {author}
  {\bibfnamefont {V.~M.}\ \bibnamefont {{Gehman}}}, \bibinfo {author}
  {\bibfnamefont {R.~A.}\ \bibnamefont {{Johnson}}}, \bibinfo {author}
  {\bibfnamefont {D.~V.}\ \bibnamefont {{Jordan}}}, \bibinfo {author}
  {\bibfnamefont {K.}~\bibnamefont {{Kazkaz}}}, \bibinfo {author}
  {\bibfnamefont {M.}~\bibnamefont {{Knapp}}}, \bibinfo {author} {\bibfnamefont
  {K.}~\bibnamefont {{Kroninger}}}, \bibinfo {author} {\bibfnamefont
  {D.}~\bibnamefont {{Lenz}}}, \bibinfo {author} {\bibfnamefont
  {L.}~\bibnamefont {{Leviner}}}, \bibinfo {author} {\bibfnamefont
  {J.}~\bibnamefont {{Liu}}}, \bibinfo {author} {\bibfnamefont
  {X.}~\bibnamefont {{Liu}}}, \bibinfo {author} {\bibfnamefont
  {S.}~\bibnamefont {{MacMullin}}}, \bibinfo {author} {\bibfnamefont {M.~G.}\
  \bibnamefont {{Marino}}}, \bibinfo {author} {\bibfnamefont {A.}~\bibnamefont
  {{Mokhtarani}}}, \bibinfo {author} {\bibfnamefont {L.}~\bibnamefont
  {{Pandola}}}, \bibinfo {author} {\bibfnamefont {A.~G.}\ \bibnamefont
  {{Schubert}}}, \bibinfo {author} {\bibfnamefont {J.}~\bibnamefont
  {{Schubert}}}, \bibinfo {author} {\bibfnamefont {C.}~\bibnamefont {{Tomei}}},
  \ and\ \bibinfo {author} {\bibfnamefont {O.}~\bibnamefont {{Volynets}}},\
  }\href {\doibase 10.1109/TNS.2011.2144619} {\bibfield  {journal} {\bibinfo
  {journal} {IEEE Transactions on Nuclear Science}\ }\textbf {\bibinfo {volume}
  {58}},\ \bibinfo {pages} {1212} (\bibinfo {year} {2011})}\BibitemShut
  {NoStop}%
\bibitem [{\citenamefont {Agostinelli}\ \emph {et~al.}(2003)\citenamefont
  {Agostinelli} \emph {et~al.}}]{geant2003}%
  \BibitemOpen
  \bibfield  {author} {\bibinfo {author} {\bibfnamefont {S.}~\bibnamefont
  {Agostinelli}} \emph {et~al.} (\bibinfo {collaboration} {Geant4
  Collaboration}),\ }\href {\doibase
  https://doi.org/10.1016/S0168-9002(03)01368-8} {\bibfield  {journal}
  {\bibinfo  {journal} {Nuclear Instruments and Methods in Physics Research
  Section A: Accelerators, Spectrometers, Detectors and Associated Equipment}\
  }\textbf {\bibinfo {volume} {506}},\ \bibinfo {pages} {250 } (\bibinfo {year}
  {2003})}\BibitemShut {NoStop}%
\bibitem [{\citenamefont {Ponkratenko}\ \emph {et~al.}(2000)\citenamefont
  {Ponkratenko}, \citenamefont {Tretyak},\ and\ \citenamefont
  {Zdesenko}}]{Ponkratenko2000}%
  \BibitemOpen
  \bibfield  {author} {\bibinfo {author} {\bibfnamefont {O.~A.}\ \bibnamefont
  {Ponkratenko}}, \bibinfo {author} {\bibfnamefont {V.~I.}\ \bibnamefont
  {Tretyak}}, \ and\ \bibinfo {author} {\bibfnamefont {Y.~G.}\ \bibnamefont
  {Zdesenko}},\ }\href {\doibase 10.1134/1.855784} {\bibfield  {journal}
  {\bibinfo  {journal} {Physics of Atomic Nuclei}\ }\textbf {\bibinfo {volume}
  {63}},\ \bibinfo {pages} {1282} (\bibinfo {year} {2000})}\BibitemShut
  {NoStop}%
\bibitem [{\citenamefont {Aalseth}\ \emph {et~al.}(2018)\citenamefont {Aalseth}
  \emph {et~al.}}]{mjd2018}%
  \BibitemOpen
  \bibfield  {author} {\bibinfo {author} {\bibfnamefont {C.~E.}\ \bibnamefont
  {Aalseth}} \emph {et~al.} (\bibinfo {collaboration} {Majorana
  Collaboration}),\ }\href {\doibase 10.1103/PhysRevLett.120.132502} {\bibfield
   {journal} {\bibinfo  {journal} {Phys. Rev. Lett.}\ }\textbf {\bibinfo
  {volume} {120}},\ \bibinfo {pages} {132502} (\bibinfo {year}
  {2018})}\BibitemShut {NoStop}%
\bibitem [{\citenamefont {Buuck}(2019)}]{buuck2019}%
  \BibitemOpen
  \bibfield  {author} {\bibinfo {author} {\bibfnamefont {M.}~\bibnamefont
  {Buuck}},\ }\emph {\bibinfo {title} {A Radiogenic Background Model for the
  \textsc{Majorana Demonstrator}}},\ \href@noop {} {Ph.D. thesis},\ \bibinfo
  {school} {University of Washington} (\bibinfo {year} {2019})\BibitemShut
  {NoStop}%
\bibitem [{\citenamefont {{Akaike}}(1974)}]{Akaike1974}%
  \BibitemOpen
  \bibfield  {author} {\bibinfo {author} {\bibfnamefont {H.}~\bibnamefont
  {{Akaike}}},\ }\href {\doibase 10.1109/TAC.1974.1100705} {\bibfield
  {journal} {\bibinfo  {journal} {IEEE Transactions on Automatic Control}\
  }\textbf {\bibinfo {volume} {19}},\ \bibinfo {pages} {716} (\bibinfo {year}
  {1974})}\BibitemShut {NoStop}%
\bibitem [{\citenamefont {Guinn}\ \emph {et~al.}(2020)\citenamefont {Guinn}
  \emph {et~al.}}]{Guinn_2020}%
  \BibitemOpen
  \bibfield  {author} {\bibinfo {author} {\bibfnamefont {I.~S.}\ \bibnamefont
  {Guinn}} \emph {et~al.} (\bibinfo {collaboration} {Majorana Collaboration}),\
  }\href {\doibase 10.1088/1742-6596/1468/1/012115} {\bibfield  {journal}
  {\bibinfo  {journal} {Journal of Physics: Conference Series}\ }\textbf
  {\bibinfo {volume} {1468}},\ \bibinfo {pages} {012115} (\bibinfo {year}
  {2020})}\BibitemShut {NoStop}%
\bibitem [{\citenamefont {Rolke}\ \emph {et~al.}(2005)\citenamefont {Rolke},
  \citenamefont {López},\ and\ \citenamefont {Conrad}}]{Rolke2005}%
  \BibitemOpen
  \bibfield  {author} {\bibinfo {author} {\bibfnamefont {W.~A.}\ \bibnamefont
  {Rolke}}, \bibinfo {author} {\bibfnamefont {A.~M.}\ \bibnamefont {López}}, \
  and\ \bibinfo {author} {\bibfnamefont {J.}~\bibnamefont {Conrad}},\ }\href
  {\doibase https://doi.org/10.1016/j.nima.2005.05.068} {\bibfield  {journal}
  {\bibinfo  {journal} {Nuclear Instruments and Methods in Physics Research
  Section A: Accelerators, Spectrometers, Detectors and Associated Equipment}\
  }\textbf {\bibinfo {volume} {551}},\ \bibinfo {pages} {493 } (\bibinfo {year}
  {2005})}\BibitemShut {NoStop}%
\bibitem [{\citenamefont {Morales}\ \emph {et~al.}(2008)\citenamefont
  {Morales}, \citenamefont {Morales}, \citenamefont {Núñez-Lagos},
  \citenamefont {Puimedón}, \citenamefont {Villar},\ and\ \citenamefont
  {Larrea}}]{morales1988}%
  \BibitemOpen
  \bibfield  {author} {\bibinfo {author} {\bibfnamefont {A.}~\bibnamefont
  {Morales}}, \bibinfo {author} {\bibfnamefont {J.}~\bibnamefont {Morales}},
  \bibinfo {author} {\bibfnamefont {R.}~\bibnamefont {Núñez-Lagos}}, \bibinfo
  {author} {\bibfnamefont {J.}~\bibnamefont {Puimedón}}, \bibinfo {author}
  {\bibfnamefont {J.}~\bibnamefont {Villar}}, \ and\ \bibinfo {author}
  {\bibfnamefont {A.}~\bibnamefont {Larrea}},\ }\href {\doibase
  https://doi.org/10.1007/BF02789497} {\bibfield  {journal} {\bibinfo
  {journal} {Nuovo Cim. A}\ }\textbf {\bibinfo {volume} {100}},\ \bibinfo
  {pages} {525} (\bibinfo {year} {2008})}\BibitemShut {NoStop}%
\bibitem [{\citenamefont {Maier}(1994)}]{maier1994}%
  \BibitemOpen
  \bibfield  {author} {\bibinfo {author} {\bibfnamefont {B.}~\bibnamefont
  {Maier}} (\bibinfo {collaboration} {Heidelberg-Moscow Collaboration}),\
  }\href {\doibase https://doi.org/10.1016/0920-5632(94)90276-3} {\bibfield
  {journal} {\bibinfo  {journal} {Nuclear Physics B - Proceedings Supplements}\
  }\textbf {\bibinfo {volume} {35}},\ \bibinfo {pages} {358 } (\bibinfo {year}
  {1994})}\BibitemShut {NoStop}%
\bibitem [{\citenamefont {Barabash}\ \emph
  {et~al.}(1995{\natexlab{b}})\citenamefont {Barabash}, \citenamefont {Derbin},
  \citenamefont {Popeko},\ and\ \citenamefont {Umatov}}]{barabash1995:2}%
  \BibitemOpen
  \bibfield  {author} {\bibinfo {author} {\bibfnamefont {A.~S.}\ \bibnamefont
  {Barabash}}, \bibinfo {author} {\bibfnamefont {A.~V.}\ \bibnamefont
  {Derbin}}, \bibinfo {author} {\bibfnamefont {L.~A.}\ \bibnamefont {Popeko}},
  \ and\ \bibinfo {author} {\bibfnamefont {V.~I.}\ \bibnamefont {Umatov}},\
  }\href {\doibase https://doi.org/10.1007/BF01298913} {\bibfield  {journal}
  {\bibinfo  {journal} {Z. Phys. A}\ }\textbf {\bibinfo {volume} {352}},\
  \bibinfo {pages} {231} (\bibinfo {year} {1995}{\natexlab{b}})}\BibitemShut
  {NoStop}%
\bibitem [{\citenamefont {Coello~P\'erez}\ \emph {et~al.}(2018)\citenamefont
  {Coello~P\'erez}, \citenamefont {Men\'endez},\ and\ \citenamefont
  {Schwenk}}]{menendez2018}%
  \BibitemOpen
  \bibfield  {author} {\bibinfo {author} {\bibfnamefont {E.~A.}\ \bibnamefont
  {Coello~P\'erez}}, \bibinfo {author} {\bibfnamefont {J.}~\bibnamefont
  {Men\'endez}}, \ and\ \bibinfo {author} {\bibfnamefont {A.}~\bibnamefont
  {Schwenk}},\ }\href {\doibase 10.1103/PhysRevC.98.045501} {\bibfield
  {journal} {\bibinfo  {journal} {Phys. Rev. C}\ }\textbf {\bibinfo {volume}
  {98}},\ \bibinfo {pages} {045501} (\bibinfo {year} {2018})}\BibitemShut
  {NoStop}%
\bibitem [{\citenamefont {Barea}\ \emph {et~al.}(2015)\citenamefont {Barea},
  \citenamefont {Kotila},\ and\ \citenamefont {Iachello}}]{barea2015}%
  \BibitemOpen
  \bibfield  {author} {\bibinfo {author} {\bibfnamefont {J.}~\bibnamefont
  {Barea}}, \bibinfo {author} {\bibfnamefont {J.}~\bibnamefont {Kotila}}, \
  and\ \bibinfo {author} {\bibfnamefont {F.}~\bibnamefont {Iachello}},\ }\href
  {\doibase 10.1103/PhysRevC.91.034304} {\bibfield  {journal} {\bibinfo
  {journal} {Phys. Rev. C}\ }\textbf {\bibinfo {volume} {91}},\ \bibinfo
  {pages} {034304} (\bibinfo {year} {2015})}\BibitemShut {NoStop}%
\bibitem [{\citenamefont {Toivanen}\ and\ \citenamefont
  {Suhonen}(1997)}]{Toivanen1997}%
  \BibitemOpen
  \bibfield  {author} {\bibinfo {author} {\bibfnamefont {J.}~\bibnamefont
  {Toivanen}}\ and\ \bibinfo {author} {\bibfnamefont {J.}~\bibnamefont
  {Suhonen}},\ }\href {\doibase 10.1103/PhysRevC.55.2314} {\bibfield  {journal}
  {\bibinfo  {journal} {Phys. Rev. C}\ }\textbf {\bibinfo {volume} {55}},\
  \bibinfo {pages} {2314} (\bibinfo {year} {1997})}\BibitemShut {NoStop}%
\bibitem [{\citenamefont {Schwieger}\ \emph {et~al.}(1998)\citenamefont
  {Schwieger}, \citenamefont {Simkovic}, \citenamefont {Faessler},\ and\
  \citenamefont {Kami\ifmmode~\acute{n}\else \'{n}\fi{}ski}}]{Schwieger1998}%
  \BibitemOpen
  \bibfield  {author} {\bibinfo {author} {\bibfnamefont {J.}~\bibnamefont
  {Schwieger}}, \bibinfo {author} {\bibfnamefont {F.}~\bibnamefont {Simkovic}},
  \bibinfo {author} {\bibfnamefont {A.}~\bibnamefont {Faessler}}, \ and\
  \bibinfo {author} {\bibfnamefont {W.~A.}\ \bibnamefont
  {Kami\ifmmode~\acute{n}\else \'{n}\fi{}ski}},\ }\href {\doibase
  10.1103/PhysRevC.57.1738} {\bibfield  {journal} {\bibinfo  {journal} {Phys.
  Rev. C}\ }\textbf {\bibinfo {volume} {57}},\ \bibinfo {pages} {1738}
  (\bibinfo {year} {1998})}\BibitemShut {NoStop}%
\bibitem [{\citenamefont {Raduta}\ and\ \citenamefont
  {Raduta}(2007)}]{Raduta2007}%
  \BibitemOpen
  \bibfield  {author} {\bibinfo {author} {\bibfnamefont {C.~M.}\ \bibnamefont
  {Raduta}}\ and\ \bibinfo {author} {\bibfnamefont {A.~A.}\ \bibnamefont
  {Raduta}},\ }\href {\doibase 10.1103/PhysRevC.76.044306} {\bibfield
  {journal} {\bibinfo  {journal} {Phys. Rev. C}\ }\textbf {\bibinfo {volume}
  {76}},\ \bibinfo {pages} {044306} (\bibinfo {year} {2007})}\BibitemShut
  {NoStop}%
\bibitem [{\citenamefont {Unlu}(2014)}]{Unlu2014}%
  \BibitemOpen
  \bibfield  {author} {\bibinfo {author} {\bibfnamefont {S.}~\bibnamefont
  {Unlu}},\ }\href {\doibase 10.1088/0256-307x/31/4/042101} {\bibfield
  {journal} {\bibinfo  {journal} {Chinese Physics Letters}\ }\textbf {\bibinfo
  {volume} {31}},\ \bibinfo {pages} {042101} (\bibinfo {year}
  {2014})}\BibitemShut {NoStop}%
\bibitem [{\citenamefont {Hyv\"arinen}\ and\ \citenamefont
  {Suhonen}(2016)}]{Hyvarinen2016}%
  \BibitemOpen
  \bibfield  {author} {\bibinfo {author} {\bibfnamefont {J.}~\bibnamefont
  {Hyv\"arinen}}\ and\ \bibinfo {author} {\bibfnamefont {J.}~\bibnamefont
  {Suhonen}},\ }\href {\doibase 10.1103/PhysRevC.93.064306} {\bibfield
  {journal} {\bibinfo  {journal} {Phys. Rev. C}\ }\textbf {\bibinfo {volume}
  {93}},\ \bibinfo {pages} {064306} (\bibinfo {year} {2016})}\BibitemShut
  {NoStop}%
\bibitem [{\citenamefont {Menéndez}\ \emph {et~al.}(2009)\citenamefont
  {Menéndez}, \citenamefont {Poves}, \citenamefont {Caurier},\ and\
  \citenamefont {Nowacki}}]{menendez2009}%
  \BibitemOpen
  \bibfield  {author} {\bibinfo {author} {\bibfnamefont {J.}~\bibnamefont
  {Menéndez}}, \bibinfo {author} {\bibfnamefont {A.}~\bibnamefont {Poves}},
  \bibinfo {author} {\bibfnamefont {E.}~\bibnamefont {Caurier}}, \ and\
  \bibinfo {author} {\bibfnamefont {F.}~\bibnamefont {Nowacki}},\ }\href
  {\doibase https://doi.org/10.1016/j.nuclphysa.2008.12.005} {\bibfield
  {journal} {\bibinfo  {journal} {Nuclear Physics A}\ }\textbf {\bibinfo
  {volume} {818}},\ \bibinfo {pages} {139 } (\bibinfo {year}
  {2009})}\BibitemShut {NoStop}%
\end{thebibliography}%

\end{document}